\documentclass[pra, notitlepage, showpacs, floatfix, superscriptaddress, reprint, nobalancelastpage]{revtex4-1}
\usepackage{amsmath}
\usepackage{amssymb}
\usepackage{amsthm}
\usepackage{amsfonts}
\usepackage{verbatim}
\usepackage{listings}
\usepackage{enumerate}
\usepackage{latexsym}
\usepackage{psfrag}
\usepackage{bm}
\usepackage{dsfont}
\usepackage[all]{xy}
\usepackage{graphicx}
\usepackage{color}
\usepackage{mathtools}
\usepackage{eufrak}
\usepackage[percent]{overpic}
\usepackage{epsfig,slashed}
\usepackage{epstopdf}
\usepackage{lipsum}
\usepackage{float}
\usepackage{mathtools}
\allowdisplaybreaks
\usepackage[colorlinks=true]{hyperref}  
\hypersetup{
    bookmarks=true,         
    unicode=false,          
    pdftoolbar=true,        
    pdfmenubar=true,        
    pdffitwindow=false,     
    pdfstartview={FitH},    
    pdftitle={Thermal Hall effect in square-lattice spin liquids: A Schwinger-boson mean-field study},    
    pdfauthor={Rhine Samajdar et al.},     
    pdfsubject={},   
    pdfcreator={},   
    pdfproducer={}, 
    pdfkeywords={} {} {}, 
    pdfnewwindow=true,      
    colorlinks=true,       
    linkcolor=blue, 
    citecolor=blue,        
    filecolor=blue,      
    urlcolor=blue           
} 

\usepackage{setspace}
\usepackage{natbib}
\usepackage{subfigure}
\usepackage[mathscr]{euscript}
\usepackage[export]{adjustbox}
\usepackage[bottom]{footmisc}
\usepackage{textcomp}
\usepackage{braket}
\makeatletter
\newsavebox\myboxA
\newsavebox\myboxB
\newlength\mylenA

\newcommand*\xoverline[2][0.75]{%
    \sbox{\myboxA}{$\m@th#2$}%
    \setbox\myboxB\null
    \ht\myboxB=\ht\myboxA%
    \dp\myboxB=\dp\myboxA%
    \wd\myboxB=#1\wd\myboxA
    \sbox\myboxB{$\m@th\overline{\copy\myboxB}$}
    \setlength\mylenA{\the\wd\myboxA}
    \addtolength\mylenA{-\the\wd\myboxB}%
    \ifdim\wd\myboxB<\wd\myboxA%
       \rlap{\hskip 0.5\mylenA\usebox\myboxB}{\usebox\myboxA}%
    \else
        \hskip -0.5\mylenA\rlap{\usebox\myboxA}{\hskip 0.5\mylenA\usebox\myboxB}%
    \fi}
\makeatother

\usepackage{latexsym}
\DeclareMathOperator{\Tr}{Tr}
\def \beq{\begin{eqnarray}}
\def \eeq{\end{eqnarray}}

\def \mc{\mathcal}

\def \Q{{\bm Q}}
\def \S{{\bm S}}
\def \r{{\bm r}}

\def \n{{\bm n}}
\def \k{{\mathbf{k}}}

\newcommand{\nn}{\nonumber \\}
\newcommand{\TM}{\mathscr{T}} 

\newcommand{\pdagger}{{\phantom{\dagger}}}
\renewcommand{\vec}[1]{\boldsymbol{#1}}
\newcommand{\figref}[1]{Fig.~\ref{#1}}
\newcommand{\equref}[1]{Eq.~(\ref{#1})}
\newcommand{\equsref}[2]{Eqs.~(\ref{#1}) and (\ref{#2})}
\newcommand{\secref}[1]{Sec.~\ref{#1}}
\newcommand{\refcite}[1]{Ref.~\onlinecite{#1}}

\DeclareGraphicsExtensions{.png}

\begin{document}
\title{{\textbf{Thermal Hall effect in square-lattice spin liquids: A Schwinger boson mean-field study}}}
\author{Rhine Samajdar}
\affiliation{Department of Physics, Harvard University, Cambridge, MA
02138, USA}
\author{Shubhayu Chatterjee}
\affiliation{Department of Physics, Harvard University, Cambridge, MA
02138, USA}
\affiliation{Department of Physics, University of California, Berkeley, California 94720, USA.}
\author{Subir Sachdev}
\affiliation{Department of Physics, Harvard University, Cambridge, MA
02138, USA}
\affiliation{Perimeter Institute for Theoretical Physics, Waterloo, Ontario N2L 2Y5, Canada}
\author{Mathias S. Scheurer}
\affiliation{Department of Physics, Harvard University, Cambridge, MA
02138, USA}

\date{\today \\
}

\begin{abstract}
Motivated by recent transport measurements in high-$T_c$ cuprate superconductors in a magnetic field, we study the thermal Hall conductivity in materials with topological order, focusing on the contribution from neutral spinons. Specifically, different Schwinger boson mean-field ans\"{a}tze for the Heisenberg antiferromagnet on the square lattice are analyzed. We allow for both Dzyaloshinskii-Moriya interactions, and additional terms associated with scalar spin chiralities that break time-reversal and reflection symmetries, but preserve their product.
It is shown that these scalar spin chiralities, which can either arise spontaneously or are induced by the orbital coupling of the magnetic field, can lead to spinon bands with nontrivial Chern numbers and significantly enhanced thermal Hall conductivity. Associated states with zero-temperature magnetic order, which is thermally fluctuating at any $T>0$, also show a similarly enhanced thermal Hall conductivity. 
\end{abstract}

\maketitle

\section{Introduction}
The Wiedemann-Franz (WF) law is a paradigmatic property of a metal that relates its electrical conductivity tensor $\hat{\sigma}$ to its thermal conductivity tensor $\hat{\kappa}$ at temperature $T$ as $\hat{\kappa}/T = L_0 \hat{\sigma}$, where $L_0 = \pi^2 k_B^2/(3 e^2)$ is the Lorenz number \cite{ziman1960EPbook}. Recent studies of the metallic state of high-$T_c$ cuprate superconductors, such as La$_{1.6-x}$Nd$_{0.4}$Sr$_x$CuO$_4$ (Nd-LSCO), obtained by suppressing superconductivity using magnetic fields, indicate a very interesting trend in the thermal Hall coefficient \cite{grissonnanche2019giant} as a function of doping. On the overdoped side, with a hole doping of $p > p^*$, where $p^*$ corresponds to the doping value where the pseudogap temperature vanishes, the thermal Hall conductivity $\kappa^{}_{xy}$ obeys the WF law for low $T$. However, for hole doping $p < p^*$, corresponding to the pseudogap phase, the thermal Hall conductivity changes sign and becomes negative, while $\sigma_{xy}$ remains positive. Further, the magnitude of $\kappa^{}_{xy}/(T \sigma_{xy})$ at low temperatures significantly exceeds $L_0$, thus signaling a comprehensive breakdown of the WF law.

A possible explanation of this observation is the presence of charge-neutral spin-carrying excitations in the pseudogap phase. By virtue of being electrically neutral, they do not couple to the external electromagnetic field and, by association, do not contribute to $\sigma_{xy}$; however, they give rise to a thermal Hall current leading to the violation of the WF law in Hall conductivities.
The large $\kappa_{xy}$ observed at dopings with and without N\'eel order suggests that magnons are not responsible for this phenomenon. 
Further, \citet{grissonnanche2019giant} argue that the observed magnitude of $\kappa^{}_{xy}$ at low temperatures is too large to be explained by spin-scattered phonons. This prompts the rather intriguing possibility of emergent neutral excitations that are responsible for this unusual behavior.

\begin{figure}[b]
    \centering
    \includegraphics[width=0.7\linewidth,trim={0cm 0cm 0cm 0cm}, clip]{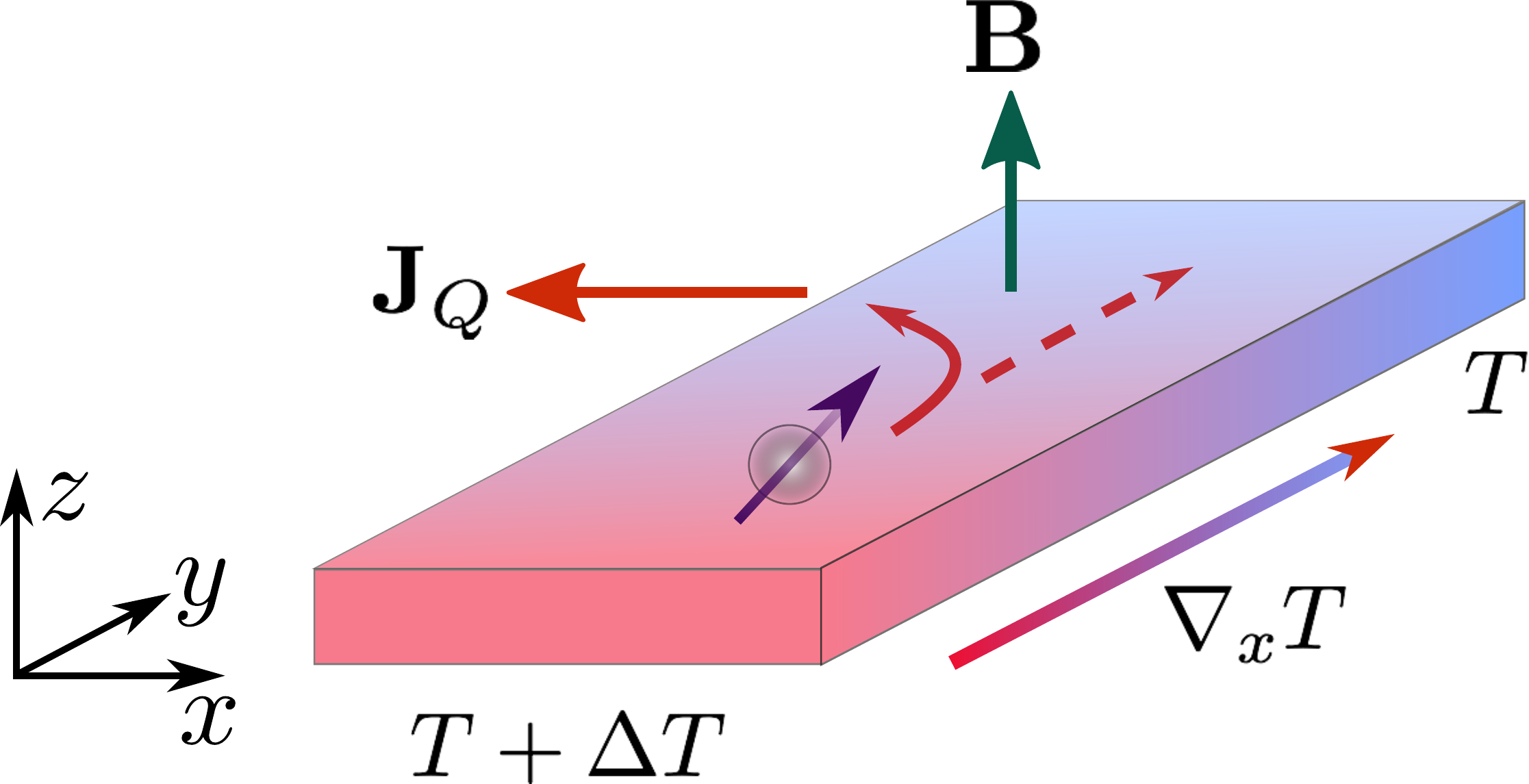}
    \caption{Schematic depiction of the thermal Hall effect in an insulator with topological order, where the heat current is carried by fractionalized $S = 1/2$ spinons.}
    \label{fig:THschem}
\end{figure}

In this paper, we investigate the thermal Hall conductivity (see \figref{fig:THschem}) of phases where the electron fractionalizes into an electrically charged gapless fermionic chargon and a gapped charge-neutral spin-carrying spinon \cite{SachdevZ2Rev}. Such a phase of matter has topological order \cite{Pnas2018}, and has been previously discussed in the context of the pseudogap metal \cite{CSE17,CSS17,SS09,SCSS17,scheurer2018orbital,PhysRevX.8.021048}. Indeed, model calculations of the longitudinal conductivities and the electrical Hall conductivity in these fractionalized phases \cite{CSE17} are consistent with experimental observations in the metallic phases of several cuprates. However, Ref.~\onlinecite{grissonnanche2019giant} shows that the large negative $\kappa^{}_{xy}$ persists even in the insulating phase as the doping $p \rightarrow 0$. This is the extreme limit of breakdown of the WF law, as $\sigma_{xy} = 0$. Motivated by this observation, we restrict our focus to Mott insulators with gapped chargons and topological order, analogous to the phases discussed in Refs.~\onlinecite{CSS17,scheurer2018orbital}, and compute the contribution to the thermal Hall effect from deconfined, charge-neutral, spinons.

Our first set of results is related to 
the thermal Hall conductivity in square-lattice spin-liquid states with nonzero scalar spin chiralities, $\chi^{}_{ijk}=\vec{S}_i \cdot (\vec{S}_j \times \vec{S}_k)$, where $\vec{S}_i$ is the spin operator on site $i=(i_x,i_y)\in \mathbb{Z}^2$ of the square lattice, 
but {\it without\/} any spin-orbit coupling; these results are presented in Section~\ref{sec:1O}. Note that, by virtue of being odd under time reversal and spin rotation invariant, $\chi_{ijk}$ can couple to bond-current operators and, hence, these states are in general associated with nonzero loop currents.
A recent paper \cite{scheurer2018orbital} classified four likely patterns (labeled A,B,C,D)  of time-reversal and mirror-plane symmetry breaking in spin liquids with nonzero $\chi_{ijk}$ and associated loop currents. Among these, only pattern D has a nonzero $\kappa^{}_{xy}$ and hence, will be the center of our attention. 
We will find that spin liquids of pattern D, which breaks square-lattice and time-reversal symmetries down to $\frac{4}{m}m'm'$, do indeed lead to values of $\kappa^{}_{xy}/T$ of order $k_B^2/\hbar$ at temperatures above the spin gap; below the spin gap, $\kappa^{}_{xy}/T$ vanishes exponentially as $T \rightarrow 0$ [see Eq.~\eqref{eq:kappaVsBatLowT}]. The reduction of the symmetry to $\frac{4}{m}m'm'$ could either be spontaneous, or simply due to the presence of an applied magnetic field. We note that, in the latter case, no hysteresis in the thermal Hall conductance is expected. As we review in Appendix~\ref{app:orbital}, the orbital coupling of the field in a Hubbard-type model induces a coupling between the magnetic field and the uniform scalar spin chirality.  

We also probe the thermal Hall conductivity of the associated magnetically ordered states which break spin rotation symmetry at $T=0$. In two spatial dimensions, spin rotation invariance is restored at any nonzero temperature by thermal fluctuations, and this allows us to treat such states with the same formalism as that used for spin liquids. For such thermally disordered descendants of magnetically ordered states we also find values of $\kappa^{}_{xy}/T$ of order $k_B^2/\hbar$, but $\kappa^{}_{xy}/T$ vanishes as a power of $T$ as $T \rightarrow 0$ [see Eq.~\eqref{kappaMagneticOrder}].

Although these results appear to be an attractive model of the observations on the cuprates, it is important to keep a caveat in mind. In the limit where full square lattice and time-reversal symmetries are restored, our Schwinger boson states can undergo phase transitions to a variety of possible magnetically ordered states, and the observed N\'eel state is only one among a continuum of possibilities; see Appendix~\ref{MagneticOrder}.
At least at the mean-field level, there is no selection mechanism for the N\'eel state when the time-reversal symmetry breaking to $\frac{4}{m}m'm'$ is turned off. Nevertheless, a weakly distorted N\'eel state is indeed one of the possible states leading to a large $\kappa^{}_{xy}/T$.

The second set of conclusions in this paper pertain to the influence of the spin-orbit coupling, which induces Dzyaloshinskii-Moriya (DM) terms in the spin Hamiltonian. We study the DM term in spin liquids connected to the N\'eel state, and find that it induces a significantly smaller value of $\kappa^{}_{xy}/T$, as described in Section~\ref{sec:DM}.

Our starting point is a Mott insulator where the low-energy degrees of freedom are the $S = 1/2$ spins of the Cu atoms located on a square lattice, with a Hamiltonian of the form
\begin{align}
\label{eq:GeneralSpinHamWithDMTerm}\begin{split}
H_{\mathrm{spin}} &= \frac{1}{2}\sum_{i,j} \left ( J_{ij}\, \mathbf{S}_i\cdot\mathbf{S}_j + \vec{D}^\textsc{m}_{ij} \cdot \mathbf{S}_i \times \mathbf{S}_j \right) \\ &\quad - \sum_i \vec{B}_Z\cdot  \mathbf{S}_i + H_{\chi}.
\end{split}\end{align}
The Heisenberg couplings $J_{ij}$ are taken to be positive, $J_{ij}>0$, and spatially local. 
The orientation of the external magnetic field is assumed to be perpendicular to the lattice plane (see Fig.~\ref{fig:THschem}). For the Zeeman field, we have $\vec{B}_Z = B_z \hat{z}$, where we have absorbed the Bohr magneton $\mu_B$ in the definition of $B_z$. The associated orbital coupling is described by $H_{\chi}$ which involves third-order (and higher-order) powers in $\mathbf{S}_i$ (see Appendix~\ref{app:orbital}). We also include a spin-orbit-induced Dzyaloshinskii-Moriya (DM) term, which is allowed when certain spatial symmetries are broken. The precise orientations of the DM coupling vectors $\vec{D}^\textsc{m}_{ij}$ will be described below.

To treat $H_{\mathrm{spin}}$, we adopt a Schwinger boson mean-field approach, which is capable of describing both spin-liquid phases and ordered antiferromagnets
\cite{auerbach1994interacting, Auerbach2011}. This approach, as detailed later, provides us with a mean-field ansatz, and the projective action of lattice or time-reversal symmetries on the ansatz describes the particular spin-liquid state under consideration \cite{WenSqLattice,yang2016schwinger}.  Among the different ans\"atze we consider, only one, for which all in-plane reflection symmetries are broken (pattern D in Ref.~\onlinecite{scheurer2018orbital}), leads to spinon bands with nonzero Chern numbers.

In previous literature, the thermal Hall effect has been widely investigated on the kagom\'{e} \cite{owerre2017topological, owerre2017topological2, mook2016spin, seshadri2018topological}, pyrochlore \cite{li2016weyl}, and honeycomb \cite{owerre2016magnon, owerre2016topological, owerre2017topological3, lu2018topological, Zhang2018spin} lattices for insulating phases with and without long-range magnetic order and in the presence of additional electric field gradients \cite{nakata2017magnonic1}. However, it is strongly constrained by no-go theorems on the square lattice owing to the geometry thereof; the fluctuation of the scalar spin chirality averaged over nearby elementary plaquettes in the square lattice vanishes for a generic phase \cite{katsura2010theory, ideue2012effect}. For our model, the discrete broken symmetries are carefully chosen such that the associated loop current pattern corresponds to a net addition of spin chirality on neighboring triangular plaquettes \cite{scheurer2018orbital}. This enables our model to overcome the symmetry barriers associated with the square lattice. This can be achieved since we consider a Schwinger boson mean-field ansatz (illustrated schematically in \figref{SBAnsatzPatternDOneObital}) that is not smoothly connected to that of the usual N\'eel state (which has topologically trivial bands). Rather, our ansatz can be viewed as a perturbation to the symmetric bosonic $\pi$-flux spin liquid \cite{yang2016schwinger}. As we show in the paper, these perturbations can indeed induce nonzero Chern numbers and lead to a much larger $\kappa^{}_{xy}$ compared to other phases with topologically trivial spinon bands. At the same time, as already noted above, the associated magnetically ordered phase can still be (a small deformation of) the N\'eel state.

We begin in Sec.~\ref{sec:formalism} by setting up the Schwinger-boson mean-field formalism and its computation of the thermal Hall conductivity. Section~\ref{sec:1O} evaluates the thermal Hall effect in spin liquids with nontrivial magnetic point groups but full SU(2) spin-rotation invariance (SRI). The DM term is not included in these analyses, but is considered separately in Sec.~\ref{sec:DM} (without the additional time-reversal symmetry-breaking terms of \secref{sec:1O}). Finally, \secref{Conclusion} summarizes the results and four Appendices, \ref{app:orbital}--\ref{3OrbitalModel}, detail our calculations.

\section{Formalism}
\label{sec:formalism}

In order to compute the thermal Hall conductivity, one needs to first know the nature of the low-energy excitations above the quantum ground state of $H_{\mathrm{spin}}$. An approximate method to treat this problem is provided by Schwinger boson mean-field theory (SBMFT) in which the Hamiltonian is written in terms of Schwinger bosons \cite{auerbach1994interacting, Auerbach2011}, whereupon an appropriate mean-field decoupling renders it quadratic. We briefly review this formalism in the context of the thermal Hall effect below.

\subsection{Schwinger-boson mean-field theory}

The spin operator can be represented at each site $i=(i_x,i_y)\in \mathbb{Z}^2$ of the square lattice (we set $a=1$ for the lattice constant) using a pair of bosons $(b_{i \uparrow}, b_{i \downarrow})$ as
\begin{equation}
\label{eq:SB}
\mathbf{S}_i = \frac{1}{2} \sum_{\sigma,\sigma'} b^\dagger_{i \sigma}\, \vec{\sigma}^{}_{\sigma \sigma'}\, b^{}_{i \sigma'},
\end{equation}
where $\vec{\sigma}=(\sigma_1,\sigma_2,\sigma_3)^T$ is a vector of Pauli matrices. These operators satisfy the standard bosonic commutation relations $[b_{i \sigma}, b^\dagger_{j \sigma'}] = \delta_{ij} \delta_{\sigma \sigma'}$. This construction enlarges the on-site Hilbert space; to remain within the physical space, Eq.~\eqref{eq:SB} has to be supplemented with the local holonomic constraint
\begin{equation}
\label{eq:constraint }
\hat{n}_i = \sum_{\sigma} b^\dagger_{i\sigma}b^{}_{i\sigma} = 2\mathrm{S}, 
\end{equation}
which enforces that $\mathbf{S}_i^2 = \mathrm{S}\, (\mathrm{S}+1)$.

In this fashion, 
the reformulated Hamiltonian $H_{\mathrm{spin}}$ contains only quadratic, quartic, and sextic terms in the bosonic operators. Now, we perform a mean-field decoupling of $H_{\mathrm{spin}}$ into quadratic operators. We neglect here the DM interactions, which will be analyzed in \secref{sec:DM} and Appendix \ref{sec:DM_Calc}, and the orbital coupling $H_{\chi}$, which will be discussed in Appendix~\ref{app:orbital}; for now, we concentrate on terms that preserve SRI. The only such operators are the spin singlets
\begin{alignat}{2}
\hat{\mc{A}}_{i,j} &= \frac{1}{2} \sum_{\sigma,\sigma'} b_{i \sigma} (\mathrm{i}\sigma_2)_{\sigma\sigma'} b_{j \sigma'}; \quad &&\hat{\mc{A}}_{j,i} = - \hat{\mc{A}}_{i,j}, \label{AOperator}\\
\hat{\mc{B}}_{i,j} &= \frac{1}{2} \sum_\sigma \,\,b^{}_{i \sigma}\, b^\dagger_{j \sigma}; \quad &&\hat{\mc{B}}^{}_{j,i} = \hat{\mc{B}}^\dagger_{i,j}, \label{BOperator}
\end{alignat}and their adjoints. Here and in the following, we use $\mathrm{i}$ to denote the imaginary unit. The expectation values, $\{\mc{A}_{i,j}, \mc{B}_{i,j} \}$, of the operators in \equsref{AOperator}{BOperator} collectively define the parameters of the mean-field \textit{ansatz}. 

First, let us examine the antiferromagnetic Heisenberg exchange term \cite{bauer2017schwinger} in a simple spin Hamiltonian:
\begin{equation}
\label{eq:H1}
H^{(1)} =  \sum_{i > j} J_{ij} \mathbf{S}_i\cdot\mathbf{S}_j; \quad J_{ij}>0.
\end{equation}
Using the identity
\begin{equation}
\mathbf{S}_i\cdot\mathbf{S}_j =\, :\hat{\mc{B}}^\dagger_{i,j} \hat{\mc{B}}^{}_{i,j} : - \hat{\mc{A}}^\dagger_{i,j} \hat{\mc{A}}^{}_{i,j} = \hat{\mc{B}}^\dagger_{i,j} \hat{\mc{B}}^{}_{i,j} - \hat{\mc{A}}^\dagger_{i,j} \hat{\mc{A}}^{}_{i,j} -\frac{1}{4} \hat{n}_i,
\end{equation}
with $: \,:$ denoting normal ordering, Eq.~\eqref{eq:H1} can be reduced to a mean-field quadratic bosonic Hamiltonian preserving SU$(2)$ spin-rotation invariance. This is achieved by neglecting bond operator fluctuations and replacing $\langle \hat{\mc{A}}_{i,j} \rangle$ and $\langle \hat{\mc{B}}_{i,j} \rangle$ by complex bond parameters $\mc{A}_{i,j}$ and $\mc{B}_{i,j}$, respectively:
\begin{alignat}{2}
\label{eq:H1mf}
&H_{\textsc{mf}}^{(1)} = \sum_{i > j, \sigma} \bigg[\frac{J_{ij}}{2} \left(\mc{B}_{i,j}^* b^{}_{i \sigma}\, b^\dagger_{j \sigma} - \mc{A}_{i,j}^*\sigma \,b^{}_{i \sigma}\, b_{j -\sigma} + \mathrm{H.c.} \right) \nonumber  \\
 & + J_{ij} \left( \lvert \mc{A}_{i,j} \rvert^2 - \lvert \mc{B}_{i,j} \rvert^2 \right)\bigg] + \lambda \sum_{i } \left(b^\dagger_{i \sigma} b^{}_{i \sigma} -  2 S \right). 
\end{alignat}
At the mean-field level, the local constraint \eqref{eq:constraint } is enforced
only on average, namely, $\langle \hat{n}_i \rangle = \kappa$ via the Lagrange multiplier $\lambda$. 
One could, in principle,  search for an optimal $\mc{A}_{i,j}$ and $\mc{B}_{i,j}$ by self-consistently solving for the stationary points of the mean-field free energy; however, for the purpose of this work, we simply treat  them as free (complex) parameters. The only constraints thereon come from the upper bounds \cite{messio2013time} on the moduli $\lvert \mc{A} \rvert \le \mathrm{S}+1/2$, $\lvert \mc{B} \rvert \le \mathrm{S}$, which must be obeyed for any self-consistent ansatz in SBMFT.

In the presence of a nonzero transverse magnetic field, spin-rotation invariance \textit{is} broken by the additional Zeeman term in the Hamiltonian:
\begin{equation}
\label{eq:H2mf}
H^{(2)} = - B_z \sum_{i} S_i^z = -\frac{B_z}{2} \sum_{i\, \sigma,\sigma'} b^\dagger_{i\sigma} (\sigma_3)_{\sigma\sigma'}b_{i\sigma'} = H^{(2)}_\textsc{mf}.\end{equation}
This term is already quadratic and thus requires no further decoupling.

Since we will discuss spin liquid phases with certain discrete broken symmetries, to be precise, let us clarify when a given ansatz breaks a symmetry. The physical spin operator is invariant under a local U(1) gauge transformation $b_j \rightarrow \mathrm{e}^{\mathrm{i}\varphi(j)} b_j$. Under such a gauge transformation, the mean-field ansatz transforms as \begin{equation}
\mc{A}_{i,j} \rightarrow \mathrm{e}^{\mathrm{i}[ \varphi(i) + \varphi(j)] }\mc{A}_{i,j}, ~~~ \mc{B}_{i,j} \rightarrow \mathrm{e}^{\mathrm{i}( \varphi(i) - \varphi(j)) }\mc{B}_{i,j}.
\end{equation}
Therefore, a symmetry $g$ is preserved as long as there is a gauge transformation, $b_j \rightarrow \mc{G}_g(j) b_j$, $ \mc{G}_g(j)=\mathrm{e}^{\mathrm{i}\varphi_g(j)}$, that leaves the ansatz invariant when combined with the action of the symmetry operation. Contrarily, if no such gauge transformation exists or, equivalently, there is some gauge-invariant operator that transforms nontrivially under $g$ and has a finite (nonzero) expectation value in the phase under consideration, then the symmetry $g$ is broken.

\subsection{Diagonalization of bosonic quadratic Hamiltonians}
\label{sec:DiagBH}

The mean-field Schwinger boson Hamiltonian can be diagonalized by the Bogoliubov-Valatin canonical transformation \cite{bogoliubov1947theory, valatin1958comments}. For illustrative purposes, consider a general quadratic bosonic Hamiltonian
\begin{equation}
\label{eq:modelH}
H = \frac{1}{2}\,\Psi^\dagger\, M\,\Psi;  \quad \Psi^\dagger = \left(b_1^\dagger,\,\ldots ,\, b_N^\dagger,\, b^{}_1,\,\ldots,\,b^{}_N \right).
\end{equation}
Generically, the index $n = 1, \ldots, N$ on $b^{}_n$ and $b_n^\dagger$ could label momentum, spin, or some other degrees of freedom. 
To find the eigenmodes corresponding to $M$, we introduce new annihilation (creation) operators $\gamma^{}_{m}$ ($\gamma^\dagger_{m}$) such that 
\begin{equation}
\Psi = \TM\,\Gamma; \quad \Gamma^\dagger \equiv \left(\gamma_1^\dagger,\,\ldots ,\, \gamma_N^\dagger,\, \gamma^{}_1,\,\ldots,\,\gamma^{}_N \right).
\end{equation} The standard bosonic commutation relations for both the $\Psi$ and $\Gamma$ fields are conveniently encapsulated in the matrix equation
\begin{equation}
\left[\Psi_i,\Psi_j^\dagger \right]  = \left[\Gamma_i, \Gamma_j^\dagger \right] = \left(\rho^{}_3\right)_{ij}; \quad \rho^{}_3 \equiv 
\left(
\begin{array}{cc}
\mathds{1}_{N \times N} & 0\\
0 & -\mathds{1}_{N\times N} 
\end{array}
\right).
\end{equation}
We choose $\TM$ such that the Hamiltonian \eqref{eq:modelH} becomes
\begin{equation}
\label{eq:ph}
H = \frac{1}{2}\,\Gamma^\dagger \,\TM^\dagger \,M\, \TM\, \Gamma; \quad 
\TM^\dagger \,M\, \TM = 
\left(
\begin{array}{cccc}
    \omega_{1} & 0 & \cdots & 0 \\
    0 & \omega_2 & \dots & 0\\
    \vdots & \vdots & \ddots & \vdots\\
    0 &   0& \cdots     & \omega^{}_{2N}
\end{array}
\right),
\end{equation}
for $\omega_i \in \mathbb{R}$. Meanwhile, to safeguard the bosonic statistics of the system, the transformation matrix must fulfill the necessary condition
\begin{equation}
\TM \,\rho^{}_3 \,\TM^\dagger = \rho^{}_3,
\end{equation}
or, in other words, $\TM$ is paraunitary \cite{colpa1978diagonalization}. The elements of the transformation $\TM$ can be obtained from the eigenvectors of the dynamic matrix $K = \rho^{}_3 M$, which defines the Heisenberg equation of motion for $\Psi$. All the eigenvalues of the dynamic matrix (when diagonalizable) are real and appear in pairs. Then, $\TM$, conventionally referred to as the derivative matrix, consists of all the eigenvectors of $K$ 
\begin{equation}
\label{eq:makeT}
\TM = \left[V\, (\omega^{}_1), \,\ldots, \,V \,(\omega^{}_N), \,V\, (-\omega^{}_1), \,\ldots, \,V\, (-\omega^{}_N) \right],
\end{equation}
with the eigenvectors $V$ ordered as 
\begin{equation}
V^\dagger (\omega_i) \,\rho^{}_3 \,V\, (\omega_i) = 1,\quad V^\dagger\, (-\omega_i) \,\rho^{}_3 \,V\,(-\omega_i) = -1
\end{equation}
for each set $\left(V\,(\omega_i), V\,(-\omega_i) \right)$. Thus, each eigenvalue of $K$ is counted up to its multiplicity and the $N$ dynamic mode pairs are separated and arranged sequentially as columns in $\TM$ such that its left (right) half is filled with eigenvectors of positive (negative) unit norms \cite{xiao2009theory}. Consequently,
\begin{alignat}{2}
\label{eq:diag}
\TM^{-1} K\, \TM &= \mathrm{diag}\,\left(\omega^{}_1,\, \ldots, \,\omega^{}_N,\, - \omega^{}_1,\, \ldots,\, -\omega^{}_N \right),\\
\TM^{\dagger} M\, \TM &= \mathrm{diag}\,\left(\omega^{}_1, \,\ldots, \,\omega^{}_N, \,\omega^{}_1,\, \ldots, \,\omega^{}_N \right),
\label{eq:dgn}
\end{alignat}
i.e. both $M$ and $K$ are simultaneously diagonalized. Borrowing fermionic terminology for Eq.~\eqref{eq:ph}, we refer to the bands with indices $n = 1, \ldots, N$ ($n = N + 1, , \ldots,2 N$) as the particle (hole) bands. 

\subsection{Berry curvature and thermal Hall conductivity}
\label{sec:procedure}

The prescription outlined above can be straightforwardly applied to the Hamiltonians in the sections hereafter, the only difference being that the matrices $\mathcal{H}({\k})$---associated with the mean-field Hamiltonian $H = \sum_{\k} (\Psi_{\k}^\dagger\, \mathcal{H}({\k})\, \Psi_{\k})/2 $---and $\TM_{\bf k}$ therein are momentum-dependent. Suppose $\varepsilon^{}_{n \mathbf{k}} > 0$ is the $n^\mathrm{th}$ band energy after such a diagonalization procedure; accordingly,
\begin{equation}
H = \sum_{\bf k} \sum_{n=1}^N \varepsilon^{}_{n \mathbf{k}} \bigg(\gamma^\dagger _{n {\bf k}}\gamma^{}_{n {\bf k}} + \frac{1}{2} \bigg).
\end{equation}
Then, within SBMFT, the thermal Hall conductivity in the clean limit  is given by \cite{matsumoto2014thermal}
\begin{equation}
\label{eq:k_xy}
\kappa^{}_{xy} = - \frac{k_B^2\,T}{\hbar\, V} \sum_{\mathbf{k}}\sum_{n=1}^N \left \{ c_2 \left[ n_B \left(\varepsilon_{n \mathbf{k}} \right)\right] - \frac{\pi^2}{3}\right\} \Omega_{n \mathbf{k}},
\end{equation}
where the sum on $n$ runs only over the particle bands. Here, $n_B (\varepsilon)$ is the Bose distribution function, and
\begin{alignat}{1}
\label{eq:c2}
c_2 (x) &\equiv \int_0^x \mathrm{d}\,t\,\left(\ln \frac{1+t}{t} \right)^2 \\
\nonumber &= (1+x)\left(\ln \frac{1+x}{x} \right)^2 - (\ln x)^2 - 2\, \mathrm{Li}_2 (-x),
\end{alignat}
which is monotonically increasing with $x$: it has a minimum value of $0$ at $x =0$ and, in the opposite limit, tends to $\pi^2/3$ as $x \rightarrow \infty$. $\Omega_{n\mathbf{k}}$ in \equref{eq:k_xy} is the Berry curvature in momentum space \cite{shindou2013topological}, which, for bosonic systems, is given by
\begin{equation}
\label{eq:berryc}
\Omega_{n \mathbf{k}} \equiv \mathrm{i}\,\epsilon_{\mu \nu} \left[\rho^{}_3 \,\frac{\partial\, \TM_{\bf k}^\dagger}{\partial\, k_\mu} \,\rho^{}_3\, \frac{\partial\, \TM_{\bf k}}{\partial\, k_\nu} \right]_{nn}; \quad n = 1, \ldots, N.
\end{equation}
The integral of the Berry curvature over the Brillouin zone (BZ) is the first Chern integer \cite{thouless1982quantized, kohmoto1985m}
\begin{equation}
C_n = \frac{1}{2\pi} \int_{\textsc{bz}} \mathrm{d} {\bf k}\,\, \Omega_{n {\bf k}} \in \mathbb{Z}.
\end{equation}
In addition to being integer valued, $C_n$ further obeys the constraint
\begin{equation}
\label{eq:C_constraint}
\sum_{n=1}^N C_n = \sum_{n=N+1}^{2N} C_n = 0,
\end{equation}
i.e., the sum of the Chern numbers over all particle and hole bands is individually zero \cite{shindou2013topological}. 
Since the expression in \equref{eq:k_xy} for $\kappa^{}_{xy}$ entails the summation over all particle bands and the momentum sum (or integral in the thermodynamic limit) is taken over a closed surface (the first Brillouin zone), Eq.~\eqref{eq:C_constraint} dictates that 
\begin{equation*}
    - \frac{k_B^2\,T}{\hbar\, V} \sum_{\mathbf{k}}\sum_{n=1}^N \left \{  - \frac{\pi^2}{3}\right\} \Omega_{n \mathbf{k}} = 0.
\end{equation*}
For this reason, we can neglect the additional $-\pi^2/3$ piece in the momentum sum in \equref{eq:k_xy} in the following.

It is worth noting that the derivation of the formula \eqref{eq:k_xy}, with the Berry curvature defined as in \equref{eq:berryc}, assumes that $\mathcal{H}({\k})$ has been chosen to satisfy the particle-hole symmetry 
\begin{equation}
\mathcal{H}({\k}) = \rho^{}_1 \left( \mathcal{H}({-\k})\right)^T  \rho^{}_1; \quad \rho^{}_1 \equiv 
\left(
\begin{array}{cc}
0 & \mathds{1}_{N \times N} \\
\mathds{1}_{N\times N} & 0
\end{array}
\right).
\end{equation}
As it will be useful below, we point out that, as a consequence, the Berry curvatures of the particle and hole bands are related as \cite{murakami2016thermal}
\begin{equation}
\label{eq:relate}
\Omega_{n+N, -{\bf k}} = - \Omega_{n {\bf k}}; \quad 1 \le n \le N.
\end{equation}

Before proceeding with the analysis of different spin-liquid states, a few general statements on the behavior of $\kappa^{}_{xy}$ are in order. First, if the temperature is much larger than the maximum energy of the $m^\mathrm{th}$ particle band so that $n_B \left(\varepsilon_{m \mathbf{k}} \right)\gg 1$, the contribution of this band to \equref{eq:k_xy} is related to its Chern number $C_m$ as
\begin{equation}
\left[\kappa^{}_{xy} \right]_m \approx \frac{\pi^2\,k_B^2\,T }{ 3\,\hbar} \int_\textsc{bz} \frac{\mathrm{d} \mathbf{k}}{4 \pi^2} \,\,\Omega_{m \mathbf{k}} = \frac{\pi \,k_B^2\,T }{ 6\,\hbar} C_m.
\end{equation}
Conversely, if $T$ lies far below the minimum of the $m^\mathrm{th}$ band, then $n_B \left(\varepsilon_{m \mathbf{k}} \right) \approx 0$ and its contribution to \equref{eq:k_xy} is exponentially small in the spinon gap divided by temperature (see also \equref{eq:c2Lims} below). 

As $\Omega_{n {\bf k}}$ is weighted by $c_2 \left[ n_B \left(\varepsilon_{n \mathbf{k}} \right)\right]$ in \equref{eq:k_xy}, there is a nonvanishing thermal Hall conductivity at finite temperatures even if all bands have zero Chern numbers.
The overall magnitude of $\kappa^{}_{xy}$, however, hinges on whether $C_n=0$ or $C_n\neq 0$. For a trivial band, the momentum-space average of the Berry curvature is itself zero and we generically expect that  ${\displaystyle \left[\kappa^{}_{xy} \right]_{m\vert\, C_m = 0} \ll \left[\kappa^{}_{xy} \right]_{m \vert\, C_m \ne 0}}$. As a result, the \textit{total} $\kappa^{}_{xy}$ is expected to be much smaller for a system with $C_n = 0\, \forall\, n$ than for one with nonzero Chern numbers. This is evident upon comparing Figs.~\ref{fig:1OTH} and \ref{fig:Hall_DM}, which correspond to conductivities arising from $C\ne 0$ and $C=0$ bands, respectively; for a similar set of parameters, the former are a thousandfold larger. We note that, in principle, it is possible that the Berry-curvature has significant energy dependence and, hence, $\kappa_{xy}$ is large even for $C_n=0$; however, such a situation was not realized for any of the ans\"atze we considered in this work.

\section{Spin liquid Ans\"{a}tze with time-reversal symmetry breaking}
\label{sec:1O}
Having established the necessity of Chern numbers for a sizable thermal Hall conductivity, we study spin liquid models that can yield such topologically nontrivial band structures within SBMFT. Inspired by the recent work of Ref.~\onlinecite{scheurer2018orbital} in the context of possible broken symmetries in cuprates, we examine states with nontrivial magnetic point groups. By breaking time-reversal symmetry while preserving SRI, the ans\"atze we discuss are naturally associated with nonzero scalar spin chiralities.

The simplest class of symmetry-breaking spin liquids of \refcite{scheurer2018orbital} are described by ans\"{a}tze that, while preserving all translational symmetries of the square lattice, have magnetic point group $m' mm$; this means that two-fold rotation perpendicular to the plane, $C_2$, and time-reversal symmetry, $\Theta$, are broken, but the product $\Theta C_2$ is preserved. Depending on whether the reflection symmetry along a Cu-O bond or along a diagonal Cu-Cu bond is present, these states are referred to as patterns A and B in \refcite{scheurer2018orbital}; they also appeared in studies of $\mathbb{Z}_2$ spin liquids using bosonic \cite{CSS17,SCSS17} and fermionic \cite{thomson2018fermionic} spinons.
However, as will be shown below, both these ans\"{a}tze lead to spinon bands which are topologically trivial, prompting the consideration of other patterns to procure nonzero Chern numbers.

To this end, we analyze a translationally invariant spin liquid phase, referred to as pattern D in \refcite{scheurer2018orbital}, that has magnetic point group $\frac{4}{m} m'm'$; this means that time-reversal symmetry and the point group $C_{4v}$ have been broken down to the symmetry group generated by fourfold rotation perpendicular to the plane, $C_4$, and $\Theta\mathcal{R}_x$ (the product of time-reversal $\Theta$ and reflection symmetry $\mathcal{R}_x$ at the $xz$ plane). Unlike the earlier cases, \textit{all} mirror symmetries are broken by this ansatz and the sum of all scalar spin chiralities within the unit cell does not add up to zero. As evidenced in this section, we find that nonzero Chern integers can indeed be realized. 
Note that the magnetic symmetries of the state we consider are the same as those of an orbital magnetic field. Consequently, if the ansatz emerges spontaneously, we find an anomalous contribution to $\kappa^{}_{xy}$, i.e., a thermal Hall response in the
absence of an external magnetic field. This, however, also means that the symmetry-breaking terms of the ansatz can be induced by the orbital coupling $H_{\chi}$. In the latter case, there is no anomalous contribution.

\subsection{One-orbital model with trivial bands}
Throughout this section, we direct our attention to the one-orbital model of the cuprate superconductors, which only involves the Cu-$d$ orbitals forming a square lattice as shown in \figref{SBAnsatzPatternDOneObital}. 
The general form of the mean-field Hamiltonian, only involving spin-rotation invariant terms, reads as
\begin{alignat}{2}
\label{eq:H1O}
\nonumber H_{\textsc{mf}} &= \frac{J}{2} && \sum_{i,j, \,\sigma} \left(\mc{B}_{i,j}\, b^\dagger_{i\sigma}b^\pdagger_{j\sigma} - \mc{A}^*_{i,j}\, \sigma\, b_{i\sigma} b_{j-\sigma} + \text{H.c.}  \right) \\
&+ \lambda && \sum_{i \sigma} \left(  b^\dagger_{i \sigma} b^\pdagger_{i \sigma}  - S \right).
\end{alignat}
One can write down a suitable ansatz consistent with all the $m' m m$ symmetries to describe pattern A as
\begin{subequations}\label{TwoSimpleAnasaetze}
\begin{alignat}{2}
\nonumber    \mc{A}_{i,i+\hat{x}}&=\mc{A}_{i,i+\hat{y}}=\mc{A}_1, \quad \mc{B}_{i,i+\hat{x}}=\mc{B}_{i,i+\hat{y}}=\mathrm{i}\mc{B}_1, \\ \mc{A}_{i,i+\hat{x}+\hat{y}}
&=\mc{A}_{i,i-\hat{x}+\hat{y}}=\mc{A}_2,
\label{eq:PatternA}
\end{alignat}
and all others terms set to zero, where $\hat{x}=(1,0)$ and $\hat{y}=(0,1)$ have been introduced. Similarly, for pattern B,
\begin{align}
\nonumber    \mc{A}_{i,i+\hat{x}}&=\mc{A}_{i,i+\hat{y}}=\mc{A}_1, \quad \mc{B}_{i,i+\hat{x}}=\mc{B}_{i,i+\hat{y}}=\mathrm{i}\mc{B}_1, \\ 
    \mc{A}_{i,i+\hat{x}+\hat{y}}&=\mc{A}_2.
    \label{eq:PatternB}
\end{align}\end{subequations}

\begin{figure}[htb]
\subfigure[]{\label{fig:PatternA}\includegraphics[width = 0.4925\linewidth]{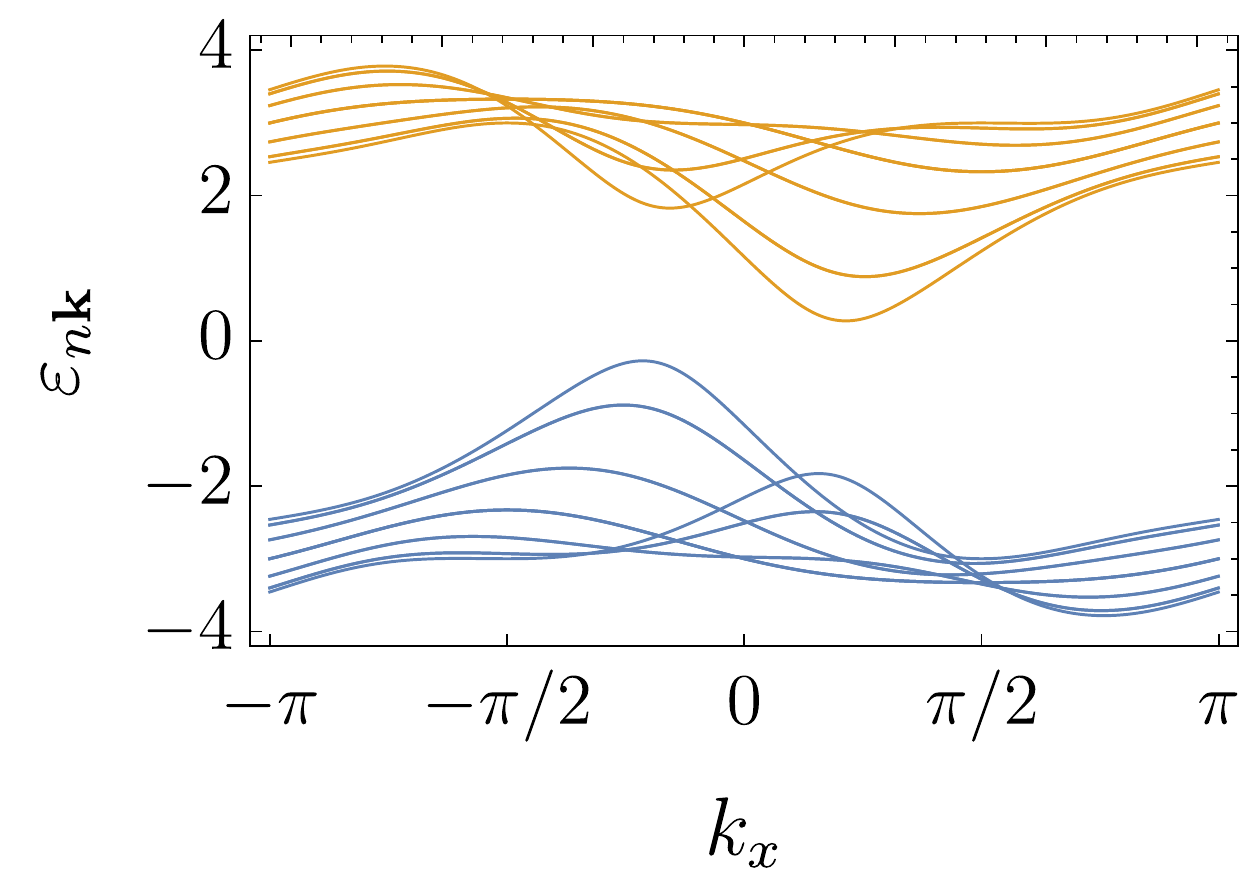}}
\subfigure[]{\label{fig:PatternB}\includegraphics[width = 0.4925\linewidth]{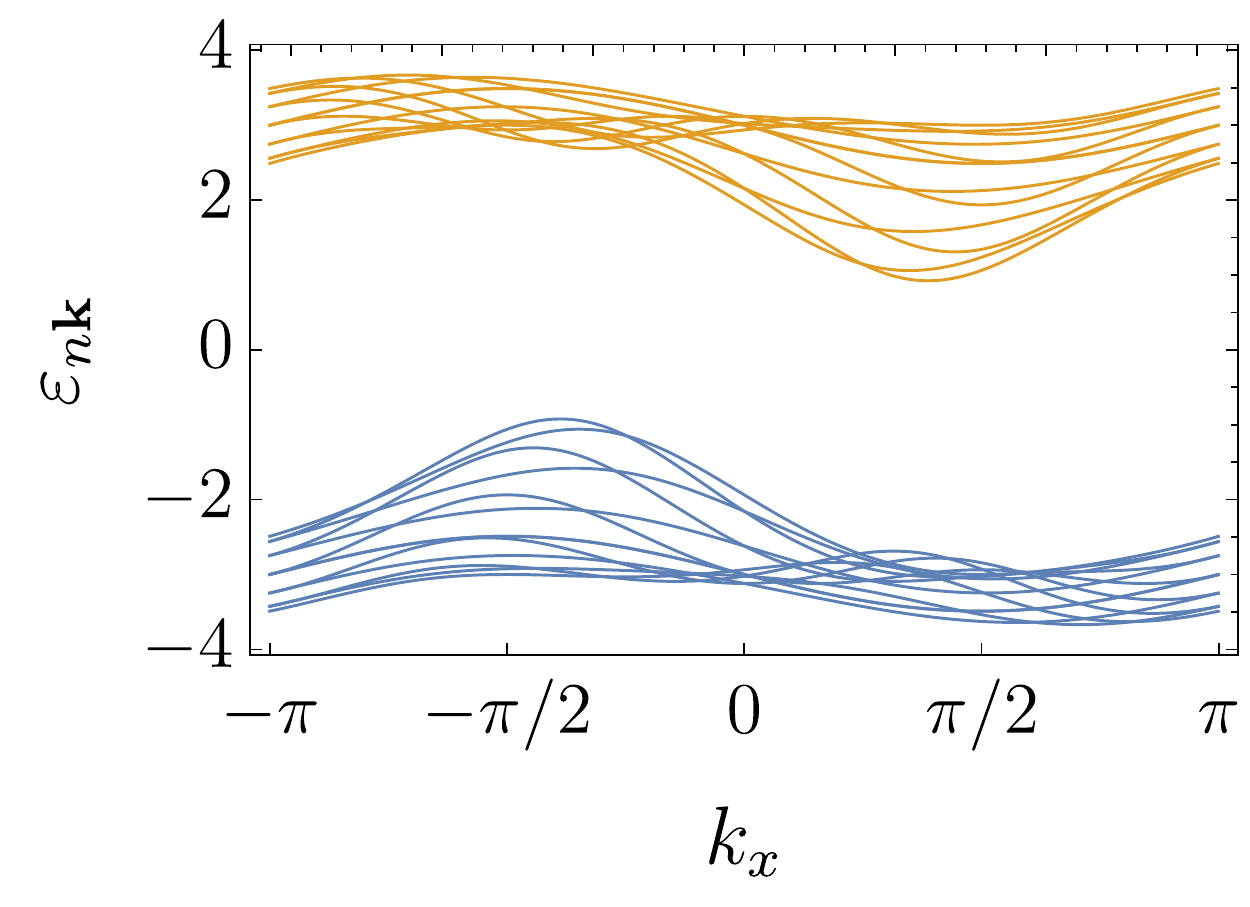}}
\caption{Schwinger boson band structure (in units of $J\mc{A}_1$) for the ansatz of (a) Eq.~\eqref{eq:PatternA} (pattern A), and (b) Eq.~\eqref{eq:PatternB} (pattern B), with $\mc{A}_2 = 0.75$, $\mc{B}_1 = 0.5$, $B_z = 0$, and $\lambda = 3$. For clarity, the eigenvalues of the dynamic matrix are shown; the energies of the actual bosonic bands are just the absolute values of the same and are strictly positive. The different lines for each of the two colors refer to distinct values of $k_y= -\pi, -\pi+\pi/6,\ldots, \pi$.  The dispersion minima are at $\pm (\pi/2, \pi/2)$ for $\mc{A}_2 = 0$, but shift to $\pm (\mathbf{K}, \mathbf{K})$, with $\mathbf{K}$ incommensurate, when $\mc{A}_2 \ne 0$. The states can thus be smoothly connected to the antiferromagnet by tuning $\mc{A}_2$.}\label{SpectrumOfPatternAandB}
\end{figure}

By tuning $|\mathcal{B}_1|$ and $|\mathcal{A}_2|$ to sufficiently small values, the ans\"atze in \equref{TwoSimpleAnasaetze} can be brought arbitrarily close to that of the conventional two-sublattice N\'eel state and its quantum-disordered partner [for which only $\mathcal{A}_1$ is nonzero in \equref{TwoSimpleAnasaetze}]. Accordingly, the concomitant magnetically ordered state is a smooth deformation of the N\'eel state and happens to be a conical spiral \cite{CSS17,yoshida2012conical}.

Since the spectrum for $|\mathcal{B}_1|,|\mathcal{A}_2| \neq 0$, illustrated in \figref{SpectrumOfPatternAandB}, retains its gap upon continuously tuning $\mathcal{B}_1$ and $\mathcal{A}_2$ to zero, the Chern numbers must be $C_n =0$ (exactly like those of the N\'eel state), wherefore these ans\"atze are not expected to be a good starting point for obtaining a sizable thermal Hall response.

\subsection{Chern numbers and thermal Hall conductivity}\label{AnsatzWithChernNumbers}
\begin{figure}[tb]
    \centering
    \includegraphics[width = 0.85\linewidth]{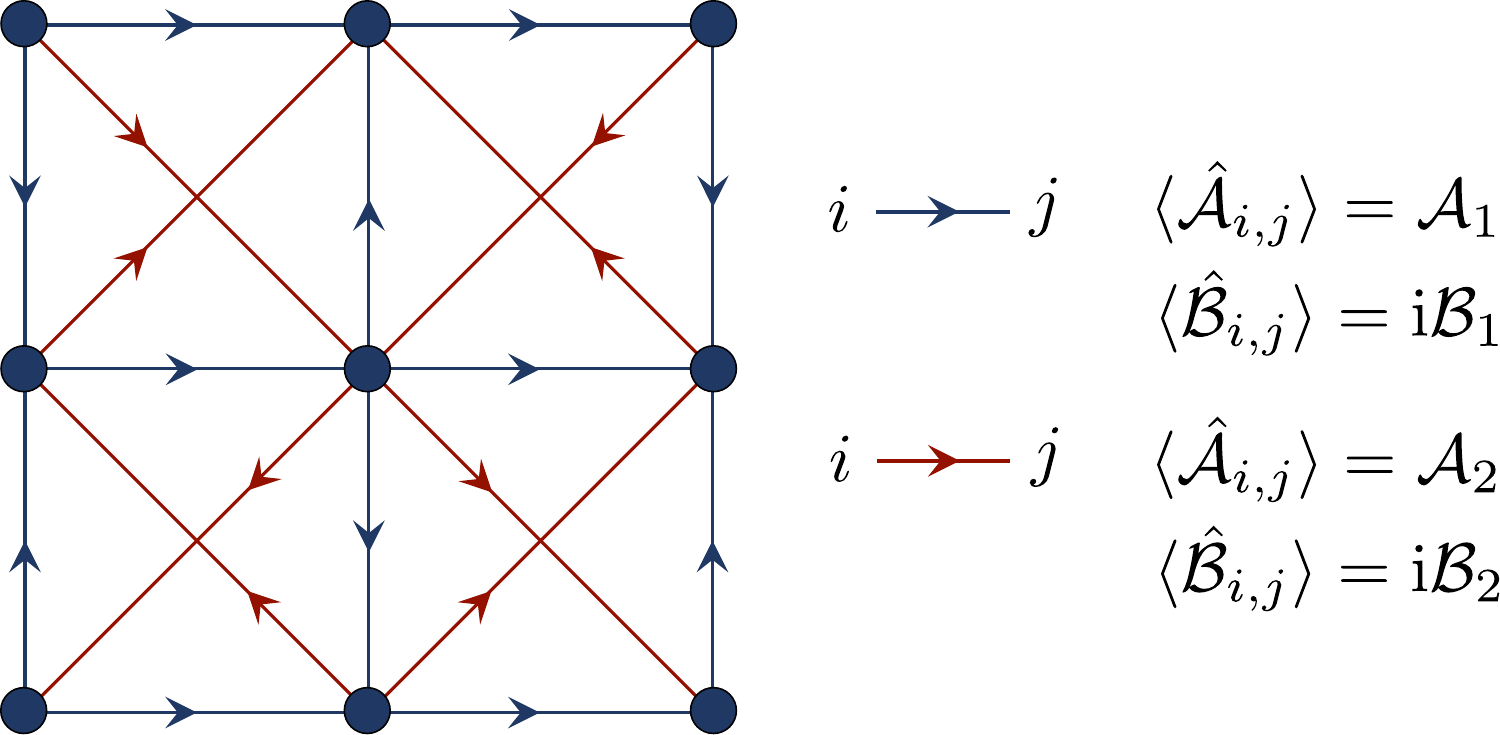}
    \caption{Schwinger-boson mean-field ansatz for the one-orbital model defined by Eqs.~\eqref{eq:subeqns} and \eqref{FullAnsatz1}. The Cu atoms in the CuO$_2$ plane are depicted here as dark blue circles. The arrows indicate the sign conventions: along the (next-)nearest-neighbor bond from site $i$ to site $j$, the bond operators have the expectation values $\langle \hat{\mc{A}}_{i,j} \rangle$ = $\mc{A}_{1(2)}$, $\langle \hat{\mc{B}}_{i,j} \rangle$ = $\mathrm{i}\mc{B}_{1(2)}$; due to $\hat{\mc{A}}_{j,i} = - \hat{\mc{A}}_{i,j}$ and $\hat{\mc{B}}_{j,i} = \hat{\mc{B}}^\dagger_{i,j}$, the bonds are directed and associated with blue (red) arrows in the figure.}
    \label{SBAnsatzPatternDOneObital}
\end{figure}

The considerations above seem to suggest looking instead at ans\"{a}tze that are not adiabatically connected to that of the conventional antiferromagnet with only $\mathcal{A}_1$ nonzero. Motivated by the recent study \cite{scheurer2018orbital} of spin-liquid states with orbital loop currents, we next consider an ansatz with magnetic point group $\frac{4}{m} m'm'$. A minimal choice, yielding this point group while preserving translations, $T_{x}$, $T_y$, is \begin{subequations}\label{eq:subeqns}
\begin{alignat}{2}
\label{MinimalAnsatzOneOrbitalModel}
    \mc{A}_{i,i+\hat{x}} &= \mc{A}_1, \quad \mc{A}_{i,i+\hat{y}} = (-1)^{i_x+i_y} \mc{A}_1,\\
    \mc{B}_{i,i+\vec{\eta}_\mu} &= \mathrm{i}\, s_\mu (-1)^{i_x+i_y} \mc{B}_2 , 
    \label{MinimalAnsatzOneOrbitalModel2}
\end{alignat}
\end{subequations}
with second-nearest-neighbor vectors $\vec{\eta}_\mu= \hat{x} + (-1)^\mu \hat{y}$. The relative signs of $s_\mu \in \{+1,-1\}$ can be read off Fig.~\ref{SBAnsatzPatternDOneObital}, and are chosen so as to attain the correct magnetic point group. 
Obviously, the ansatz is not explicitly invariant under the symmetry generators $T_{x}$, $T_y$, $C_4$, and $\Theta \mathcal{R}_x$. However, since the symmetries act projectively, it \textit{is} invariant under the respective symmetry operations when they are applied in conjunction with the following gauge transformations:
\begin{subequations}
\label{eq:gauge_rep}
\begin{align}
\mc{G}^{}_{T_{\mu}}(j) &= (-1)^{j_y}, \label{BehaviorUnderTransl}; \quad \mu=x,y,  \\
\mc{G}^{}_{\Theta \mc{R}_{\mu}}(j) &= \mathrm{i}(-1)^{j_x+j_y},\label{BehaviorUnderTR} \\
\mc{G}^{}_{C_2}(j) &= (-1)^{j_x},\\
\mc{G}^{}_{C_4}(j) &= 
\begin{cases}
\vspace*{0.2cm}
{\displaystyle \cos\left(\frac{\pi}{2} ( j_x +j_y)\right)}; \quad j \in \alpha,\\
{\displaystyle \sin\left(\frac{\pi}{2} ( j_x +j_y)\right)}; \quad j \in \beta.
\end{cases}
\end{align}
\end{subequations}
At the same time, one can indeed construct explicit gauge-invariant fluxes which are odd under $\Theta$ or $\mc{R}_\mu$ \cite{scheurer2018orbital}, and our ansatz does break these symmetries. 

It turns out that the ansatz of Eq.~(\ref{eq:subeqns}a--b) alone proves to be insufficient to yield bands with nonzero Chern numbers, so we add on top the additional operator expectation values: 
\begin{subequations}
\begin{alignat}{2}
 \label{AdditionToAnsatz1}
        \mc{B}_{i,i+\hat{x}} &= \mathrm{i} \mc{B}_1, \quad \mc{B}_{i,i+\hat{y}} = \mathrm{i} (-1)^{i_x+i_y} \mc{B}_1,\\
         \mc{A}_{i,i+\vec{\eta}_\mu} &= \, s_\mu (-1)^{i_x+i_y} \mc{A}_2. \label{AdditionToAnsatz}
\end{alignat}\label{FullAnsatz1}\end{subequations}
It is straightforward to check that Eqs.~\eqref{eq:subeqns} and \eqref{FullAnsatz1}, in totality, preserve both translation and $\frac{4}{m} m'm'$ by applying the gauge transformations in \equref{eq:gauge_rep}. From this point onward, the term ``one-orbital model''  always implicitly refers to this combined ansatz for pattern D. For completeness, the three-orbital model of the cuprates, also taking into account the oxygen $p$ orbitals, is discussed in Appendix~\ref{3OrbitalModel}; the conclusions are similar in spirit.

The generalization in \eqref{FullAnsatz1} results in topologically nontrivial bosonic bands and, hence, a considerable thermal Hall response as we show below. As long as the interband gaps remain open, the Chern integers are invariant under smooth variations of the mean-field parameters $\{\mathcal{A}_\mu, \mc{B}_\mu\}$ in the Hamiltonian. Consequently, this state is not smoothly connected to the SBMFT of the conventional square-lattice antiferromagnet, for which the Chern numbers of all the bands are identically zero. 

A useful characterization of spin-liquid phases can be obtained by gauge invariant fluxes. Of particular importance for our study is the flux $\phi = \mc{A}_{1,2}\mc{A}^*_{2,3}\mc{A}_{3,4}\mc{A}^*_{4,1}$, where $1,2,3,$ and $4$ label the four sites of any elementary square plaquette in counterclockwise order. The limiting case $\mc{A}_2=\mc{B}_1=\mc{B}_2=0$ of the ansatz in Fig.~\ref{SBAnsatzPatternDOneObital} corresponds to the $\pi$-flux states of Yang and Wang \cite{yang2016schwinger}, which have full square-lattice and time-reversal symmetries; turning on nonzero values of $\mc{A}_2$, $\mc{B}_1$, and $\mc{B}_2$ reduces the symmetry to $\frac{4}{m}m'm'$, and leads to 
spinon bands with nonzero Chern numbers. On the other hand, the $\mathbb{C}\mathbb{P}^1$ model \cite{sachdev1990effective}, a low-energy effective field theory of quantum antiferromagnets on a square lattice,  describes the more familiar zero-flux Schwinger boson state \cite{yang2016schwinger}. It was shown in \refcite{scheurer2018orbital} that there is no quadratic perturbation to the $\mathbb{C}\mathbb{P}^1$ theory which breaks the symmetry down to $\frac{4}{m}m'm'$, and we discuss the needed perturbations further in Appendix~\ref{app:cp1}.
Our results here are consistent with these earlier results: we need to perturb a $\pi$-flux state to have nonzero Chern numbers of spinon bands in SBMFT; such nontrivial bands cannot be obtained as a perturbation of the zero-flux state. Further, the $\mathbb{C}\mathbb{P}^1$ theory can naturally describe low-energy excitations close to $\Q = (0,0)$ and $(\pi,\pi)$; in contrast the spin-liquid phase we consider has low energy excitations at $(0,\pi)$ and $(\pi,0)$ as well.

Yang and Wang \cite{yang2016schwinger} also analyzed the magnetic ordered states that appeared upon condensing bosonic spinons from the $\pi$-flux state. They found a variety of possibilities with ordered moments at wavevectors $(0, \pi)$, $(\pi, 0)$, and $(\pi, \pi)$: this included cases where the dominant moment was at the $(\pi, \pi)$ wave vector of the N\'eel state. Nonzero values of $\mc{A}_2$, $\mc{B}_1$, and $\mc{B}_2$ distort these states to also allow for a (possibly small) ferromagnetic moment at $(0, 0)$, leading to a four-sublattice magnetic order of the form (see Appendix~\ref{MagneticOrder} for details)
\begin{align}
\langle \bm{S}(j) \rangle = \n_{(0,0)} + (-1)^{j_x}\, \n_{(\pi,0)} &+ (-1)^{j_y}\, \n_{(0,\pi)} \\ \nonumber 
&+ (-1)^{j_x+j_y}\, \n_{(\pi,\pi)}.    
\end{align}
Note that this ferromagnetic moment arises without a Zeeman term in the Hamiltonian, and is a consequence of either spontaneous breaking of the symmetry to $\frac{4}{m}m'm'$, or one induced by the orbital coupling to the external field (see Appendix~\ref{app:orbital}).


One might wonder whether adding the orbital coupling of the magnetic field, $H_{\chi}$, described in leading order in $t/U$ by terms involving the triple products $\S_i \cdot (\S_j \times \S_k)$ \cite{sen1995large}, can be used to describe the symmetry reduction to the magnetic point group $\frac{4}{m} m'm'$ within SBMFT. We consider the decoupling of this triple-product term in Appendix~\ref{app:orbital}. Although we do not include this  self-consistently in our analysis, we verify that spin-liquid states with symmetry broken to $\frac{4}{m} m'm'$ do indeed lead to a nonzero 
expectation value for the triple products in the Hamiltonian, in the quadratic approximation.

\subsubsection{Spectrum and symmetries}
\label{sec:symm}
In spite of the final thermal Hall conductivity itself being a gauge-invariant quantity, any intermediate calculations require the explicit choice of a gauge. Owing to the alternating factor of $(-1)^{i_x+i_y} $, the ansatz \eqref{eq:subeqns} is translationally invariant only modulo a gauge transformation or, in other words, it is invariant under \textit{two}-site lattice translations when working in a fixed gauge. 
We therefore choose a two-sublattice unit cell with sublattice indices defined by the parity of $i_x+i_y$. In each unit cell, we denote the Schwinger boson operators by $\alpha$ (even parity) and $\beta$ (odd parity). The basis vectors for this new bipartite lattice are $\vec{\eta}_\mu$, and the reciprocal lattice vectors are $\vec{G}_\mu = \pi \,\vec{\eta}_\mu$, so the BZ can be chosen to be the conventional antiferromagnetic Brillouin zone,  $\{(k_x, k_y)\, \vert\, k_x, k_y \in [-\pi, \pi);\, \lvert k_x \rvert + \lvert k_y\rvert \le \pi\}$. 

\begin{figure*}[htb]
\subfigure[]{\label{fig:1OB1}\includegraphics[width = 0.32\linewidth]{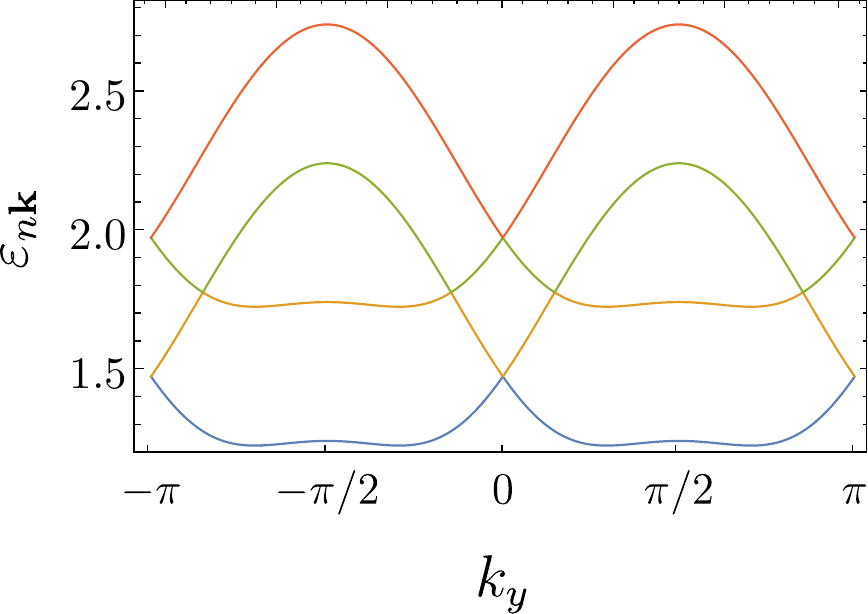}}
\subfigure[]{\label{fig:ref_demand}\includegraphics[width = 0.32\linewidth]{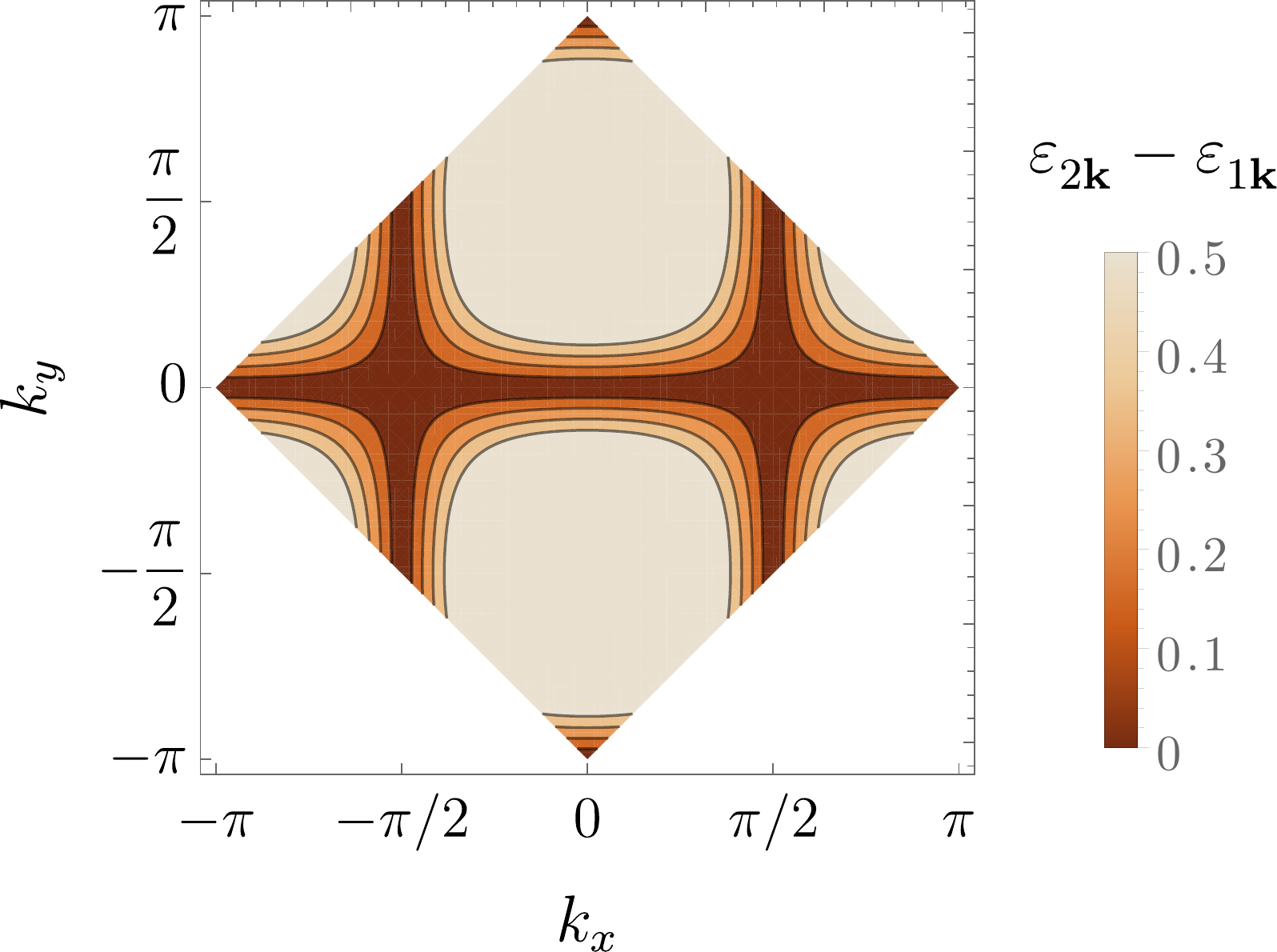}}
\subfigure[]{\label{fig:1OB2}\includegraphics[width = 0.32\linewidth]{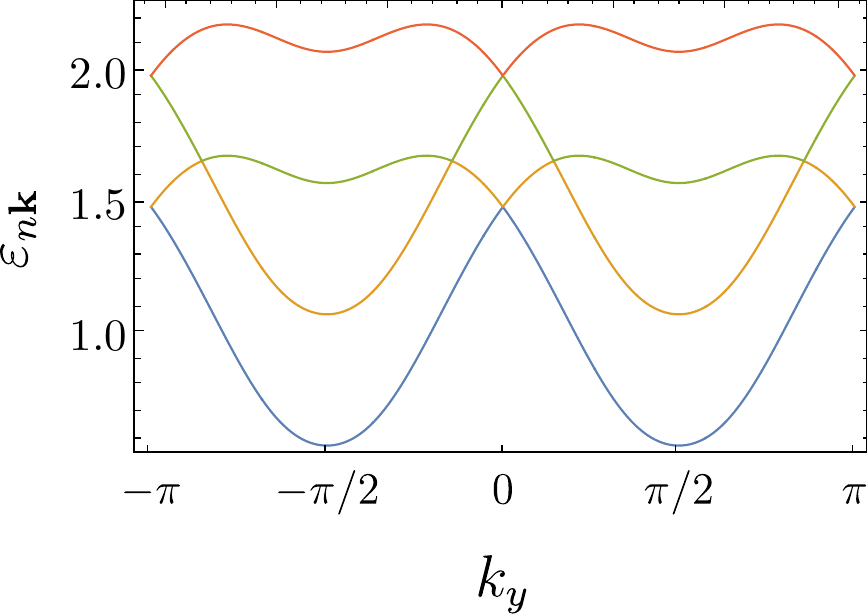}}
\subfigure[]{\includegraphics[width = 0.32\linewidth]{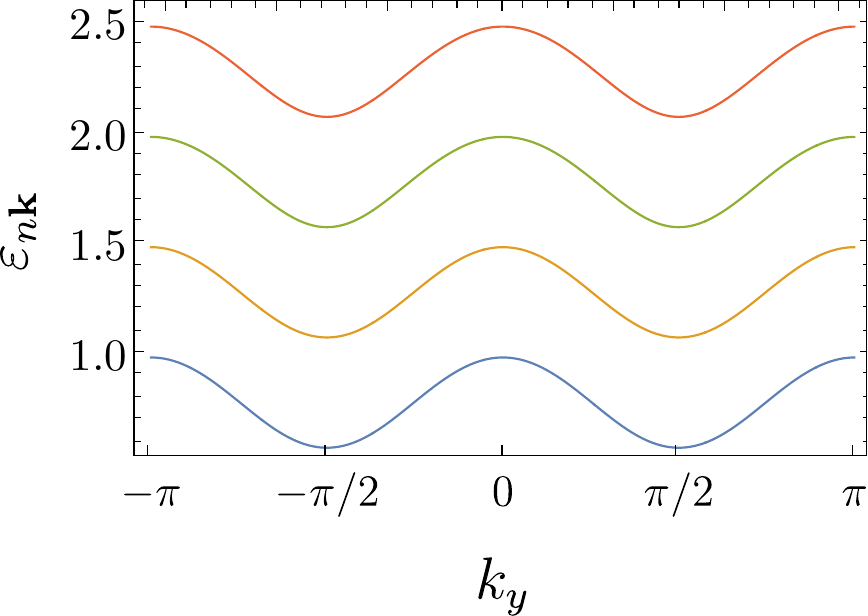}}
\subfigure[]{\label{fig:1OB3}\includegraphics[width = 0.32\linewidth]{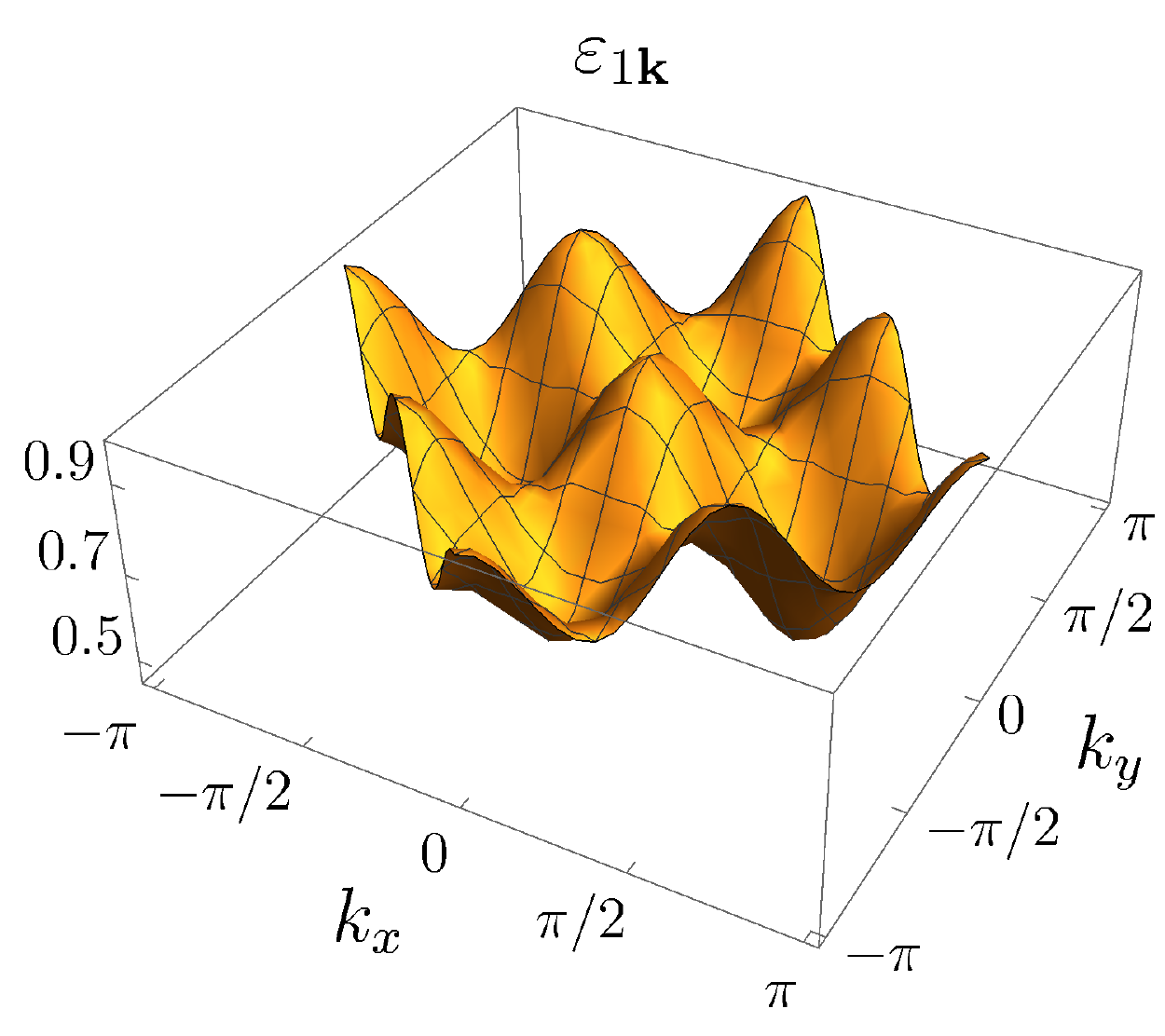}}
\subfigure[]{\label{fig:1OmegaPD}\includegraphics[width = 0.32\linewidth]{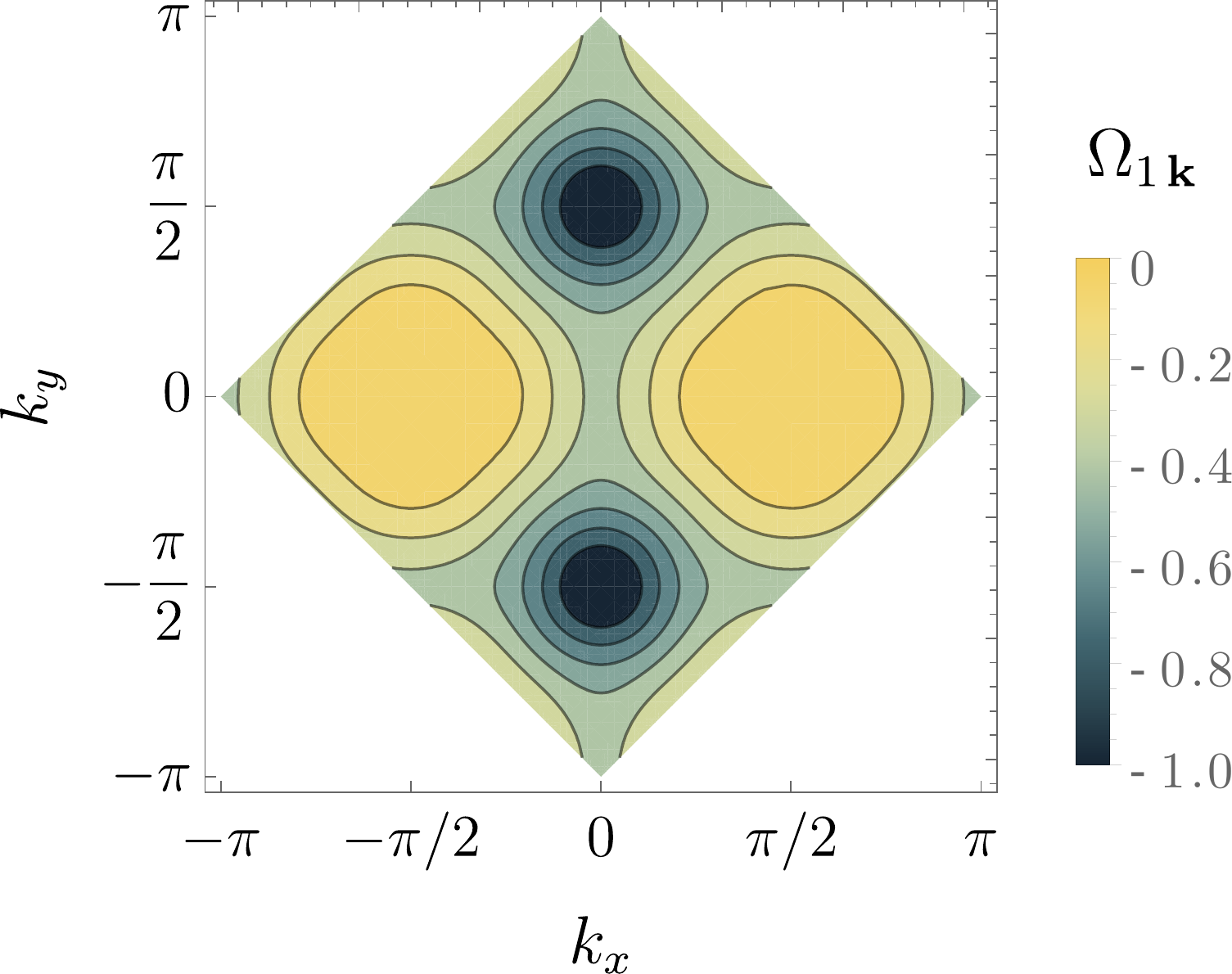}}
\caption{\label{fig:1OBands}(a) Dispersion of the Schwinger boson particle bands $\varepsilon_{n\vec{k}}$, $n=1, \ldots, 4$, shown in blue, orange, green, and red, respectively, along the line $k_x = 0$, for the one-orbital model with $\mc{A}_2 =0$, $\mc{B}_1 =0$, $\mc{B}_2 = 0.25$, $\lambda = 2$, and $B_z=0.5$, measured in units of $J\mc{A}_1$. The bands touch along lines in the BZ, as underscored by the density plot of $\varepsilon_{2 \mathbf{k}}-\varepsilon_{1 \mathbf{k}}$ in (b), and thus lack well-defined Chern numbers. (c) The intersection of the bands persists even with $\mc{A}_2 =0.75 $ on top of the parameters in (a,b). (d) The addition of a nonzero $\mc{B}_1$ (taken to be 0.5 here) is required to prevent the touching of two particle bands, necessitating the addition of Eq.~\eqref{FullAnsatz1} to the minimal ansatz. With $\mc{B}_1 \ne 0$, the bands acquire a nontrivial Chern number. (e) The dispersion of the lowest-energy band in (d) exhibits minima at ${\k} =(\pm \pi/2,0)$, indicating anisotropic antiferromagnetic order in the corresponding confined phase. (f)  Berry curvature for the particle band displayed in (e); it is seen that $\Omega_{1\mathbf{k}} = 0$ at the global minima of the dispersion. The first Chern integers are $C_n = -1$ ($+1$) for the $n =1, 2$ ($n =3, 4$) bands. The curvatures are ill-defined at $B_z=0$, for which all the particle bands are pairwise degenerate. }
    \label{fig:BandsPD}
\end{figure*} 

As sketched in Appendix~\ref{sec:BdG}, the mean-field Hamiltonian can be represented in terms of the eight-component spinor  $\Psi^\dagger_\mathbf{k} = ( \alpha^\dagger_{\k\uparrow}\, \beta^\dagger_{\k\uparrow}\, \alpha^\dagger_{\k \downarrow}\, \beta^\dagger_{\k\downarrow}\,\alpha^{}_{-\k\uparrow}\, \beta^{}_{-\k\uparrow}\,\alpha^{}_{-\k\downarrow}\, \beta^{}_{-\k\downarrow})$ with $H_{\textsc{mf}} = \sum_{\k} (\Psi_{\k}^\dagger\, \mathcal{H}({\k})\, \Psi_{\k})/2 $. 
The associated band structures upon diagonalization are plotted in Fig.~\ref{fig:1OBands}. At each momentum $\mathbf{k}$, the dynamic matrix $K$ has eight eigenvalues, four positive and four negative; we label the former (latter) by $n=1, \ldots, 4$ ($n =5, \ldots, 8$) in ascending (descending) order. The energies of the actual bosonic bands are simply the absolute values of these and are necessarily positive.

Additionally, the Hamiltonian $\mathcal{H}({\k})$ harbors another symmetry that is somewhat less apparent. Although the particle bands are generically distinct, they become pairwise degenerate when there is no Zeeman field, $B_z=0$. We emphasize that this degeneracy is \textit{not} the same as the trivial redundancy described in Eq.~\eqref{eq:dgn}, which arises due to the pairwise occurrence of the eigenvalues of the dynamic matrix. Despite the seeming lack of an \textit{a priori} reason, the degeneracy of these eigenvalues stems from an effective antiunitary symmetry, which we scrutinize more carefully later in Appendix~\ref{sec:TRS}.

\begin{figure*}[htb]
\centering
\subfigure[]{\label{fig:1OK1} \includegraphics[height=0.255\linewidth,clip]{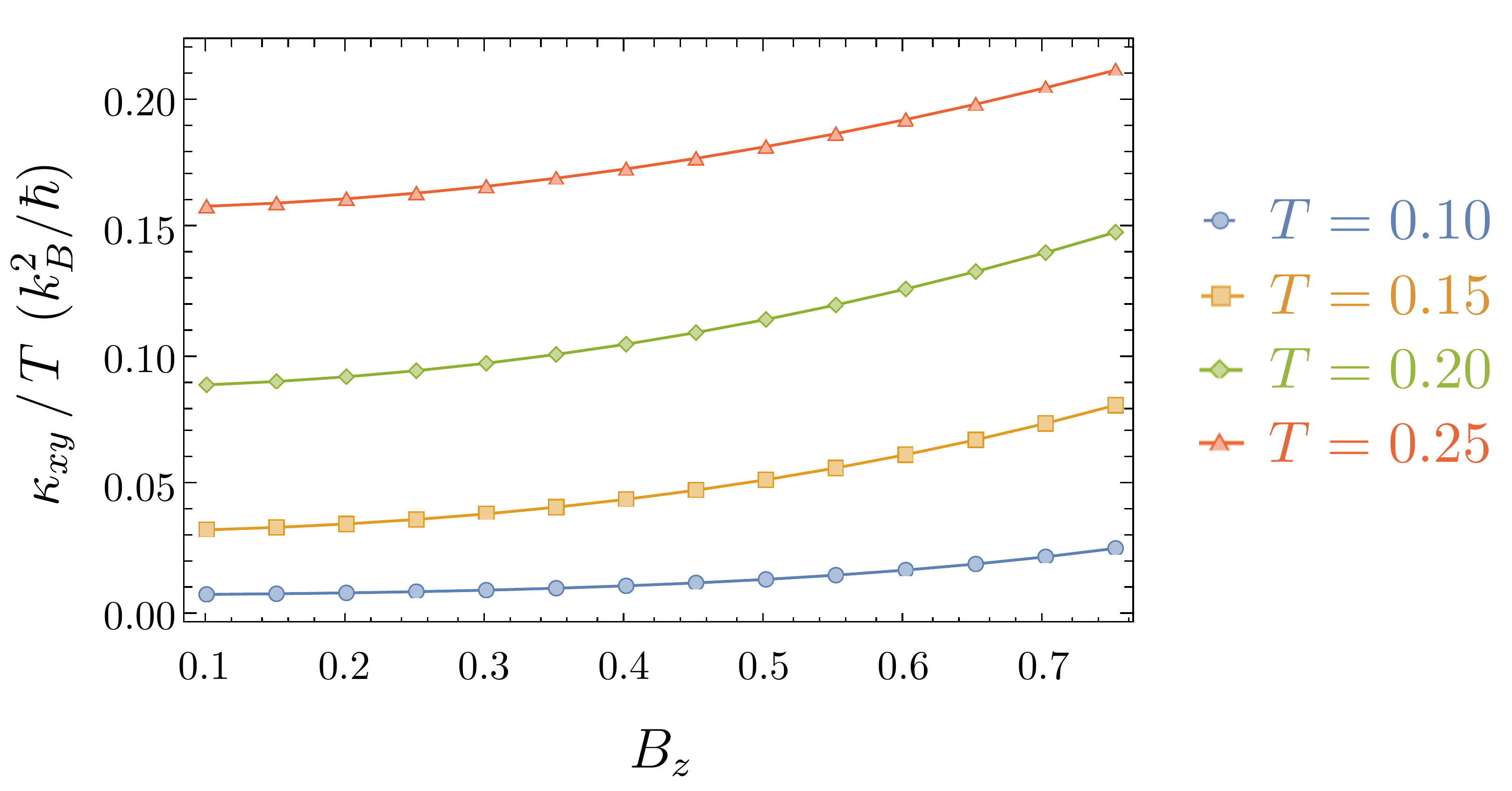}
}
\subfigure[]{\label{fig:1OK2} \includegraphics[height=0.255\linewidth,clip]{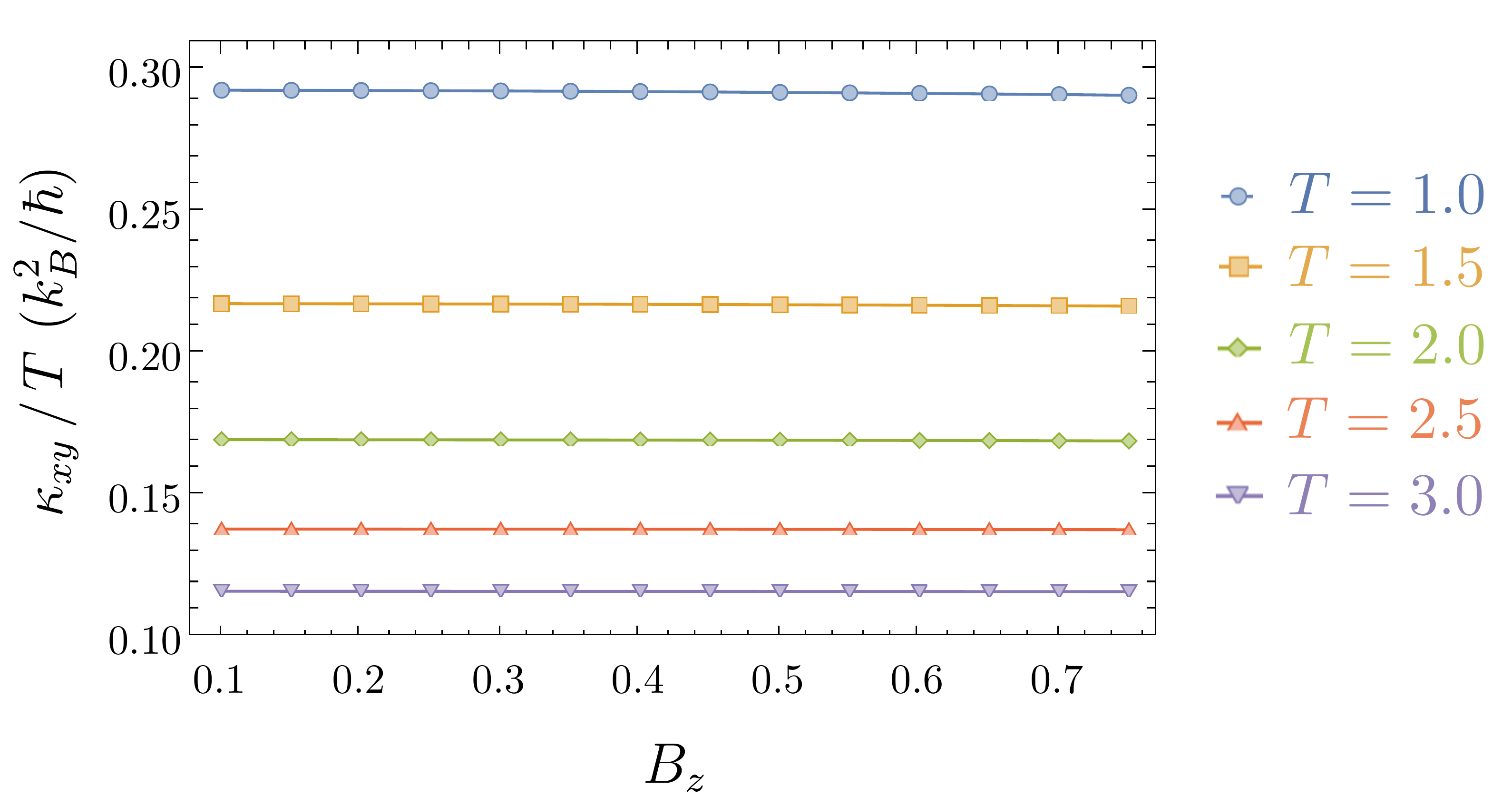}}\\
\subfigure[]{\label{fig:1OK3} \includegraphics[height=0.32\linewidth]{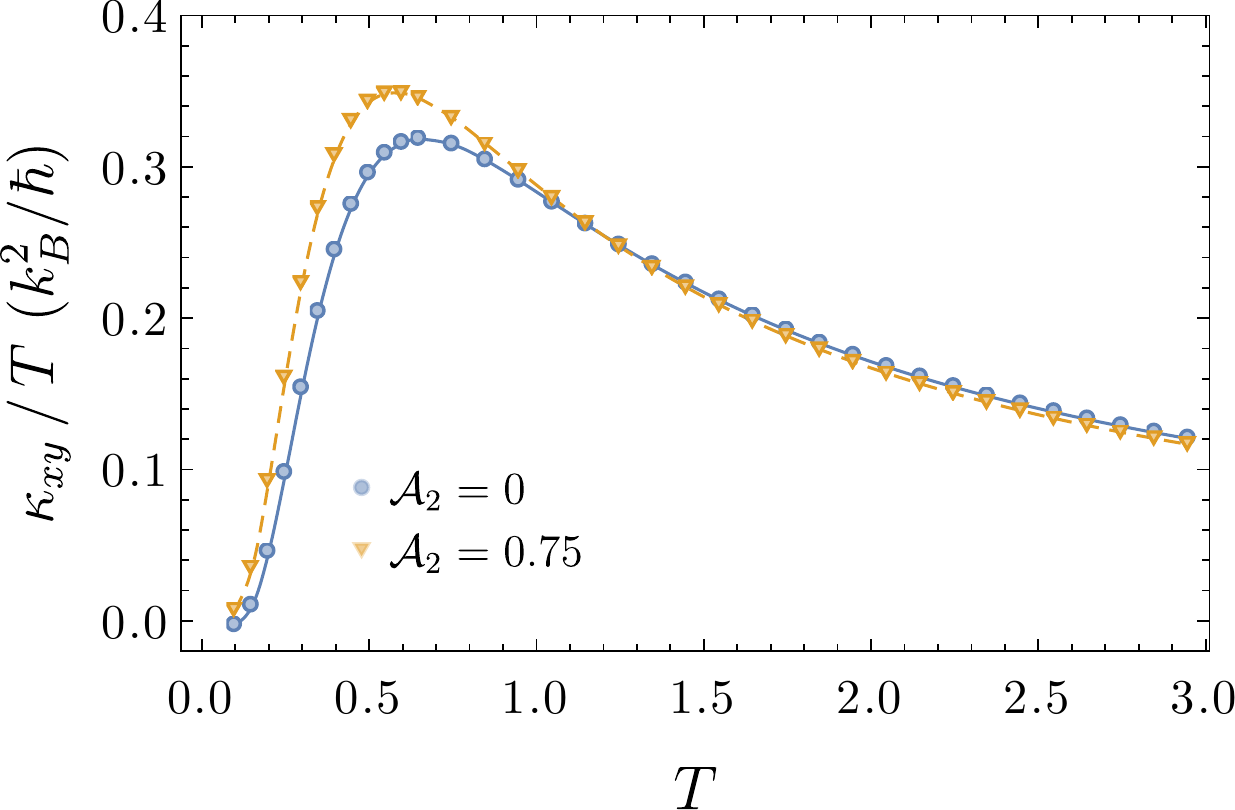}
\begin{picture}(0,0)
\put(-110,91){\includegraphics[height=2.3cm]{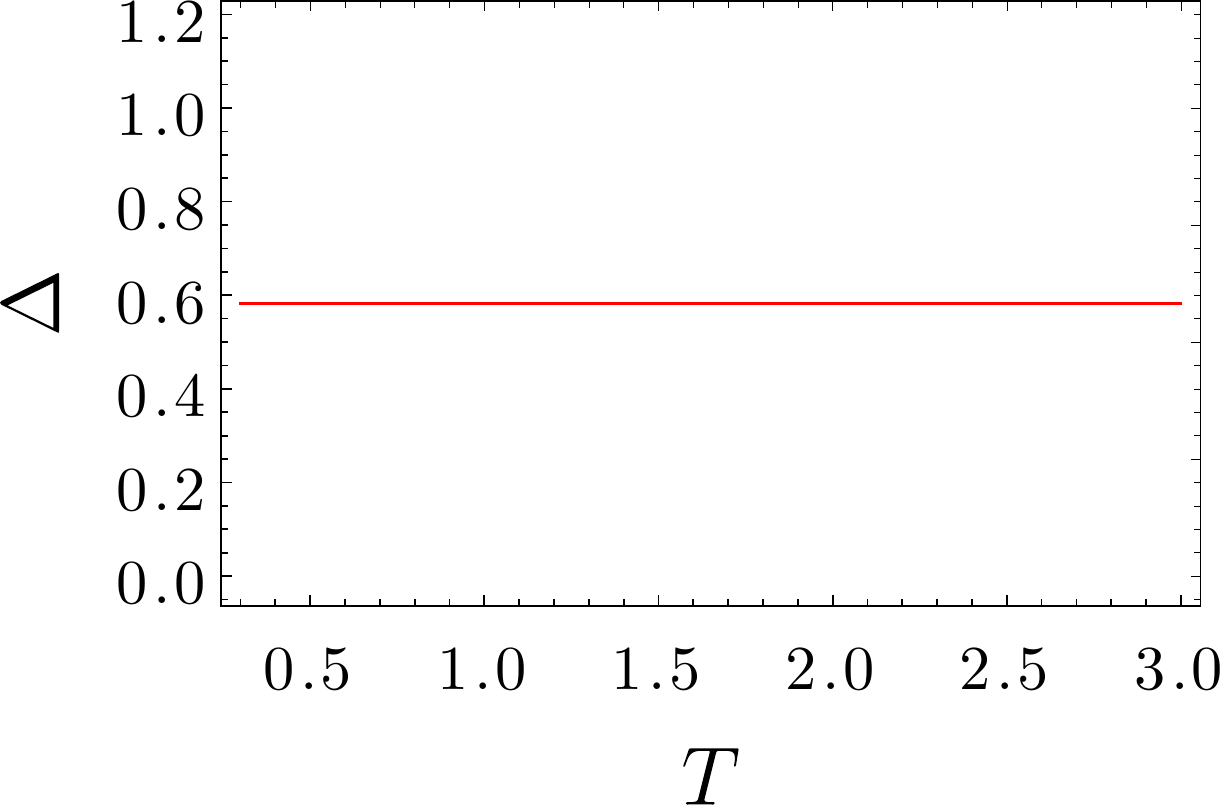}}
\end{picture}}\hfill
\subfigure[]{\label{fig:1OK4} \includegraphics[height=0.32\linewidth]{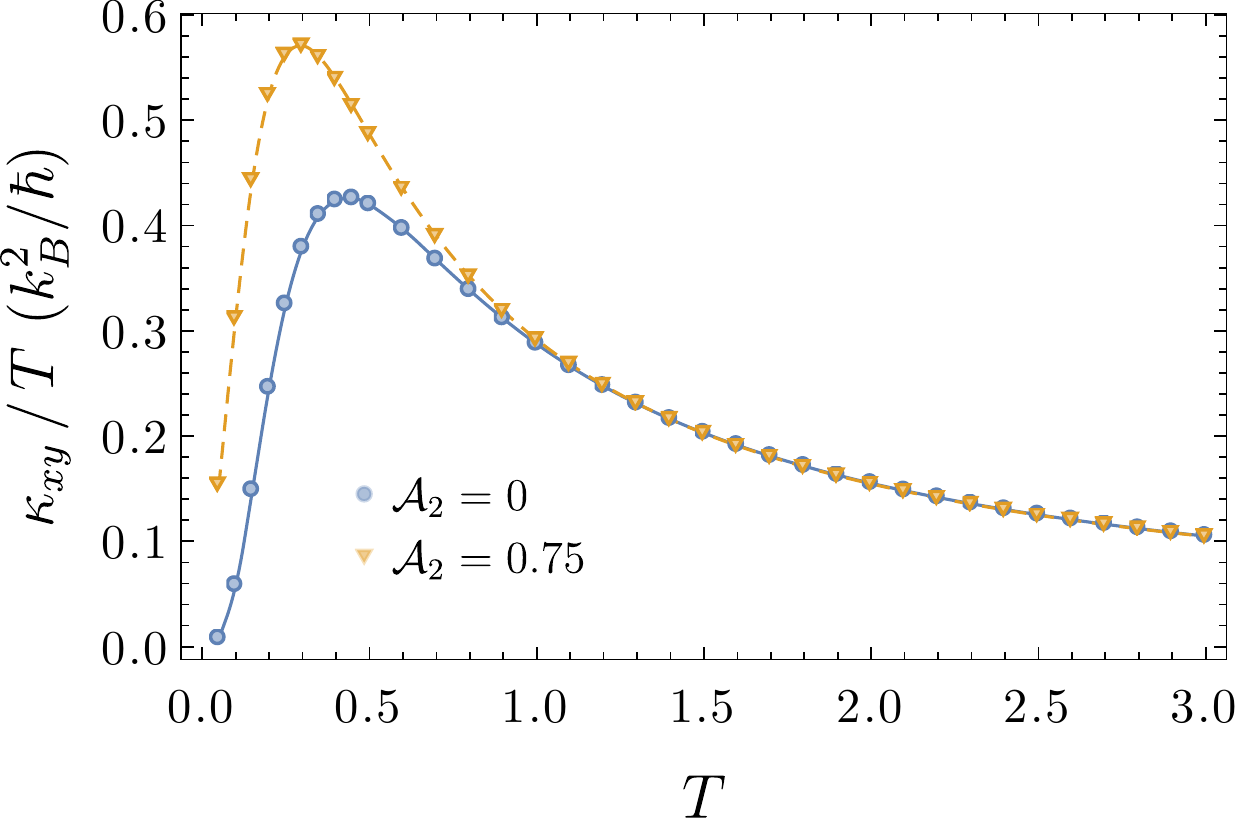}
\begin{picture}(0,0)
\put(-125,78){\includegraphics[height=2.75cm]{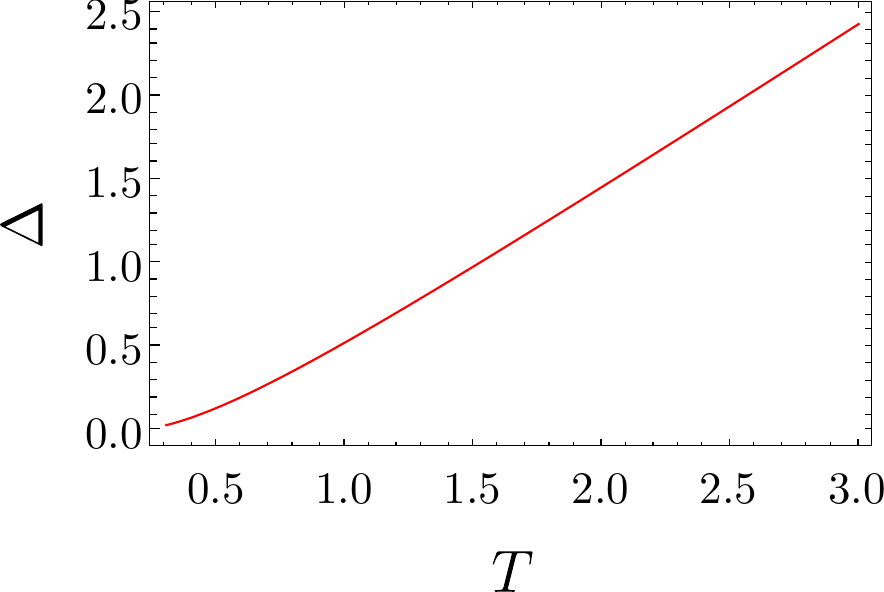}}
\end{picture}}
 \caption{Thermal Hall conductivity in the one-orbital model with the parameters $\mc{A}_2 = 0.75$, $\mc{B}_1 = 0.5$, $\mc{B}_2 = 0.25$, and $\lambda = 2$, as a function of Zeeman field at (a) low, and (b) high temperatures. In the second case, there is almost no dependence on $B_z$. We emphasize that we only show the dependence of $\kappa^{}_{xy}/T$ on the Zeeman field at constant orbital coupling. The latter enters indirectly through the parameters, $\mathcal{A}_{1,2}$, $\mathcal{B}_{1,2}$, of the ansatz. (c) The variation of $\kappa^{}_{xy}/T$ with temperature at a constant  $B_z = 0.25$ for which the spinon gap (inset) is $\Delta = 0.582$. The parameter $\mc{A}_2$ can be used to tune the strength of the response. $\kappa^{}_{xy}/T$ decays as $(\Delta/T)^2 \exp(-\Delta/T)$ and $1/T$ (with $1/T^3$ corrections) at low and high temperatures, respectively. (d) The same as in (c) but with the gap now varied as $\Delta (T) = T \exp(-m /T); m = 0.2 \pi $, so that it is exponentially small with temperature. As in the figures above, all energies are measured in units of $J\mc{A}_1$.}
    \label{fig:1OTH}
\end{figure*}

From the paraunitary matrix $\TM_{\bf k}$, one can calculate the Berry curvatures of the bands. However, the Berry connection, defined as 
\begin{equation}
A_{j, \mu} (\mathbf{k}) \equiv \mathrm{i} \Tr \left[\Gamma_j\, \rho^{}_3\, \TM^\dagger_{\bf k} \,\rho^{}_3 \left(\partial_{k_\mu} \TM_{\mathbf{k}} \right) \right], \label{DefinitionGaugeField}
\end{equation}
where $\Gamma_j$ is a diagonal matrix with $(\Gamma_j)_{ab} = \delta_{ja}\delta_{jb}$, cannot be smoothly specified over the entire BZ and the phases of the eigenvectors that constitute $\TM_{\bf k}$ must be chosen accordingly.  The resolution lies in decomposing the BZ into two overlapping regions $H_1$ and $H_2$ with $H_1 \cup H_2$ = BZ, and $H_1 \cap H_2 = \partial H_1 = - \partial H_2 $ \cite{shindou2013topological}. These regions are chosen such that $[\TM_{\bf k}]_{m_{\nu},j}$ is never zero within the region $H_\nu$, where $\nu = 1,2$, and $m_{\nu} = 1, \ldots, 8$. The phase of the $j^{\mathrm{th}}$ eigenvector can then be uniquely defined by choosing a gauge in region $H_1$ ($H_2$) such that $[\TM_{\bf k}]_{m_{1},j}$ ($[\TM_{\bf k}]_{m_{2},j}$) is always real and positive. The two gauge choices, which are related by a U(1) transformation, are patched together to cover the entire BZ. This construction enables us to unambiguously calculate the Chern number  \cite{fukui2005chern, fukui2007quantum} as
\begin{equation}
C_j = \frac{1}{2\pi} \oint_{\partial H_1} \mathrm{d}\,{\bf k} \cdot \,\left(\bm{A}_j^{(1)}-\bm{A}_j^{(2)} \right),
\end{equation}
where $(\bm{A}_j^{(\nu)})_\mu$ is the gauge field [\equref{DefinitionGaugeField}] of band $j$ in the patch $\nu$.
Inspecting the eigenstructure of $\TM_\mathbf{k}$, we find a suitable partition to be $H_1 = \{ \mathbf{k}: k_y \le 0,\,  \lvert k_x \rvert + \lvert k_y\rvert \le \pi \}$ and $H_2 = BZ\backslash H_1$. The resultant Berry curvatures for the particle bands are illustrated in Fig.~\ref{fig:1OmegaPD}. The final thermal Hall conductivity, which involves contributions from all four bands, is plotted in Fig.~\ref{fig:1OTH}. 

\subsubsection{Parameter dependence of $\kappa_{xy}$}
\label{sec:asymptotic}
In this subsection, we discuss the parameter dependence of the thermal Hall conductivity in \figref{fig:1OTH} in detail and compare with asymptotic analytical considerations.

First, note that while $\kappa^{}_{xy}$ is always positive in the plots of \figref{fig:1OTH}, its sign is actually determined by that of the parameters $\mc{A}_\mu$ and $\mc{B}_\mu$ of the ansatz; under the simultaneous reversal of $\mc{A}_\mu \rightarrow -\mc{A}_\mu$ and $\mc{B}_\mu \rightarrow -\mc{B}_\mu$, the Hall conductivity also changes sign as $\kappa^{}_{xy} \rightarrow -\kappa^{}_{xy}$. This is required by symmetry as the global sign reversal of $\mc{A}_\mu$ and $\mc{B}_\mu$ is equivalent [modulo gauge transformation $\mathcal{G}(j)=\mathrm{i}$] to performing a time-reversal transformation.

Next, we turn to the temperature and field dependence. $\kappa^{}_{xy}/T$ tends to zero at high temperatures, where all bands are equally occupied, as well as very low temperatures, below the spinon gap, when all bands are nearly empty: intuitively, $c_2(n_B)$ is the same constant for any band for both high and low $T$; factoring it out, we are left with the sum of the Chern numbers of all the particle bands and these add up to zero.  
To determine how $\kappa^{}_{xy}/T$ decays for low and high $T$, we use the asymptotic expansions for the $c_2$ function defined in \equref{eq:c2}: 
\begin{alignat}{1}
\label{eq:c2Lims}
c_2 (x) &\rightarrow 
\begin{cases}
{\displaystyle \frac{\pi^2}{3} - \frac{1}{x} + \frac{1}{2x^2} + \mathcal{O}\left( \frac{1}{x^3}\right)}; \text{ for } x \rightarrow \infty, \\
 \left(2 - \ln(x) + \ln^2(x)\right)x + \mathcal{O}(x^2 \ln(x));\text{ for } x \rightarrow 0.
\end{cases}
\end{alignat}
For simplicity, consider the contribution to $\kappa^{}_{xy}$ for a single pair of particle bands that have equal Berry curvatures (ergo, Chern numbers)---the existence of such a pair is guaranteed by the effective antiunitary symmetry 
in the one-orbital model discussed above. Without loss of generality, let these be labeled by $n=1, 2$; the discussion here can be easily extended to include the $n=3, 4$ bands for the specific case of the one-orbital model. At zero external magnetic field, the bands in the pair are degenerate energetically, i.e. $\varepsilon^{}_{1\k} = \varepsilon^{}_{2\k} \equiv E_\k$, and have the same curvatures $\Omega_{1\k} = \Omega_{2\k}$. A finite uniform Zeeman field splits their energies to $E_\k \pm B_z/2$. The Zeeman term is proportional to the identity in the dynamical matrix $K$ of Eq.~\eqref{eq:Kmatrix}. Therefore, it leaves the spinon wave functions, which are determined by the dynamic matrix $K$ rather than the Hamiltonian, unchanged. Hence, the Berry curvature remains unaffected, whereby we still have $\Omega_{1\k} = \Omega_{2\k}$.

At temperatures much larger than the band maximum, it is reasonable to approximate the Bose distribution function by $n_B(E) \sim k_B T/E$ for $k_B T \gg E$. Using Eq.~\eqref{eq:k_xy}, the thermal Hall conductivity then follows as
\begin{alignat}{2}
\nonumber \frac{\kappa^{}_{xy}}{T} &= - \frac{k_B^2}{\hbar\, V} \sum_{\mathbf{k}}\sum_{n=1,2} \left \{ c_2 \left[ n_B \left(\varepsilon^{}_{n \mathbf{k}} \right)\right] - \frac{\pi^2}{3}\right\} \Omega_{n \mathbf{k}}, \\
\nonumber
& \approx \frac{k_B^2}{\hbar\, V} \sum_{\mathbf{k}} \left( \frac{\Omega_{1\k}} {n_B(\varepsilon^{}_{1\k})} + \frac{\Omega_{2\k}}{ n_B(\varepsilon^{}_{2\k})} \right) + \mc{O}\left( \frac{1}{n_B^2(\varepsilon^{}_{n\k})}\right)\\ 
\nonumber & = \frac{k_B^2}{\hbar\, V} \sum_{\mathbf{k}} \Omega_{1\k} \left( \frac{E_\k - B_z/2}{k_B T}  +  \frac{E_\k + B_z/2}{k_B T} \right) \\
& = \left(\frac{2}{T}\right) \frac{k_B}{\hbar\, V} \sum_{\mathbf{k}} \Omega_{1\k} \, E_\k \approx \frac{k_B\, \zeta\, C_1}{\pi\,\hbar\, T},
\label{eq:linB}
\end{alignat}
where $C_1$ is the Chern number of the $n=1$ band, and $\zeta$ is a measure of the average band energy without the magnetic field. We stress that Eq.~\eqref{eq:linB} is a consequence of the effective antiunitary symmetry explicated in Appendix~\ref{sec:TRS}, and, in particular, of Eq.~\eqref{eq:omega_symm}, which ensures the equality of the Berry curvatures for the two bands. Therefore, to first order, $\kappa^{}_{xy}$ is independent of $B_z$ at high temperatures, in consistence with Fig.~\ref{fig:1OK2}. In particular, there is an anomalous thermal Hall response, i.e., $\kappa^{}_{xy} \neq 0$ for $B_z=0$. This is expected based on the symmetries of the ansatz that are identical to those of the orbital magnetic field.

Going beyond leading order in the $1/T$ expansion incorporates a subleading term
\begin{equation}
\label{eq:next}
\frac{\kappa^{}_{xy}}{T} = \frac{k_B\, \zeta\, C_+}{\pi\,\hbar\, T} \left(1 - \frac{3B_z^2 + 4\, \zeta^2}{72\, k_B^2\,T^2} \right) + \mc{O} \left(\frac{1}{T^4} \right).
\end{equation}
This term is of the opposite sign but it is parametrically small, and being of $\mc{O}(B_z^2/T^3)$, negligible at high $T$. Hence, the decrease of $\kappa{}_{xy}$ with $B_z$ is hardly observable in Fig.~\ref{fig:1OK2}. Note, however, that in reality, the parameters of the ansatz itself might be magnetic field dependent; this is not accounted for in the present calculation, and might yield a rather different dependence of $\kappa^{}_{xy}$ on the magnetic field.

Equation~\eqref{eq:next} also specifies that $\kappa^{}_{xy}/T$ goes to zero as $1/T$ at large temperatures (with $1/T^3$ corrections), which is indeed confirmed by Fig.~\ref{fig:1OK3} for $T \gtrsim 0.5$.

On the contrary, at $T$ much smaller than the spinon gap $\Delta$, the bosonic band occupancies are almost zero, and we can approximate $n_B(E) \approx \mathrm{e}^{-E/k_B T}$ for all bands. For the leading contribution, we need only consider the dominant term in the small-$x$ expansion of $c_2(x)$ from \equref{eq:c2Lims}, which goes as $x \ln^2(x)$. The net result in the $T \ll \Delta$ limit is
\begin{align}
\label{eq:lowT}
\nonumber \frac{\kappa^{}_{xy}}{T} &= - \frac{k_B^2}{\hbar\, V} \sum_{\mathbf{k}}\sum_{n=1,2}  c_2 \left[ n_B \left(\varepsilon^{}_{n \mathbf{k}} \right)\right]  \Omega_{n \mathbf{k}} \\
\nonumber & \approx -\frac{k_B^2}{\hbar\, V} \sum_{\mathbf{k}} \left(\varepsilon_{1\k}^2 \mathrm{e}^{-\varepsilon^{}_{1\k}/k_B T} +\varepsilon_{2\k}^2 \mathrm{e}^{\varepsilon^{}_{2\k}/k_B T} \right)  \frac{\Omega_{1\k} }{(k_B T)^2} \\
& \approx \frac{C_1}{2\pi\,\hbar\,T^2} \mathrm{e}^{-\Delta/k_B T} \left( \Delta^2 +\mathrm{e}^{-B_z/k_B T} (\Delta+B_z)^2 \right). 
\end{align}
In moderate magnetic fields $B_z > T$, $\kappa^{}_{xy}/T$ decays exponentially as $(\Delta/ T)^2 \exp(-\Delta/k_B T)$ at low temperatures, in agreement with the regime of $T \lesssim 0.5$ in Fig.~\ref{fig:1OK3}. 
Concurrently, Eq.~\eqref{eq:lowT} tells us about the dependence of $\kappa^{}_{xy}$ on the external magnetic field. Recognizing that the spinon gap $\Delta$ at a finite field $B_z$ is related to the zero-field gap $\Delta_0$ as $\Delta = \Delta_0 - B_z/2$, we find that
\begin{align}
\label{eq:kappaVsBatLowT}
\nonumber \frac{\kappa^{}_{xy}}{T} & \approx \frac{C_1}{2\pi\,\hbar\,T^2} \mathrm{e}^{-\Delta_0/k_B T} \left( \sum_{n=\pm} \mathrm{e}^{-nB_z/2k_B T} (\Delta_0+nB_z/2)^2 \right)\\
&\approx \frac{C_1}{\pi\,\hbar\,T^2} \cosh\left(\frac{B_z}{2 k_B T}\right) \mathrm{e}^{-\Delta_0/k_B T }
\end{align}
for small $B_z \ll \Delta_0$, thereby justifying the nonlinear behavior observed in Fig.~\ref{fig:1OK1}.

Another interesting limit is the intermediate temperature range when $\Delta < \max \varepsilon_{1, \mathbf{k}}\lesssim T \lesssim \min \varepsilon_{2, \mathbf{k}}$. From the aforementioned calculations, we notice that the thermal Hall conductivity is the largest in this two-band picture when the magnetic field splits the particle and hole bands---both of which have nonzero Chern numbers---such that the temperature $T$ is greater than the lower-band maximum, but smaller than the upper-band minimum.

With our formalism, we can also study phases with magnetic order at $T=0$, but with restored SRI due to thermal fluctuations at nonzero temperature. To this end, we vary the gap such that it is exponentially small with temperature; in practice, this is achieved by tuning the Lagrange multiplier $\lambda$. Instead of performing a self-consistent calculation, we assume a functional form  $\Delta (T) = T \exp(-m /T)$, $m = 2\pi \rho_s$ (with spin stiffness $\rho_s$), in analogy with the two-dimensional (2D) antiferromagnetic Heisenberg model \cite{chn1988, chn1989, sachdev2011quantum}. The variation of $\kappa^{}_{xy}/T$ with this choice of $\Delta (T)$ is conveyed by Fig.~\ref{fig:1OK4}. Despite always being in the regime $\Delta < T$, $\kappa^{}_{xy}/T$ does not diverge as $T \rightarrow 0$, but instead tends to zero. To understand this, we focus on the contribution from the lowest band and momenta close to the dispersion minima $\pm \mathbf{k}_0$. Near $\pm \mathbf{k}_0$, the momentum dependence of the energy is quadratic, while that of the Berry curvature is empirically observed to be quartic. Accordingly, assuming $\Delta = 0$,
\begin{alignat}{1}
&\frac{\kappa^{}_{xy}}{T} \approx  -\frac{k_B^2}{\hbar\, V} \sum_{\mathbf{k}} \frac{\Omega_{1\mathbf{k}}}{n_B(\varepsilon_{1 \mathbf{k}})}\\
\nonumber
&\approx{-\frac{k_B^2}{\hbar}}\hspace*{-0.3cm} \int\displaylimits_{\lvert \mathbf{k} - \mathbf{k}_0 \rvert < \Lambda } \hspace*{-0.2cm}\frac{\mathrm{d}^2\mathbf{k}}{(2\pi)^2}\, \Omega_0 (\mathbf{k} - \mathbf{k}_0)^4\, c^{}_2 \left(\frac{1}{\mathrm{e}^{(\mathbf{k} - \mathbf{k}_0 )^2/2m^* T}-1} \right)\\
\nonumber
&-{\frac{k_B^2}{\hbar}} \hspace*{-0.3cm}\int\displaylimits_{\lvert \mathbf{k} + \mathbf{k}_0 \rvert < \Lambda }\hspace*{-0.2cm} \frac{\mathrm{d}^2\mathbf{k}}{(2\pi)^2} \,\Omega_0 (\mathbf{k} +\mathbf{k}_0)^4 \,c^{}_2 \left(\frac{1}{\mathrm{e}^{(\mathbf{k} + \mathbf{k}_0 )^2/2m^* T}-1} \right).
\end{alignat} 
As $T \rightarrow 0$, we may rescale $\mathbf{k} \pm \mathbf{k}_0 = \mathbf{y} \sqrt{2 m^* T}$ and extend the upper limit of $\mathbf{y}$ integration to infinity, to obtain
\begin{eqnarray}
\frac{\kappa^{}_{xy}}{T} &=& - \frac{2 k_B^2 (2 m^* T)^3 \Omega_0}{\hbar } \int_0^\infty \frac{y^5 dy}{2 \pi} c_2 \left(\frac{1}{e^{y^2} - 1} \right) \nn
&=& - \frac{2 k_B^2 (2 m^* T)^3 \Omega_0}{\hbar } \,(5.78117\ldots)
\label{kappaMagneticOrder}
\end{eqnarray}
So we find that $\kappa^{}_{xy}/T \sim T^3$ as $T \rightarrow 0$ with $\Delta \ll T$.

\section{Antiferromagnet with Dzyaloshinskii-Moriya interactions}
\label{sec:DM}
So far, our discussion has been confined exclusively to spin-rotation-invariant spin liquids. In this section, we will extend the analysis to include spin-orbit coupling, i.e.~spin-rotations are not independent symmetry operations any more, but are coupled with real-space symmetry transformations. In terms of the underlying spin model, this corresponds to including DM interactions \cite{dzyaloshinsky1958thermodynamic, moriya1960anisotropic, moriya1960new} as described by the term proportional to $\vec{D}^\textsc{m}_{ij}$ in Eq.~\eqref{eq:GeneralSpinHamWithDMTerm},
\begin{equation}
\label{eq:H2}
H^{(3)} = \sum_{\langle i, j \rangle} \vec{D}^\textsc{m}_{ij} \cdot \left(  \mathbf{S}_i \times \mathbf{S}_j \right) .
\end{equation}

The thermal transport properties of a spin Hamiltonian with DM coupling were studied on the kagom\'{e} lattice in Ref.~\onlinecite{Lee2015} for the magnetically ordered phase using both Holstein-Primakoff bosons and Schwinger bosons; in particular, the latter approach featured a large thermal Hall coefficient at $B_z, T \sim J$. On the square lattice, however, it is strongly constrained by no-go theorems \cite{katsura2010theory,ideue2012effect}. In a recent spin-wave analysis, Ref.~\onlinecite{kawano2018thermal} demonstrated that a thermal Hall effect can be realized in an inversion-symmetry-broken square-lattice antiferromagnet with DM couplings. Here, we move away from the magnon description, which necessarily requires long-range magnetic order, and probe the influence of the DM interactions relevant to the cuprate superconductors in a spin-liquid phase using Schwinger bosons. We will show that some of these DM vectors can lead to a nonzero Berry curvature, $\Omega_{n {\bf k}}\neq 0$, and, in turn, a nonzero thermal Hall coefficient, albeit with much smaller magnitude than in the ansatz of \secref{AnsatzWithChernNumbers}. This is related to the fact that the Chern number vanishes for each band in the models with DM interactions that we study here.

\begin{figure}[tb]
    \centering
    \includegraphics[width = 0.9\linewidth]{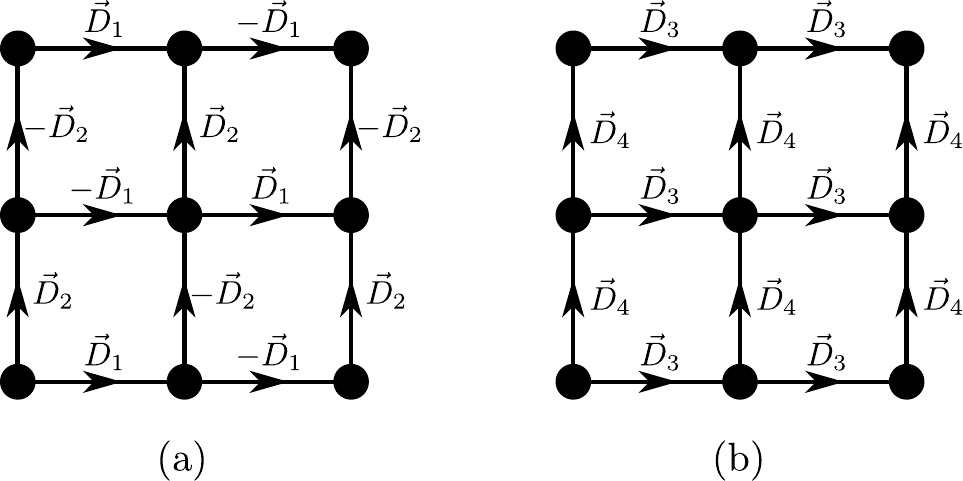}
    \caption{Illustration of the DM coupling vectors \cite{coffey1990effective, Coffey} for (a) orthorhombic La$_2$CuO$_4$ and (b) YBCO, where the black dots represent the Cu atoms of the CuO$_2$ planes and $\vec{D}_{1}=(d_1,d_2,0)^T$, $\vec{D}_{2}=(-d_2,-d_1,0)^T$, $\vec{D}_{3}=(d_3,0,0)^T$, $\vec{D}_{4}=(0,d_3,0)^T$ with real constants $d_j$ (not determined by symmetry). Given that $\vec{D}^\textsc{m}_{ij}=-\vec{D}^\textsc{m}_{ji}$, the DM coupling vector $\vec{D}^\textsc{m}_{ij}$ corresponds to a directed bond, which is indicated by the arrows in the figure. The different DM textures are due to the different symmetries: in YBCO, the Cu atoms are not centers of inversion, which allows a spatially constant DM coupling vector; in La$_2$CuO$_4$, it must alternate in sign since the Cu atoms are inversion centers, which is permitted because of the broken translational symmetry.}
    \label{DMVectorsOfCuprates}
\end{figure}

We will focus here on the Zeeman coupling of the magnetic field and neglect orbital effects. In this case, only a certain class of DM coupling vectors can lead to $\kappa^{}_{xy}\neq 0$ due to symmetry constraints. For instance, consider global spin rotations by angle $\lvert \vec{\varphi}\rvert$ along axis $\vec{\varphi}/|\vec{\varphi}|$. Under these transformations, it holds that $J_{ij}\rightarrow J_{ij}$, $\vec{B}_Z\rightarrow R_{\vec{\varphi}}\vec{B}_Z$, and $\vec{D}^\textsc{m}_{ij}\rightarrow R_{\vec{\varphi}}\vec{D}^\textsc{m}_{ij}$, where $R_{\vec{\varphi}}$ is the vector representation of the spin rotation. As for any spin-rotation-invariant observable, the thermal Hall conductivity $\kappa^{}_{xy}$ satisfies
\begin{equation}
    \kappa^{}_{xy}[J_{ij},\vec{D}^\textsc{m}_{ij},\vec{B}_Z] = \kappa^{}_{xy}[J_{ij},R_{\vec{\varphi}}\vec{D}^\textsc{m}_{ij},R_{\vec{\varphi}}\vec{B}_Z]. \label{Sym1}
\end{equation}
Being odd under time-reversal, it further obeys
\begin{equation}
    \kappa^{}_{xy}[J_{ij},\vec{D}^\textsc{m}_{ij},\vec{B}_Z] = -\kappa^{}_{xy}[J_{ij},\vec{D}^\textsc{m}_{ij},-\vec{B}_Z]. \label{Sym2}
\end{equation}
Consequently, if the DM coupling vectors are collinear, i.e. $\vec{D}^\textsc{m}_{ij}\propto \hat{\vec{d}}$, and $\hat{\vec{d}} \cdot \vec{B}_Z=0$, the combination of Eqs.~\eqref{Sym1} and \eqref{Sym2}, with $\vec{\varphi}=\pi \hat{\vec{d}}$, implies $\kappa^{}_{xy}=0$. To wit, this is the case for $\vec{D}^\textsc{m}_{ij}=D_0\hat{x}$, or for the potentially more relevant (spatially alternating) DM coupling vector of the tetragonal phase of La$_2$CuO$_4$ \cite{Coffey}. 

It is also easily seen that $\kappa^{}_{xy}$ vanishes for the DM coupling vector in the orthorhombic phase of La$_2$CuO$_4$ [\figref{DMVectorsOfCuprates}(a)]: the spatial reflection symmetry $\mathcal{R}_y$ with action $(x,y)\rightarrow (-x,y)$, not combined with any rotation in spin space, remains a symmetry of the system also in the presence of Zeeman field along $\hat{z}$. Being odd under $\mathcal{R}_y$, $\kappa^{}_{xy}$ has to vanish.

This is different for the DM coupling vector expected to arise in the tetragonal phase of YBa$_2$Cu$_3$O$_{6+x}$ (YBCO) \cite{Coffey}, shown in \figref{DMVectorsOfCuprates}(b), which analytically corresponds to
\begin{equation}
\vec{D}^\textsc{m}_{ij} = D_{\parallel}\, \hat{d}_{ij},\, \quad \hat{d}_{ij} = d_{ij} \left(\cos \theta_{ij}\, \hat{x} + \sin \theta_{ij} \,\hat{y} \right), \label{DMYBCO}
\end{equation}
where $\hat{d}_{ij}$ is a unit vector, $d_{ij} = - d_{ji} = \pm 1$ for $j = i \pm \hat{e}_\mu$ ($\mu = x,y$), and $\theta_{ij} = 0$ ($\pi/2$) on all $x \,(y)$ bonds. Note that this form of $\vec{D}^\textsc{m}_{ij}$ respects the translational and fourfold-rotational ($C_4$) symmetries of the underlying square lattice (when accompanied by an appropriate rotation in spin space). It is not collinear and does break time-reversal symmetry [the argument in Eqs.~\eqref{Sym1}--\eqref{Sym2} does not apply]; furthermore, it also breaks all in-plane reflection symmetries in combination with a Zeeman field and will indeed give rise to a nonzero thermal Hall response as we will show next.  

\begin{figure}[htb]
\subfigure[]{\label{fig:Ea}\includegraphics[width=0.494\linewidth]{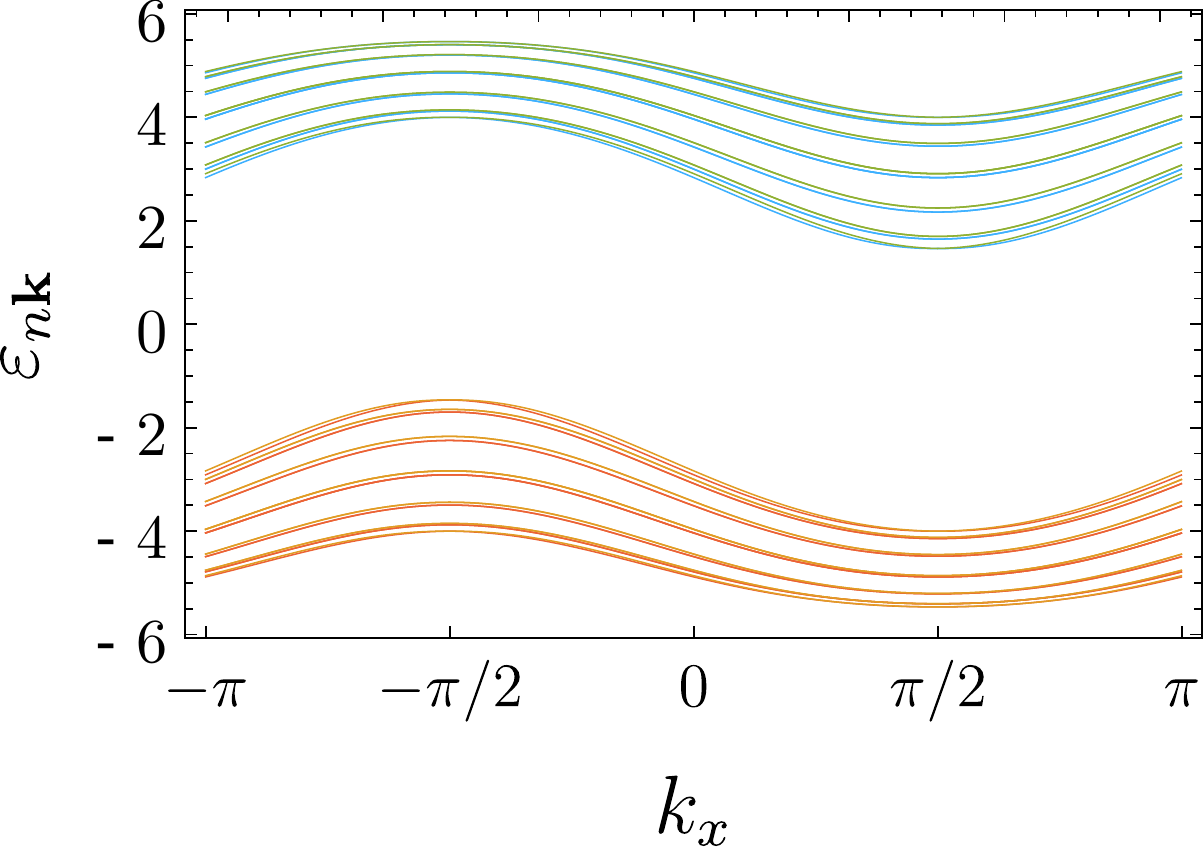}}
\subfigure[]{\label{fig:Eb}\includegraphics[width=0.494\linewidth]{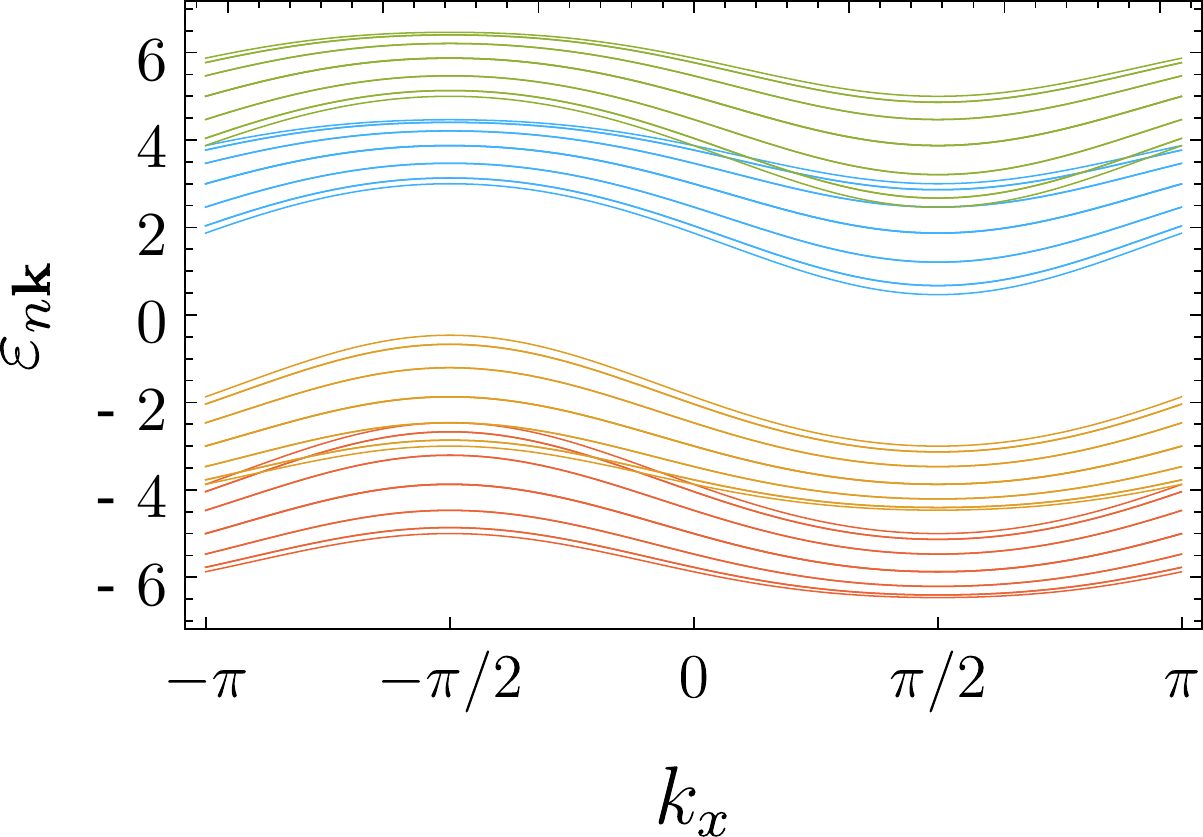}}
\subfigure[]{\label{fig:Ec}\includegraphics[width=0.494\linewidth]{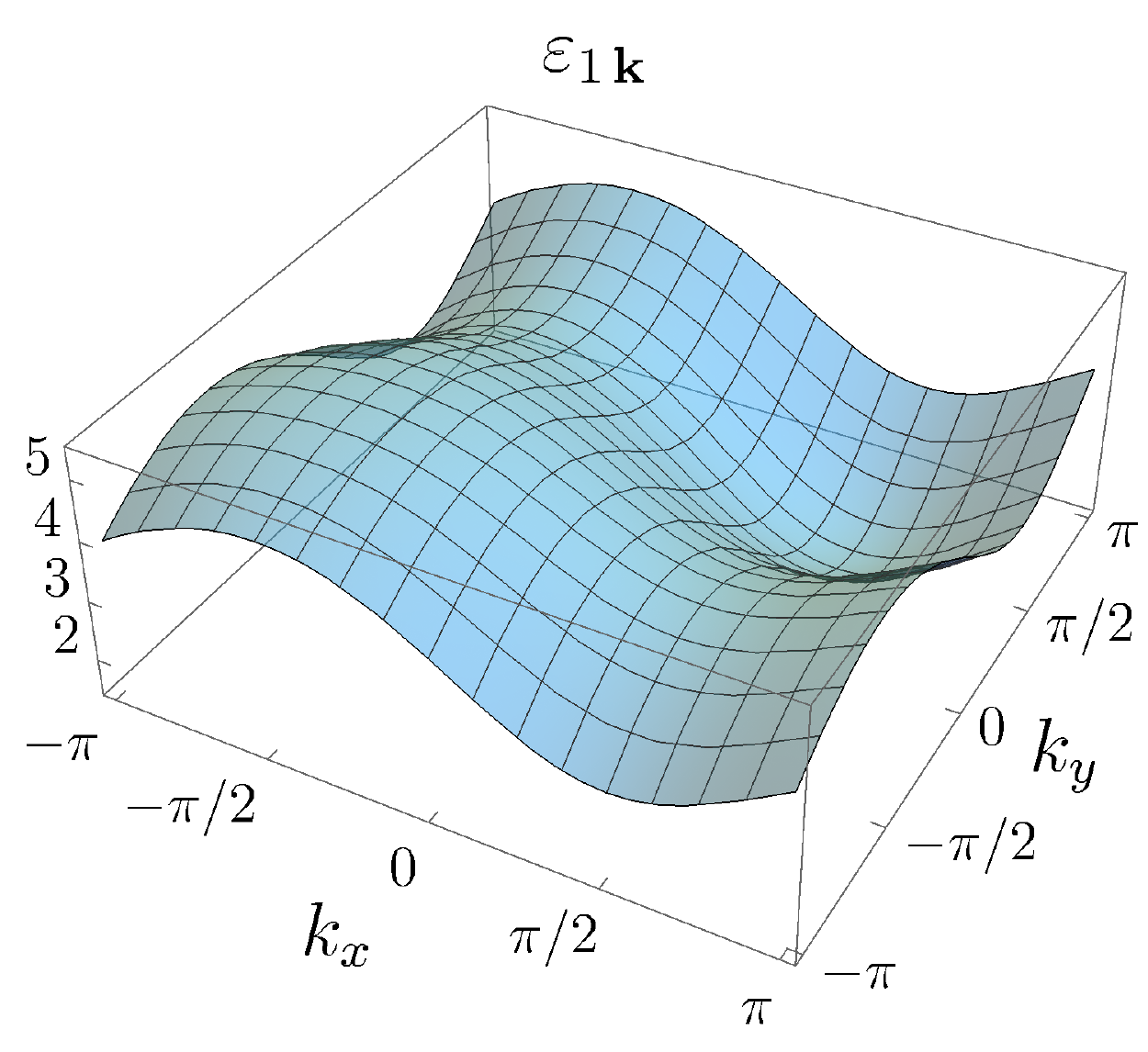}}
\subfigure[]{\label{fig:Ed}\includegraphics[width=0.494\linewidth]{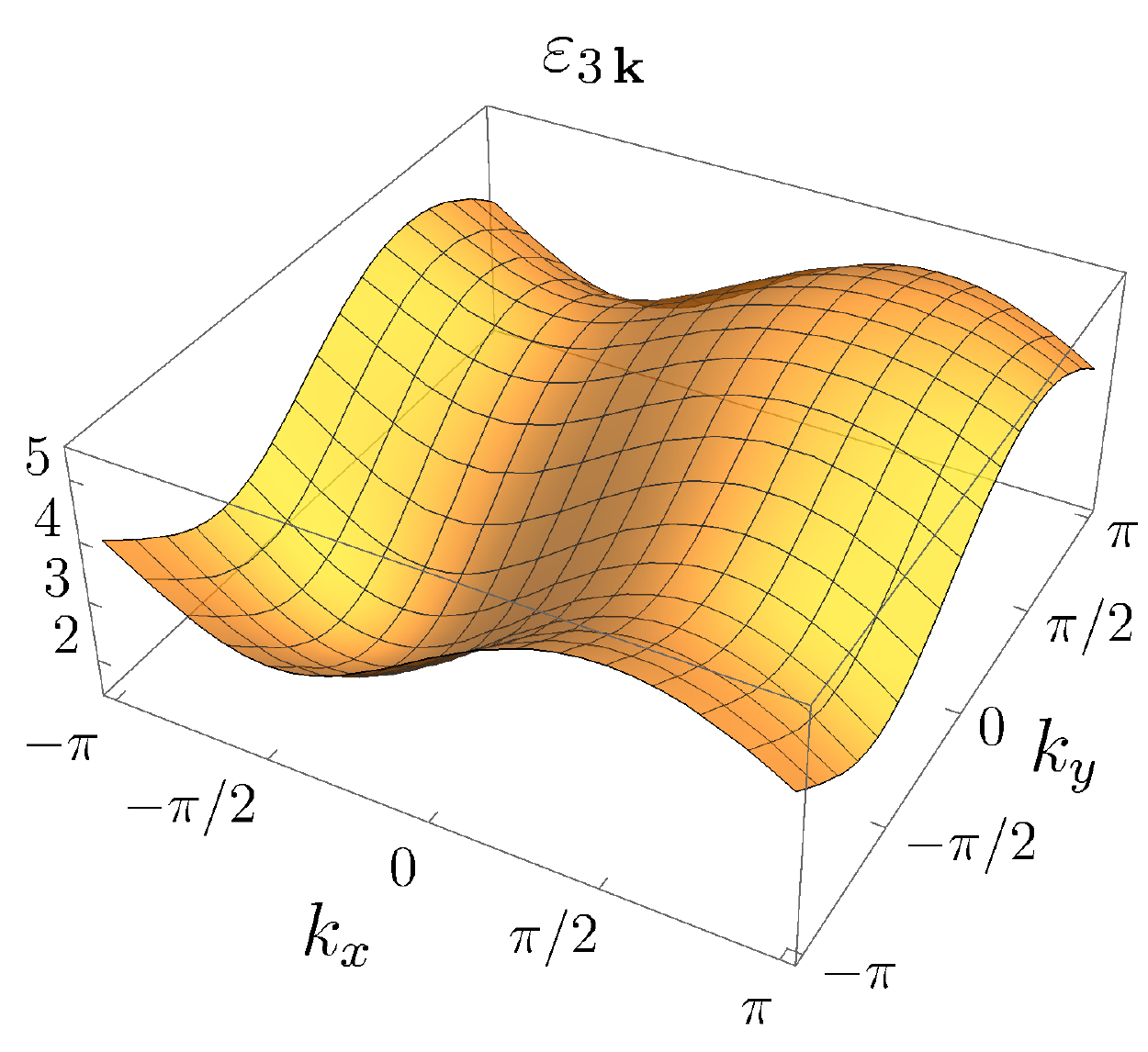}}
\subfigure[]{\label{fig:OIa}\includegraphics[width=0.494\linewidth, trim={0.5cm 0.2cm 0.5cm, 0.18cm},clip]{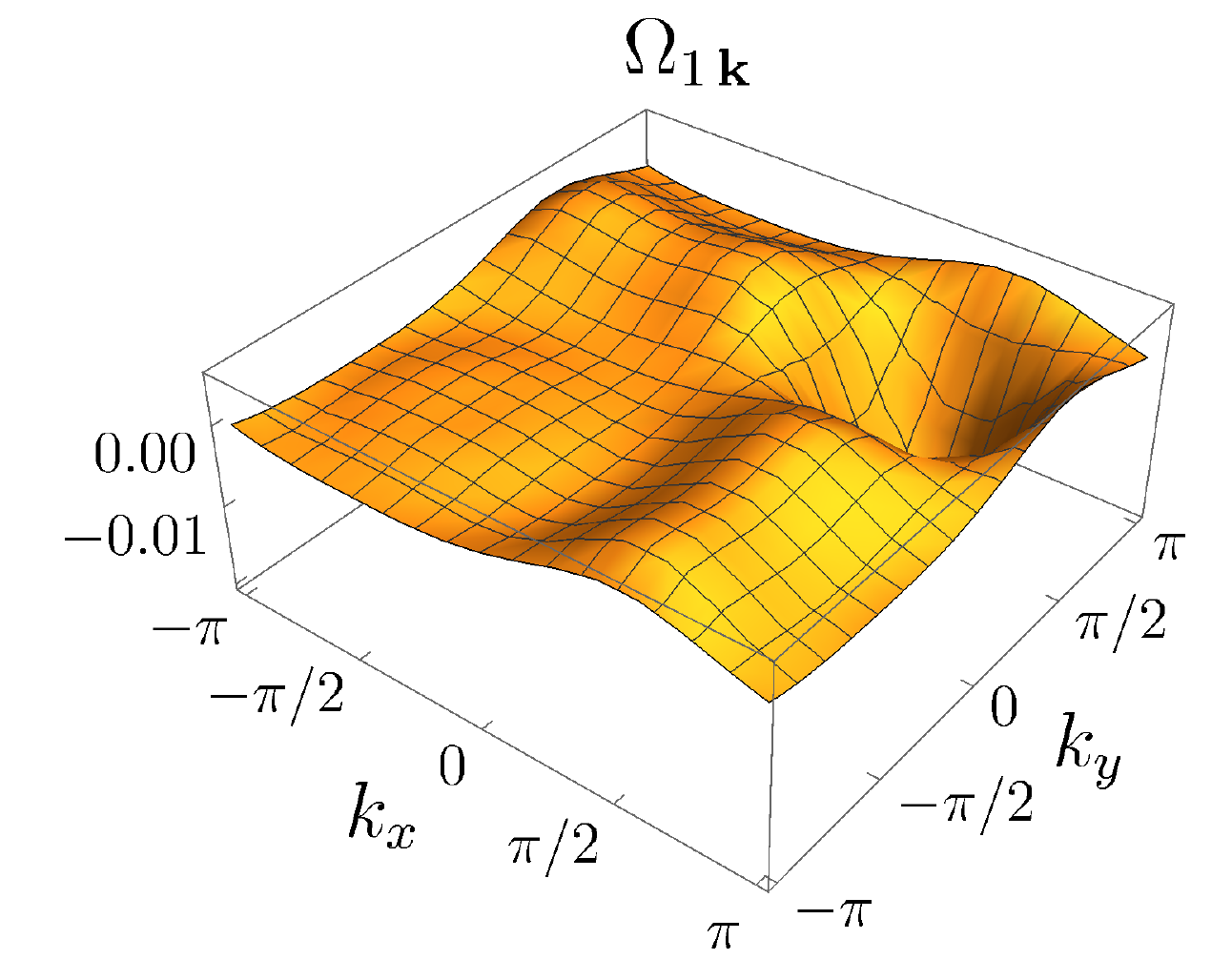}}
\subfigure[]{\label{fig:OIb}\includegraphics[width=0.494\linewidth,trim={0.5cm 0.2cm 0.5cm, 0.18cm},clip]{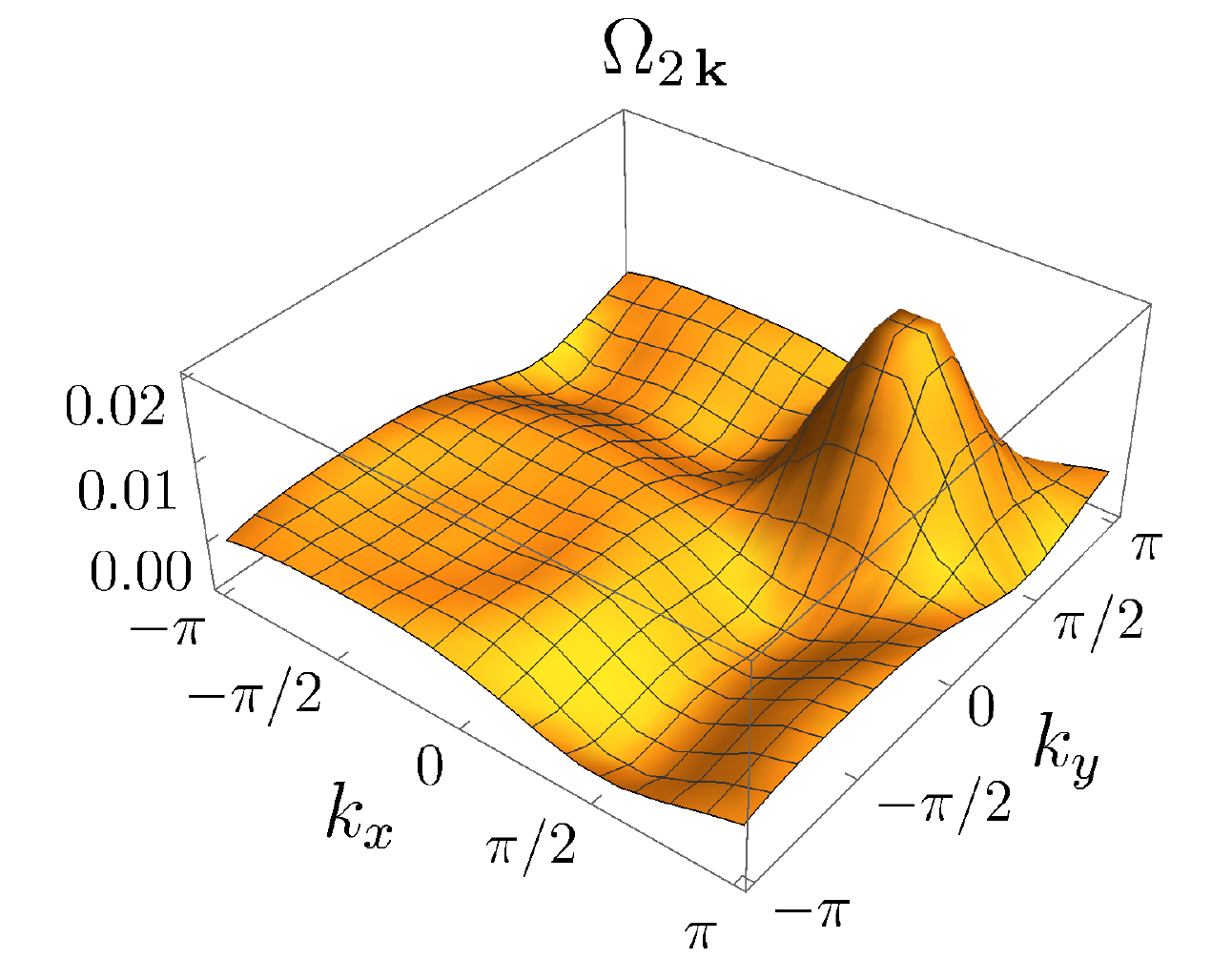}}
\caption{\label{fig:Evsk3D}(a--b): Dispersion of the Schwinger boson bands for the mean-field approximation to $H_\mathrm{spin}$ \eqref{eq:GeneralSpinHamWithDMTerm}, with $J_x\mc{A} = J_y\mc{A} = 1$, $\mc{B} = 0.5 \mathrm{i}$, and $D_{\parallel} = 0.10$, in (a) zero and (b) large ($B_z=2$) magnetic fields. Shown are the eigenvalues of the dynamic matrix---the bosonic bands have energies given by the absolute values of the same, which are always positive. In a finite magnetic field, the individual particle and hole bands become progressively well separated. Exactly as in \figref{SpectrumOfPatternAandB}, the lines refer to different values of $k_y=-\pi, -\pi+\pi/6,\ldots, \pi$. (c--d): Same as above but now plotted in the $k_x$--$k_y$ plane for the (c) $n=1$ (blue; particle) and (d) $n=3$ (yellow; hole) bands, at $B_z=0$---the two bands are nonidentical. At each point in $\mathbf{k}$-space, $\min\, (\varepsilon_{1 \mathbf{k}}, \varepsilon_{3 \mathbf{k}})$ corresponds to the lowest energy eigenmode and the band minima are at $\{(\pi/2, \pi/2),(-\pi/2, -\pi/2)\}$. Condensation of these Schwinger bosons generally leads to long-range antiferromagnetic order. (e--f) Berry curvatures of the particle bands with the same parameters as before, and a magnetic field $B_z = 0.5$.}
\end{figure}

To proceed with the Schwinger boson description of the DM interactions, we define the additional operators
\begin{alignat*}{2}
\hat{\mc{C}}^\dagger_{i,j} &= \frac{1}{2} \sum_{\mu\, \nu} b^\dagger_{i \mu} \left( \mathrm{i} \,\hat{d}_{ij}\cdot \vec{\sigma} \right)_{\mu \nu} b^{}_{j \nu} &&= \frac{\mathrm{i}}{2}\, d_{ij}\, \mathrm{e}^{-\mathrm{i}\, \sigma\,\theta_{ij}} \,b^\dagger_{i \sigma} b^{}_{j -\sigma},\\
\hat{\mc{D}}_{i,j} &= \frac{1}{2} \sum_{\mu\, \nu} b^{}_{i \mu} \left( \sigma^{}_2 \,\hat{d}_{ij}\cdot \vec{\sigma} \right)_{\mu \nu} b^{}_{j \nu} &&= -\frac{\mathrm{i}}{2}\, \sigma\,d_{ij}\, \mathrm{e}^{\mathrm{i}\, \sigma\,\theta_{ij}} \,b^{}_{i \sigma} b^{}_{j \sigma},
\end{alignat*}
whereupon the DM term can be decomposed as \cite{manuel1996heisenberg}
\begin{alignat}{1}
\nonumber\hat{d}_{ij} \cdot \left( \mathbf{S}_i \times \mathbf{S}_j \right) = \frac{1}{2} \bigg(:&\hat{\mc{B}}^\dagger_{i,j} \hat{\mc{C}}_{i,j} + \hat{\mc{C}}^\dagger_{i,j} \hat{\mc{B}}_{i,j}: \\
+ &\hat{\mc{A}}^\dagger_{i,j} \hat{\mc{D}}_{i,j} + \hat{\mc{D}}^\dagger_{i,j} \hat{\mc{A}}_{i,j} \bigg).
\label{eq:DM_identity}
\end{alignat}
Assuming only SU(2) spin-rotation-invariant operators acquire nontrivial expectation values in the mean-field decoupling (e.g., $\hat{\mc{B}}^\dagger_{i,j} \hat{\mc{C}}_{i,j}\rightarrow \braket{\hat{\mc{B}}^\dagger_{i,j}}\hat{\mc{C}}_{i,j} +\text{const.}$), the SBMFT analysis is carried out in Appendix~\ref{sec:DM_Calc} to obtain the dispersion of the bosonic bands for a zero-flux ansatz appropriate to a conventional N\'eel state \cite{yang2016schwinger}, $\mc{A}_{i,i+\mu} = \mc{A}$, $\mc{B}_{i,i+\mu} = \mc{B}$ $\forall\, i$, $\mu=\hat{x},\hat{y}$, taking the DM coupling vector defined in \equref{DMYBCO} and \figref{DMVectorsOfCuprates}(b).
Note that considering only SU(2)-invariant operators does \textit{not} mean that the resulting mean-field Hamiltonian preserves SRI since the DM term in \equref{eq:DM_identity} couples the operators, $\hat{\mc{A}}_{i,j}$ and $\hat{\mc{B}}_{i,j}$, that are spin rotation invariant, to $\hat{\mc{C}}_{i,j}$ and $\hat{\mc{D}}_{i,j}$, which are not.

\begin{figure}[htb]
\centering
     \includegraphics[height=0.61\linewidth, trim={3.2cm 19cm 9cm 2.5cm},clip]{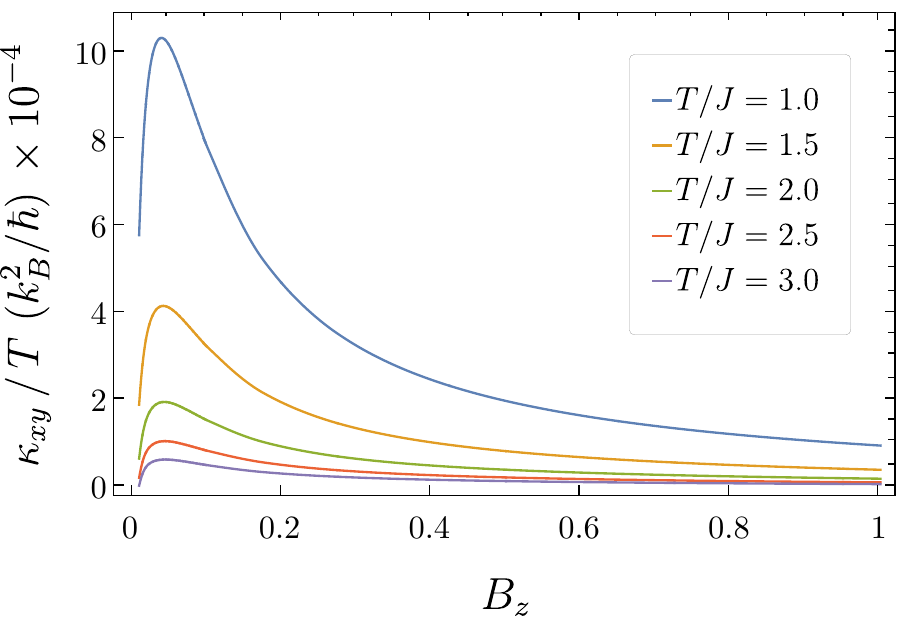}
    \includegraphics[height=0.60\linewidth,trim={3cm 19cm 9cm 2.5cm},clip]{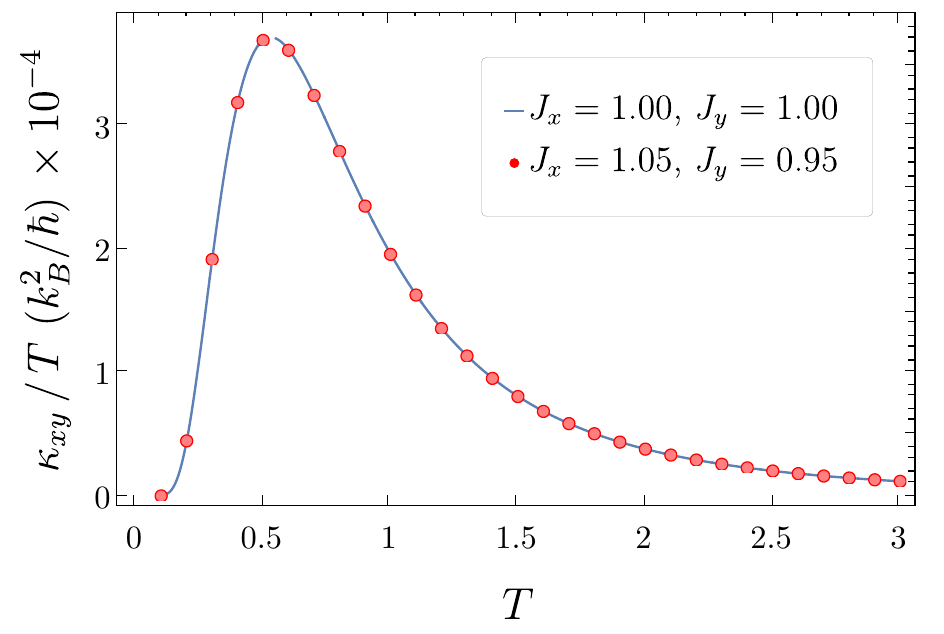}
    \caption{The thermal Hall conductivity in an antiferromagnetic Heisenberg spin model with Dzyaloshinskii-Moriya interactions, as a function of magnetic field for different constant temperatures (top) and as a function of temperature at a constant magnetic field $B_z = 0.5$ (bottom). Although not clearly visible in the numerical data, $\kappa^{}_{xy}$ has to vanish exactly at zero field (no anomalous contributions) as dictated by symmetry.  The couplings considered are $J_x\mc{A} = J_y\mc{A} = 1$ (solid lines in both plots) and $J_x\mc{A} = 1.05, J_y\mc{A} = 0.95$ (red dots), with all other parameters the same as in Fig.~\ref{fig:Evsk3D}. When $J_x \neq J_y$, $C_4$ rotational symmetry is broken. The Schwinger-boson bands do not acquire nontrivial Chern numbers in the model considered, and $\kappa^{}_{xy}$ is thus much smaller than for the spin-liquid ans\"{a}tze in Sec.~\ref{sec:1O}.}
    \label{fig:Hall_DM}
\end{figure}

Unlike previously, there is no effective antiunitary symmetry and therefore, the bands are nondegenerate even in zero fields. Nonetheless, in the absence of a magnetic field, the two particle (and hole) bands intersect at a finite number of points as can be seen in Fig.~\ref{fig:Ea}, so the Berry curvatures are well-defined only for $B_z \ne 0$. These are plotted for the Schwinger-boson particle bands in Figs.~\ref{fig:OIa} and \ref{fig:OIb}; the curvatures of the hole bands are related by Eq.~\eqref{eq:relate}. Despite a nonvanishing Berry curvature, each bands is actually topologically trivial with zero Chern number.

The ensuing thermal Hall conductivities, which can be calculated directly using the formalism of Sec.~\ref{sec:procedure}, are found to be more than two orders of magnitude smaller than for the earlier spin-liquid ans\"{a}tze that result in nonzero Chern numbers. Although the Hall coefficients are nonzero, as displayed in Fig.~\ref{fig:Hall_DM}, this is a purely thermal effect in the sense that the main contribution to $\kappa^{}_{xy}$ comes from asymmetric weighting of the Berry curvature by the thermal distribution function $n_B \left(\varepsilon_{n \mathbf{k}} \right)$ in Eq.~\eqref{eq:k_xy} because the integral of $\Omega_{n \mathbf{k}}$ over the Brillouin zone alone is identically zero.
We also remark that there is no anomalous contribution as time-reversal symmetry is preserved at zero Zeeman field, guaranteeing that $\kappa^{}_{xy}=0$.

Since the CuO$_2$ square plaquettes in YBCO are slightly distorted and form a rectangular lattice \cite{mook2000one, hinkov2004two}, we have also studied the impact of anisotropic Heisenberg exchanges $J_x$ and $J_y$ along the $\hat{x}$ and $\hat{y}$ directions, respectively; this breaks the $C_4$ rotation symmetry down to $C_2$. As demonstrated by \figref{fig:Hall_DM}, even a moderately large anisotropy has no significant impact on $\kappa^{}_{xy}$.

\section{Conclusion}\label{Conclusion}

Our primary collection of results concerns the thermal Hall effect of spin liquids on the square lattice using SBMFT in the absence of spin-orbit coupling. We have discussed different spin-rotation and translation-invariant ans\"atze that break time-reversal and certain point group symmetries; these phases exhibit nonzero scalar spin chiralities. Among the ans\"atze considered, only one, with magnetic point group $\frac{4}{m}m'm'$ and defined in \figref{SBAnsatzPatternDOneObital}, yields spinon bands with nonzero Chern numbers. As seen in Fig.~\ref{fig:1OTH}, where the Zeeman field, $B_z$, and temperature, $T$, dependence of the resulting thermal Hall conductivity $\kappa^{}_{xy}$ are shown, the nonzero Chern numbers lead to a sizable $\kappa^{}_{xy}$, of order one in units of $k_B^2/\hbar$. We derived asymptotic expressions for the dependence of $\kappa^{}_{xy}$ on $T$ and $B_z$, and established that $\kappa^{}_{xy}/T$ vanishes as $\sim \exp(-\Delta_0/T)$ at low $T$ for a spin liquid with a nonzero energy gap $\Delta_0$.

Our formalism also enables us to consider states in which spin-rotation symmetry is broken and there is magnetic order as $T \rightarrow 0$. Any broken spin rotation symmetry is restored at infinitesimal temperatures in two spatial dimensions, and within SBMFT, this can be captured by a spin liquid with a gap, $\Delta$, which vanishes as $\Delta \sim \exp(-m/T)$. In this case, we found that $\kappa^{}_{xy}/T$ acquired similarly large values (Fig.~\ref{fig:1OTH}), and vanished only as a power of $T$ as $T \rightarrow 0$.

The spin-liquid states with $\frac{4}{m}m'm'$ symmetry descend from the time-reversal-preserving 
$\pi$-flux SBMFT states of Yang and Wang \cite{yang2016schwinger}. As such, they do not have a special connection to the N\'eel state in the limit of a vanishing spin gap. However, our spin liquids do include cases in which they condense to small distortions of the N\'eel state, although there is no natural selection mechanism for such states, at least in mean-field theory. With such a selection mechanism, our results yield an attractive proposal to explain recent observations in the cuprates \cite{grissonnanche2019giant}.

The breaking of square-lattice and time-reversal symmetries to $\frac{4}{m}m'm'$ in our states could either be spontaneous, or simply induced by the orbital coupling of the applied magnetic field (see Appendix~\ref{app:orbital}). Only for the case when the symmetries are
spontaneously broken, there is an anomalous contribution to the thermal Hall effect, i.e. $\kappa^{}_{xy} \neq 0$ even when $B_z=0$. 

Finally, we also discussed whether the DM interactions relevant to the cuprates can give rise to a thermal Hall effect within a SBMFT treatment of the spin model in \equref{eq:GeneralSpinHamWithDMTerm}. We identify one DM coupling vector, defined in \equref{DMYBCO} and in \figref{DMVectorsOfCuprates}(b), which not only is expected to be realized in YBCO \cite{Coffey} but also produces a nonzero $\kappa^{}_{xy}$. However, as evinced by Fig.~\ref{fig:Hall_DM}, the thermal Hall conductivity is much weaker than that of the ansatz in \figref{SBAnsatzPatternDOneObital} with $\frac{4}{m}m'm'$ symmetry, 
due to the absence of bands with nontrivial Chern numbers. \\
{\it Notes added:} ({\it i\/}) In a recent paper with others \cite{CGSSSSX}, we have discussed the thermal Hall response of antiferromagnets using fermionic spinons.
({\it ii\/}) Han {\it et al.\/} \cite{Lee19} have described the thermal Hall response of the cuprates using a quantum spin Hall state, that could be favored by spin-orbit interactions.

\section*{Acknowledgments}

We acknowledge many insightful discussions with G.~Grissonnanche and L.~Taillefer, and thank them for sharing their results before publication. We also thank A.~Rosch and A.~Vishwanath for useful discussions. This research was supported by the National Science Foundation under Grant No.~DMR- 1664842. Research at Perimeter Institute is supported by the Government of Canada through Industry Canada and by the Province of Ontario through the Ministry of Research and Innovation. SS also acknowledges support from Cenovus Energy at Perimeter Institute. MS acknowledges support from the German National Academy of Sciences Leopoldina through grant LPDS 2016-12. SC acknowledges support from the ERC synergy grant UQUAM.

\appendix

\section{Coupling to an orbital magnetic field}
\label{app:orbital}
Aside from the Zeeman coupling \eqref{eq:H2mf}, which we focused on in the main text, there is also an orbital coupling of the magnetic field. Being odd under time reversal and spin-rotation invariant, its leading contribution in a $t/U$ expansion of the underlying Hubbard model involves the triple product of neighboring spins and is of order $t^3/U^2$. Explicitly, it reads as \cite{sen1995large} \begin{alignat}{1}
H_{\chi}  = -\Upsilon \sum_{\triangle} \sin(\Phi)\, \S_i \cdot (\S_j \times \S_k), \quad \Upsilon = \frac{24 t^2t'}{U^2},
\label{eq:chi}
\end{alignat}
where the sum involves the triangular plaquettes $\triangle$ formed by nearest-neighbor (with hopping $t$) and next-nearest-neighbor bonds (hopping $t'$), and $\Phi$ is the flux of an applied magnetic field through a single triangular plaquette. We see from \equref{eq:chi} that this orbital coupling induces uniform scalar spin chiralities and, as mentioned earlier, breaks the symmetry of the system to $\frac{4}{m}\,m'm'$. 

In this appendix, we prove that the different terms in \equref{eq:chi} cancel out exactly on the square lattice after performing a Schwinger-boson mean-field decoupling, as long as there exists a gauge where the ansatz is explicitly translation invariant. This is, for instance, certainly the case for the conventional ansatz of the antiferromagnetic state (with only $\mc{A}_{i,i+\hat{x}}=\mc{A}_{i,i+\hat{y}}=\mathcal{A}_1$), but not for the one with $\frac{4}{m}\,m'm'$ symmetry defined in \secref{AnsatzWithChernNumbers}. As we outline below, $H_{\chi}$ in \equref{eq:chi} will lead to a nonzero contribution at the mean-field level when decoupled with the parameters in \equsref{eq:subeqns}{FullAnsatz1}.

\begin{figure}
\includegraphics[width=0.75\linewidth, trim={2cm 2cm 2cm 2cm}, clip]{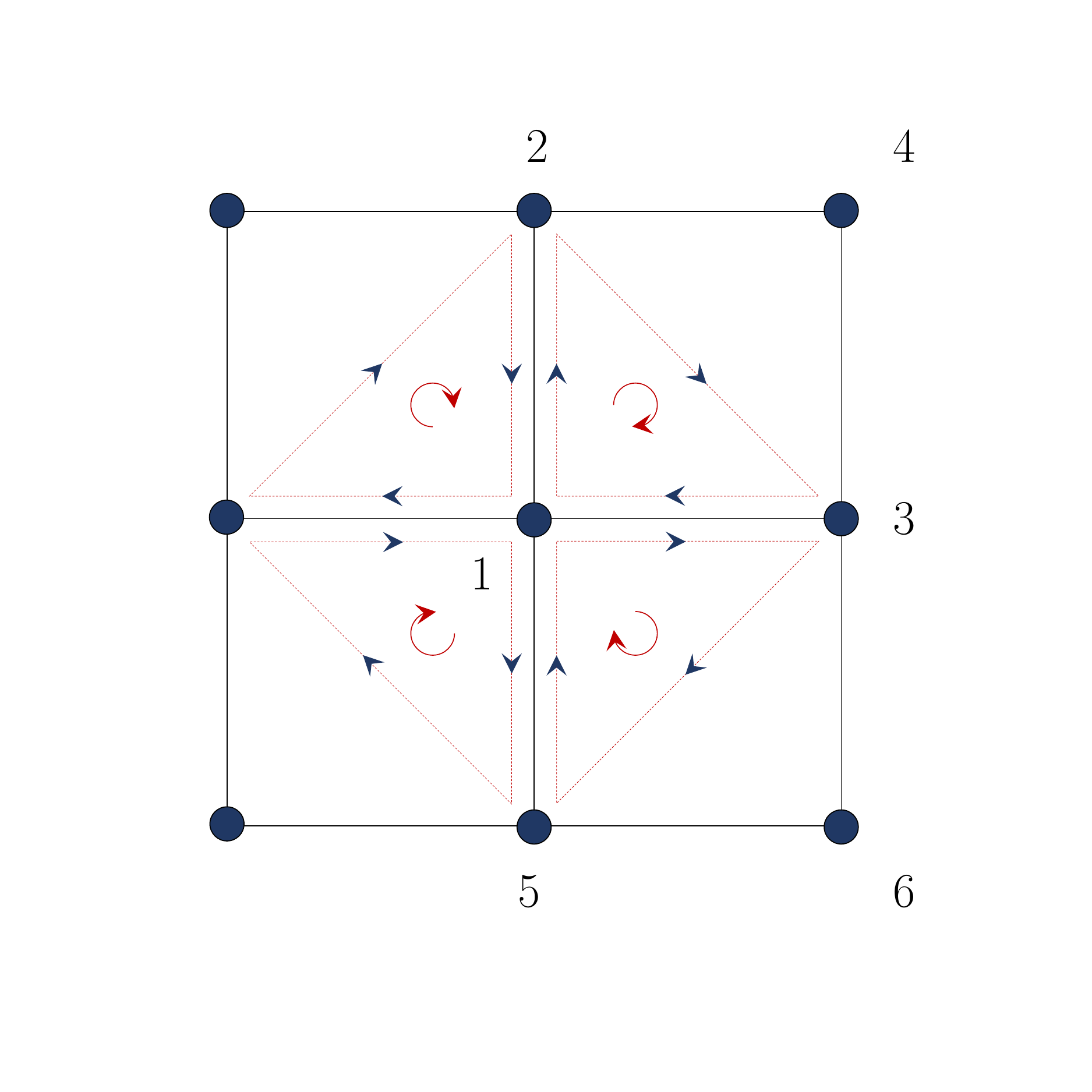}\caption{\label{fig:DMConv}Convention for the spin chirality term $\mathbf{S}_i \cdot\left(\mathbf{S}_j \times \mathbf{S}_k \right)$ in the Hamiltonian. For each triangular plaquette, the sites $i$, $j$, and $k$ are the vertices of the corresponding dashed triangle, taken succesively in a clockwise fashion. The net interaction $H_\chi$ involves the sum over all $C_4$ rotated copies of such triangles.}
\end{figure}

As a means of decoupling $H_\chi$ within SBMFT, we use the identity \cite{messio2013time}, 
\begin{alignat}{2}
\nonumber    4 : \hat{\mc{B}}_{i,j}\, \hat{\mc{B}}_{j,k}\, \hat{\mc{B}}_{k,i} :\, &= \frac{1}{2}\left(\hat{n}_i\, \mathbf{S}_j\cdot \mathbf{S}_k + \hat{n}_j\, \mathbf{S}_k\cdot \mathbf{S}_i + \hat{n}_k\, \mathbf{S}_i\cdot \mathbf{S}_j\right) \\
    &+ \frac{\hat{n}_i \hat{n}_j \hat{n}_k}{8} - \mathrm{i}\, \mathbf{S}_i \cdot \left(\mathbf{S}_j \times \mathbf{S}_k \right),
\end{alignat}
from which, it follows that
\begin{equation}
\label{eq:id}
    \mathbf{S}_i \cdot\left(\mathbf{S}_j \times \mathbf{S}_k \right) = 2 \mathrm{i}\, \left( \hat{\mc{B}}_{i,j}\, \hat{\mc{B}}_{j,k}\, \hat{\mc{B}}_{k,i} -  \hat{\mc{B}}^\dagger_{k,i}\, \hat{\mc{B}}^\dagger_{j,k} \,\hat{\mc{B}}^\dagger_{i,j} \right).
\end{equation}
In a mean-field approximation, 
\begin{alignat}{1}
\label{eq:idmf}
\hat{\mc{B}}_{i,j}\, \hat{\mc{B}}_{j,k}\, \hat{\mc{B}}_{k,i} &\simeq \langle\hat{\mc{B}}_{i,j}\rangle\, \langle\hat{\mc{B}}_{j,k}\rangle\, \hat{\mc{B}}_{k,i} + \langle\hat{\mc{B}}_{i,j}\rangle\, \hat{\mc{B}}_{j,k}\, \langle\hat{\mc{B}}_{ki,}\rangle \\
\nonumber&+ \hat{\mc{B}}_{i,j}\, \langle\hat{\mc{B}}_{j,k}\rangle\, \langle\hat{\mc{B}}_{k,i}\rangle - 2 \langle\hat{\mc{B}}_{i,j}\rangle\, \langle\hat{\mc{B}}_{j,k}\rangle\, \langle\hat{\mc{B}}_{k,i}\rangle.
\end{alignat}
Based off this simplification, we can now evaluate the quadratic terms for each individual bond. As an example, consider a bond linking sites $i$ and $i +\hat{x}$; following the labeling scheme of Fig.~\ref{fig:DMConv}, let this be numbered 1--3. The only spin chirality terms in the Hamiltonian that involve this bond are
\begin{alignat}{1}
 &\mathbf{S}_1 \cdot[\left(\mathbf{S}_2 \times \mathbf{S}_3 \right)+ \left(\mathbf{S}_4 \times \mathbf{S}_3 \right)+ \left(\mathbf{S}_3 \times \mathbf{S}_6 \right)+ \left(\mathbf{S}_3 \times \mathbf{S}_5 \right)]
\\
\nonumber &\approx \left[\hat{\mc{B}}_{1,3} \left({\mc{B}^*}^2 +\lvert \mc{B}\rvert^2\right) + \hat{\mc{B}}^\dagger_{1,3} \left(\mc{B}^2 +\lvert \mc{B}\rvert^2 \right)\right] - \mathrm{H.c.} + \ldots,
\end{alignat}
where we have isolated the terms proportional to $\hat{\mc{B}}_{1,3}$ or $\hat{\mc{B}}_{3,1}$, and those from all other bonds are grouped together in the ellipsis. However, the term enclosed in the brackets is already Hermitian so the total contribution from the 1--3 (and more generally, any horizontal or vertical) edge is always zero. An analogous statement holds for any bond in the diagonal direction as well. In this regard, let us survey the 1--4 link, which connects sites $i$ and $i+\hat{x}+\hat{y}$. The relevant spin interactions in which this bond participates are $\mathbf{S}_{3}  \cdot \left(\mathbf{S}_1 \times \mathbf{S}_{4} \right)$, and $\mathbf{S}_{2}  \cdot \left(\mathbf{S}_{4} \times \mathbf{S}_1 \right)$, and collecting the quadratic terms for Eq.~\eqref{eq:idmf}, we finally have
\begin{equation}
 \hat{\mc{B}}_{1,4} {\mc{B}^*}^2 + \hat{\mc{B}}^\dagger_{1,4} \mc{B}^2  - \mathrm{H.c.} = 0.
\end{equation}
Since this cancellation occurs on any bond on the square lattice, $H_\chi$ in Eq.~\eqref{eq:chi} does not contribute to the Hamiltonian to quadratic order and the orbital coupling to the magnetic flux necessarily vanishes in the mean-field framework.

If, instead, we use the parameters of the ansatz with symmetry $\frac{4}{m}\,m'm'$ in \equsref{eq:subeqns}{FullAnsatz1}, there is no cancellation using SBMFT. In fact, as expected from a symmetry point of view, the resultant mean-field contribution of $H_\chi$ can be absorbed by rescaling of the ansatz per se as
\begin{align}
\mc{B}_1 \quad &\longrightarrow \quad \mc{B}_1 - 4\Upsilon \sin \Phi\,\mc{B}_1 \mc{B}_2, \\
\mc{B}_2 \quad &\longrightarrow \quad \mc{B}_2 - 2\Upsilon \sin \Phi \,\mc{B}_1^2 .
\end{align}
This conveys that the parameters $\mc{B}_1$ and $\mc{B}_2$ can also be induced or enlarged by the orbital coupling to the external magnetic field.

\section{Perturbations in the $\mathbb{C}\mathbb{P}^1$ theory}
\label{app:cp1}

Quantum fluctuations about the conventional square-lattice N\'eel state are conveniently described in the Schwinger boson theory using a continuum formulation based on the $\mathbb{C}\mathbb{P}^1$ model \cite{PhysRevLett.62.1694}. Here, we discuss, following Ref.~\onlinecite{scheurer2018orbital}, the additional perturbations that are introduced into this theory from the three-spin interaction in Eq.~(\ref{eq:chi}), which is induced by the orbital effect of the applied magnetic field, and which breaks the symmetry down to $\frac{4}{m}m'm'$.

The $\mathbb{C}\mathbb{P}^1$ model is expressed in terms of a bosonic spinor $z_\sigma$ which is coupled to a U(1) gauge field $a_\mu$
($\mu = \tau, x, y$) with Lagrangian
\beq
\mathcal{L}_{\mathbb{C}\mathbb{P}} = \frac{1}{g} |(\partial_\mu - i a_\mu) z_\sigma|^2
\eeq
Perturbations with symmetry of $H_\chi$ are most conveniently expressed in terms of the gauge field $a_\mu$. In a relativistic formulation, the leading perturbation is the term \cite{PhysRevX.7.031051,scheurer2018orbital} $\epsilon_{\mu\nu\lambda} f_{\mu\nu} \partial_\rho f_{\rho\lambda}$. But, more generally, without relativistic invariance, there are two independent terms which are expressed in terms of the internal electric and magnetic fields derived from $a_\mu$ (these are unrelated to the applied external electromagnetic field):
\beq
e_i = \partial_\tau a_i - \partial_i a_\tau \quad , \quad b = \partial_x a_y - \partial_y a_x\,.
\eeq
Analysis of symmetries leads to the perturbation
\beq
\mathcal{L}_\chi = i\lambda_1 \left( e_x \partial_\tau e_y - e_y \partial_\tau e_x \right) + i\lambda_2 \, b \, \partial_i e_i
\eeq
with couplings $\lambda_{1,2}$ which are expected to be proportional to $\Upsilon \sin (\Phi)$ in Eq.~(\ref{eq:chi}).

In terms of the underlying spin-wave fluctuations, the gauge field $a_\mu$ involves terms with one gradient, and so $\mathcal{L}_\chi$ has five spatiotemporal gradients \cite{scheurer2018orbital}. As such, its effects can be expected to be quite weak.

\begin{widetext}
{\setstretch{1.075}
\section{Mean field Hamiltonian for the one-orbital model}
\label{sec:BdG}

The mean-field Hamiltonian for the one-orbital model presented in Sec.~\ref{sec:1O} is described by Eq.~\eqref{eq:H1O}. We first expand out the different terms therein with the ansatz of \equsref{eq:subeqns}{FullAnsatz1}. Labeling the two kinds of sites for a fixed gauge choice by $\alpha$ and $\beta$, this can be written as
\begin{alignat}{4}
\nonumber H_{\textsc{mf}} &=\frac{J}{2}\sum_{(u,v)\, \in\, \alpha, \, \sigma} \bigg( \mathrm{i} \mathcal{B}_1\, \alpha^{\dagger}_{(u,v)\, \sigma}\, \beta^{}_{(u,v)+\hat{x}\, \sigma} 
+ \mathrm{i} \mathcal{B}_1\, \alpha^{\dagger}_{(u,v)\, \sigma}\, \beta^{}_{(u,v)+\hat{y}\, \sigma} 
+\mathrm{i}  \mathcal{B}_2\, \alpha^{\dagger}_{(u,v)\, \sigma}\, \alpha^{}_{(u,v)+\vec{\eta}_1\, \sigma} -\mathrm{i}  \mathcal{B}_2\, \alpha^{\dagger}_{(u,v)\, \sigma}\, \alpha^{}_{(u,v)+\vec{\eta}_2\, \sigma}\\
\nonumber& 
- \mathcal{A}_1 \,\sigma\, \alpha^{}_{(u,v)\, \sigma}\, \beta^{}_{(u,v)+\hat{x}\, -\sigma}
- \mathcal{A}_1\,\sigma\, \alpha^{}_{(u,v)\, \sigma}\, \beta^{}_{(u,v)+\hat{y}\, -\sigma} -\mathcal{A}_2\, \sigma\,\alpha_{(u,v)\, \sigma}\, \alpha^{}_{(u,v)+\vec{\eta}_1\, -\sigma}
+\mathcal{A}_2\, \sigma\,\alpha^{}_{(u,v)\, \sigma}\, \alpha^{}_{(u,v)+\vec{\eta}_2\, -\sigma} \bigg) 
+ \mathrm{H.c.}\\
\nonumber &+ \frac{J}{2} \sum_{(u,v)\, \in\, \beta,\, \sigma} \bigg( \mathrm{i} \mathcal{B}_1\, \beta^{\dagger}_{(u,v)\, \sigma}\, \alpha^{}_{(u,v)+\hat{x}\, \sigma} 
- \mathrm{i} \mathcal{B}_1\, \beta^{\dagger}_{(u,v)\, \sigma}\, \alpha^{}_{(u,v)+\hat{y}\, \sigma} 
- \mathrm{i}  \mathcal{B}_2\, \beta^{\dagger}_{(u,v)\, \sigma}\, \beta^{}_{(u,v)+\vec{\eta}_1\, \sigma}+\mathrm{i}  \mathcal{B}_2\, \beta^{\dagger}_{(u,v)\, \sigma}\, \beta^{}_{(u,v)+\vec{\eta}_2\, \sigma}\\
\nonumber& 
- \mathcal{A}_1 \,\sigma\, \beta^{}_{(u,v)\, \sigma}\, \alpha^{}_{(u,v)+\hat{x}\, -\sigma}
+ \mathcal{A}_1\,\sigma\, \beta^{}_{(u,v)\, \sigma}\, \alpha^{}_{(u,v)+\hat{y}\, -\sigma}+\mathcal{A}_2\, \sigma\,\beta^{}_{(u,v)\, \sigma}\, \beta^{}_{(u,v)+\vec{\eta}_1\, -\sigma}
-\mathcal{A}_2\, \sigma\,\beta^{}_{(u,v)\, \sigma}\, \beta^{}_{(u,v)+\vec{\eta}_2\, -\sigma} \bigg) + \mathrm{H.c.}\\
& + \lambda \sum_{(u,v),\, \sigma} \left(\alpha^\dagger_{(u, v) \,\sigma}\,\alpha^{}_{(u, v) \,\sigma} + \beta^\dagger_{(u, v) \,\sigma} \, \beta^{}_{(u, v) \,\sigma}  - 2 S\right),
\end{alignat}
with $(u, v)$ running exclusively over all $\alpha$ ($\beta$) sites in the first (second) summation above. Fourier transforming to momentum space, with the convention $b_{i \sigma} = \sum_{\bf k} b_{{\bf k} \sigma} \exp(\mathrm{i}\, {\bf k}\cdot {\bf r}_i)/\sqrt{N}$, we find 
\begin{alignat}{2}
\label{eq:HMFKSp}
\nonumber H_{\textsc{mf}}  = \bigg[ \frac{J}{2}&\sum_{\k \sigma} \bigg(\mathrm{i} \mathcal{B}_1 E_+ \alpha^{\dagger}_{\k\, \sigma}\, \beta^{}_{\k\, \sigma}   + 2 \mc{B}_2 \mathrm{e}^{\mathrm{i} k_x} \mc{S}_y \alpha^{\dagger}_{\k\, \sigma}\, \alpha^{}_{\k\, \sigma} - \mathcal{A}_1\,\sigma\, (E_+)^* \,\alpha^{}_{\k\, \sigma}\, \beta^{}_{-\k\, -\sigma}
- 2 \mathrm{i} \mathcal{A}_2\,\sigma\, \mathrm{e}^{-\mathrm{i \,k_x}} \mc{S}_y \,\alpha^{}_{\k\, \sigma}\, \alpha^{}_{-\k\, -\sigma}\bigg)\\
\nonumber + \frac{J}{2} &\sum_{\k \sigma} \bigg( \mathrm{i} \mathcal{B}_1 E_- \beta^{\dagger}_{\k\, \sigma}\, \alpha^{}_{\k\, \sigma}   - 2 \mc{B}_2 \mathrm{e}^{\mathrm{i} k_x} \mc{S}_y \beta^{\dagger}_{\k\, \sigma}\, \alpha^{}_{\k\, \sigma} - \mathcal{A}_1\,\sigma\, (E_-)^*\,\beta^{}_{\k\, \sigma}\, \alpha^{}_{-\k\, -\sigma}+ 2 \mathrm{i} \mathcal{A}_2\,\sigma\, \mathrm{e}^{-\mathrm{i \,k_x}} \mc{S}_y \,\beta^{}_{\k\, \sigma}\, \beta^{}_{-\k\, -\sigma}\bigg) \bigg]+ \mathrm{H.c.} \\
+  \lambda  &\sum_{\k\,\sigma} \left (\alpha^{\dagger}_{\k\, \sigma}\, \alpha^{}_{\k\, \sigma} + \beta^{\dagger}_{\k\, \sigma} \beta^{}_{\k\, \sigma} - 2 S \right),
\end{alignat}
where we have adopted the shorthand $\mc{C}_\mu \equiv \cos (k_\mu)$, $\mc{S}_\mu \equiv \sin (k_\mu)$, and $E_{\pm} \equiv \exp(\mathrm{i} k_x) \pm \exp(\mathrm{i} k_y)$.  In real space, the positions of the $\alpha$ and $\beta$ states within the same unit cell are spatially separated, so the second-quantized Hamiltonian is invariant under $\mathbf{k} \rightarrow \mathbf{k} + \vec{G}_\mu$ only up to a gauge transformation \cite{singh2018time}. The presence of an external magnetic field $B_z$ now appends the Zeeman term \eqref{eq:H2mf} to $H_{\textsc{mf}}$. Equation~\eqref{eq:HMFKSp} is easily converted into the form $H_{\textsc{mf}} = \sum_{\k} (\Psi_{\k}^\dagger\, \mathcal{H}({\k})\, \Psi_{\k})/2$, where $\Psi$ is the eight-component spinor defined as $\Psi^\dagger_\mathbf{k} = ( \alpha^\dagger_{\k\uparrow}\, \beta^\dagger_{\k\uparrow}\, \alpha^\dagger_{\k \downarrow}\, \beta^\dagger_{\k\downarrow}\,\alpha^{}_{-\k\uparrow}\, \beta^{}_{-\k\uparrow}\,\alpha^{}_{-\k\downarrow}\, \beta^{}_{-\k\downarrow})$. This can be diagonalized in accordance with the process sketched in Sec.~\ref{sec:DiagBH} to calculate the Berry curvatures and conductivities.

More compactly though, $H_{\textsc{mf}}$ can equivalently be expressed using the reduced \textit{four}-component spinor $\psi^\dagger = ( \alpha^\dagger_{\k\uparrow}\, \beta^\dagger_{\k\uparrow}\, \alpha^{}_{-\k\downarrow}\, \beta^{}_{-\k\downarrow})$. Up to a constant, the bosonic mean-field Hamiltonian reads as
\begin{equation}
\label{eq:kernel}
\mc{H} (\k) =
\frac{1}{2}\left(
\begin{array}{cccc}
 -B +4 \mathcal{B}_2\, J \,\mathcal{C}_x\, \mathcal{S}_y+2 \lambda  & 2 \mathrm{i}\, \mathcal{B}_1 \,J \,(\mathcal{C}_y+i \mathcal{S}_x) & 4 \mathrm{i}\, \mathcal{A}_2\, J \, \mathcal{C}_x\, \mathcal{S}_y & -2 \mathcal{A}_1 \,J\, (\mathcal{C}_y+ \mathrm{i} \mathcal{S}_x) \\
 -2 \mathrm{i} \,\mathcal{B}_1\, J \,(\mathcal{C}_y- \mathrm{i} \mathcal{S}_x) & -B-4 \mathcal{B}_2\, J\,\mathcal{C}_x\,  \mathcal{S}_y+2 \lambda  & 2 \mathcal{A}_1\, J\, (\mathcal{C}_y- \mathrm{i} \mathcal{S}_x) & -4 \mathrm{i} \mathcal{A}_2 \,J\, \mathcal{C}_x\, \mathcal{S}_y \\
 -4 \mathrm{i} \mathcal{A}_2 \,J\, \mathcal{C}_x\, \mathcal{S}_y & 2 \mathcal{A}_1\, J\, (\mathcal{C}_y+ \mathrm{i} \mathcal{S}_x)\, & B-4 \mathcal{B}_2 \,J\, \mathcal{C}_x \, \mathcal{S}_y+2 \lambda  & 2 \mathcal{B}_1 \,J \,(-\mathrm{i} \mathcal{C}_y+\mathcal{S}_x) \\
 -2 \mathcal{A}_1\, J\, (\mathcal{C}_y-\mathrm{i} \mathcal{S}_x) & 4 \mathrm{i} \mathcal{A}_2 \,J\,\mathcal{C}_x\,  \mathcal{S}_y & 2 \mathcal{B}_1\, J\, (\mathrm{i} \mathcal{C}_y+\mathcal{S}_x) & B+4 \mathcal{B}_2 \,J\, \mathcal{C}_x\, \mathcal{S}_y+2 \lambda  \\
\end{array}
\right).
\end{equation}
Denoting the Pauli matrices acting in spin and sublattice space by $\sigma$ and $\tau$, respectively, \begin{alignat}{1}
\label{eq:H_pauli}
\mathcal{H} &=  \lambda \sigma^{}_0 \tau^{}_0 +  J \sigma^{}_2 (\mc{A}_1 \mc{S}_x  \tau^{}_1 +  \mc{A}_1 \mc{C}_y  \tau^{}_2 - 2 \mc{A}_2 \mc{C}_x \mc{S}_y  \tau^{}_3) -  \frac{B}{2} \sigma^{}_3 \tau^{}_0  - J \sigma^{}_3 ( \mc{B}_1 \mc{S}_x \tau^{}_1 + \mc{B}_1 \mc{C}_y  \tau^{}_2 -2\, \mc{B}_2 \mc{C}_x \mc{S}_y \tau^{}_3). \end{alignat}
This form of the kernel $ \mathcal{H}$ contains the same information as the $8 \times 8$ matrix for the full spinor $\Psi$ but is much more amenable to analytical calculations. On grounds of simplicity, it is therefore convenient to frame the discussion in the following subsections in terms of the $4 \times 4$-matrix description of the mean-field Hamiltonian $\mc{H} (\k) $. In this language, the dynamic matrix $K =\rho^{}_3 \, \mathcal{H} = \sigma^{}_3 \tau^{}_0 \, \mathcal{H}$ is
\begin{alignat}{1}
\label{eq:Kmatrix}
K &=  - \frac{B}{2} \sigma^{}_0 \tau^{}_0  -  J \mc{B}_1 \mc{S}_x \sigma^{}_0 \tau^{}_1 - J \mc{B}_1 \mc{C}_y \sigma^{}_0 \tau^{}_2 + 2 J \mc{B}_2 \mc{C}_x \mc{S}_y \sigma^{}_0 \tau^{}_3 -  \mathrm{i} J \mc{A}_1 \mc{S}_x \sigma^{}_1 \tau^{}_1 -  \mathrm{i} J \mc{A}_1 \mc{C}_y \sigma^{}_1 \tau^{}_2  + 2 \mathrm{i} J \mc{A}_2 \mc{C}_x \mc{S}_y \sigma^{}_1 \tau^{}_3 +  \lambda \sigma^{}_3 \tau^{}_0.
\end{alignat}
Diagonalizing this dynamic matrix results in two particle bands, which we list as $m = 1, 2$, and two hole bands ($m = 3, 4$). Note that one could just as well have elected to work with $\Psi$ instead of $\psi$ and the correspondence between these bands and our previous indexing scheme is $m = \{1, 2, 3, 4\} \leftrightarrow n = \{1, 3, 6, 8\}$. For the remaining $n$ bands, associated with $n =\{2,4,5,7 \} \equiv n'$, the energies and curvatures are simply related as $\varepsilon^{}_{n', \mathbf{k}} = \varepsilon^{}_{(n'+4)\,\mathrm{ mod }\, 8\,, \mathbf{k}}$ and $\Omega^{}_{n', \mathbf{k}} = -\Omega^{}_{(n'+4)\,\mathrm{ mod }\, 8 \,, -\mathbf{k}}$, but $(n'+4)\,\mathrm{ mod }\, 8 \,\in\,\{1, 3, 6, 8\}$, closing the loop between the four- and eight-component formulations.
 
\subsection{Effective antiunitary symmetry}
\label{sec:TRS}

As mentioned in Sec.~\ref{sec:symm}, the pairwise degeneracy of the particle bands in the one-orbital model (at zero Zeeman fields) is due to an effective symmetry of the Hamiltonian, which we single out here. 
To begin with, we identify an anti-unitary operator $\mc{O} = U \, C$, where $U$ is unitary and $C$ is complex conjugation such that 
\begin{alignat}{1}
\mc{O}\, K (\mathbf{k})\, \mc{O}^\dagger = - K (\mathbf{k}) \implies U\, K^* (\mathbf{k})\, U^\dagger = - K (\mathbf{k}). \label{StructureOfK}
\end{alignat} 
This implies that if $\Phi^{}_m$ is an eigenvector of $K$ with eigenvalue $\omega^{}_m$, then so is $U\, \Phi_m^*$ but with eigenvalue $-\omega^{}_m$, which is precisely the particle-hole symmetry that must be broken to lift the degeneracy of the bosonic bands. The only such operator (unique up to an additional phase factor) is $\mc {O} = \sigma_2 \tau_2 C$, i.e. $U = \sigma_2 \tau_2$. Equation~\eqref{StructureOfK} then states that
\begin{alignat}{1}
\sigma^{}_3 U \sigma^{}_3 \mc{H}^* (\mathbf{k})\, U^\dagger = - \mc{H}(\mathbf{k}). 
\end{alignat}
As $\sigma^{}_3$ and $U=\sigma^{}_2\tau^{}_2$ anticommute, this yields an effective ``time-reversal symmetry'', i.e. $\mc{H}(\mathbf{k})$ and the anti-unitary operator $\mathcal{O}$ commute,
\begin{equation}
    \mc{O} \,\mc{H} ({\bf k}) \, \mc{O}^\dagger = \mc{H} ({\bf k}). \label{GenTRS}
\end{equation}
Since $\mc{O}^2=+1$, this does \textit{not} translate to a Kramers degeneracy (in general, all eigenvalues of $\mc{H}$ are indeed nondegenerate) whereas \equref{StructureOfK} does force the spectrum of $K$ to be symmetric with respect to zero energy. It then follows that the resulting degenerate bands have opposite Chern numbers. The wave functions are the eigenvectors of $K$ and, by virtue of Eq.~\eqref{StructureOfK}, may be grouped according to the eigenvalues as $\TM_{\mathbf{k}} =  [v_1(\mathbf{k})\,\, v_2(\mathbf{k}) \,\, (U v^*_1(\mathbf{k})) \,\, (U v^*_2(\mathbf{k}))]$. More concisely,
\begin{equation}
    \TM_{\mathbf{k}} = U \TM^*_{\mathbf{k}} \sigma_1; \quad U = \sigma_2\tau_2.
\end{equation}
The implication for the Berry curvature is that
\begin{align}
\nonumber     \Omega_{m \mathbf{k}} &= \mathrm{i}\,\epsilon_{\mu \nu} \left[\sigma^{}_3 \,\frac{\partial\, \TM_{\bf k}^\dagger}{\partial\, k_\mu} \,\sigma^{}_3\, \frac{\partial\, \TM_{\bf k}}{\partial\, k_\nu} \right]_{m m} = \mathrm{i}\,\epsilon_{\mu \nu} \left[\sigma^{}_3 \sigma^{}_1 \,\frac{\partial\, \TM_{\bf k}^T}{\partial\, k_\mu} \, U^\dagger \sigma^{}_3 U\, \frac{\partial\, \TM^*_{\bf k}}{\partial\, k_\nu} \sigma_1 \right]_{m m} = \mathrm{i}\,\epsilon_{\mu \nu} \left[\sigma^{}_3  \,\frac{\partial\, \TM_{\bf k}^T}{\partial\, k_\mu} \, \sigma^{}_3 \, \frac{\partial\, \TM^*_{\bf k}}{\partial\, k_\nu}  \right]_{\overline{m}\,\overline{m}} \\
    &= -\Omega^*_{\overline{m} \mathbf{k}} = -\Omega_{\overline{m} \mathbf{k}},
    \label{eq:omega_symm}
\end{align} where $\overline{m}=3$ ($\overline{m}=4$) for $m=1$ ($m=2$), and we have used the fact that  $\Omega_{\overline{m} \mathbf{k}}$ is real in the last step. Translating back to the band index $n$, this proves that the pairs $n = (1, 2)$ and $ (3, 4)$ are indeed degenerate and also have the same curvatures modulo $\mathbf{k} \rightarrow -\mathbf{k}$. The degeneracy is split at any temperature by a uniform Zeeman field $\vec{B}_Z$, which creates a constant gap between the two bands at each momentum.

\subsection{Magnetic order}\label{MagneticOrder}
Within the Schwinger boson framework, magnetic order is obtained via the condensation of bosons, which occurs when the bosonic modes have at least one zero eigenvalue \cite{sarker1989bosonic, sachdev1992kagome}. The minima of the spinon bands are found from diagonalizing $K = \sigma^{}_3 \tau^{}_0 \mc{H} (\k)$, with $\mc{H} (\k)$ as in Eq.~\eqref{eq:kernel}, and lie at $\pm \k_0$, where $\k_0 = (\pi/2,0)$. Without an external magnetic field ($B_z = 0$), the eigenvalues, each doubly degenerate at these momenta, are
\begin{equation}
\varepsilon^{}_{\pm} = \left\lvert\sqrt{\lambda ^2-2 \mc{A}_1^2 J^2} \pm \sqrt{2} \mc{B}_1 J \right \rvert.
 \end{equation}
For $\mc{B}_1>0$, the spinon gap is set by $\varepsilon^{}_{-}$ and closes when $\sqrt{\lambda^2 - 2 \mc{A}_1^2 J^2} = \sqrt{2} \mc{B}_1 J$; $\mc{B}_2$ appears neither in this equation nor in the eigenstates below. Eliminating $\mc{B}_1$ in favor of $\mc{A}_1, \lambda$, and setting $ \xi \equiv \lambda/(\sqrt{2}\mc{A}_1J) $ for notational convenience, we find that the two zero energy eigenvectors at $\k = \k_0 = (\pi/2,0)$ are
\begin{align}
\Psi_1 =\left(\mathrm{e}^{\mathrm{i} \pi/4} \xi, \mathrm{i} \sqrt{ \xi^2 -1}, 0,1\right)^T,\quad \Psi_2 & = \left (\mathrm{i} \sqrt{ \xi^2 -1},-\mathrm{e}^{-\mathrm{i} \pi/4} \xi,1,0 \right)^T,
\end{align}
where the superscript $T$ denotes transpose.
Likewise, at $\k = -\k_0 = (-\pi/2,0)$, there are two degenerate eigenvectors when the gap closes:
\begin{align}
\Psi_3 = \left (\mathrm{e}^{-\mathrm{i} \pi/4} \xi, \mathrm{i} \sqrt{ \xi^2 -1}, 0,1\right)^T,\quad \ \Psi_4  = \left(\mathrm{i} \sqrt{ \xi^2 -1},-\mathrm{e}^{\mathrm{i} \pi/4} \xi,1,0 \right)^T.
\end{align}
The condensate in real space is a linear combination of those at $\pm \k_0$. Introducing arbitrary complex numbers $z_i$ to represent the strength thereof, we have
\begin{equation}
\begin{pmatrix}
\langle \alpha_{ \r \uparrow} \rangle \\
\langle \beta_{ \r \uparrow} \rangle \\
\langle \alpha^\dagger_{\r \downarrow} \rangle \\
\langle \beta^\dagger_{\r \downarrow} \rangle \\
\end{pmatrix} = (z_1 \Psi_1  + z_2 \Psi_2)\mathrm{e}^{\mathrm{i} \k_0 \cdot \r} +  (z_3 \Psi_3  + z_4 \Psi_4)\mathrm{e}^{-\mathrm{i} \k_0 \cdot \r},
\end{equation}
whereafter the condensate on each sublattice can be written as
\begin{align}
X_\alpha = \begin{pmatrix}
\langle \alpha_{ \r \uparrow} \rangle \\
\langle \alpha_{ \r \downarrow} \rangle 
\end{pmatrix} = \begin{pmatrix}
 \mathrm{e}^{\mathrm{i}\pi/4}z_1  \xi + \mathrm{i} z_2 \sqrt{ \xi^2-1} &  \mathrm{e}^{-\mathrm{i}\pi/4}z_3   \xi + \mathrm{i}  z_4 \sqrt{ \xi^2-1} \\
 z_4^* & z_2^*
\end{pmatrix} \begin{pmatrix}
\mathrm{e}^{\mathrm{i} \k_0 \cdot \r} \\
\mathrm{e}^{-\mathrm{i} \k_0 \cdot \r}
\end{pmatrix}, \\ \nonumber
X_\beta = \begin{pmatrix}
\langle \beta_{ \r \uparrow} \rangle \\
\langle \beta_{ \r \downarrow} \rangle 
\end{pmatrix} = \begin{pmatrix}
\mathrm{i}  z_1 \sqrt{ \xi^2-1} - \mathrm{e}^{-\mathrm{i}\pi/4}z_2  \xi &  \mathrm{i}  z_3 \sqrt{ \xi^2-1} - \mathrm{e}^{\mathrm{i}\pi/4}z_4   \xi \\
 z_3^* & z_1^*
\end{pmatrix} \begin{pmatrix}
\mathrm{e}^{\mathrm{i} \k_0 \cdot \r} \\
\mathrm{e}^{-\mathrm{i} \k_0 \cdot \r}
\end{pmatrix}.
\end{align}
The spinor $(\mathrm{e}^{\mathrm{i} \k_0 \cdot \r}, \mathrm{e}^{-\mathrm{i} \k_0 \cdot \r})^T$ is proportional to $(1,1)^T$ for even $x$, whereas for odd $x$ coordinate, it is $\propto (1,-1)^T$ [the overall U(1) phase is redundant for calculating physical spin expectation values]. This calls for further classification of the sites on the $\alpha $ and $\beta$ sublattices---defined by $(-1)^{j_x+j_y} = 1$ and $-1$, respectively, according as whether $x$ is even ($e$) or odd ($o$), creating a four-sublattice structure for the magnetic order. The expectation value of the spin at each site can then be evaluated as $\langle \bm{S}_{\mu a}(\r) \rangle  = X^\dagger_{\mu a} \bm{\sigma} X^{}_{\mu a}$ for $\mu = \{\alpha, \beta \}$ and $a = \{e,o\}$.

At this point, we note that the spin-liquid state described by the ansatz \eqref{eq:subeqns} has a gauge-invariant flux $\phi$ of $\pi$ (modulo $2 \pi$) through each elementary square plaquette \cite{tchernyshyov2005flux} (see main text for definition). Similar $\pi$-flux states on the square lattice were studied by \citet{yang2016schwinger}; the latter states are all identical in the limit of only $\mc{A}_1 \neq 0$. The corresponding magnetically ordered state was found to be a subset of the classical ground state for the $J_2/J_1 = 1/2$-Heisenberg model, and, in general, quite distinct  from N\'{e}el order. A formal route to draw a connection to the Yang-Wang $\pi$-flux ansatz is to construct a local gauge transformation mapping the one-orbital model onto it. Recall that under such a transformation, one generically has
\begin{align}
\label{eq:GT}
 b_{j \sigma} \rightarrow  \mathrm{e}^{\mathrm{i} \varphi(j)} b_{j \sigma}, \quad
 \mc{A}_{i,j} \rightarrow  \mathrm{e}^{\mathrm{i} (\varphi(i)+ \varphi(j))} \mc{A}_{i, j}\,
 \quad
 \mc{B}_{i, j} \rightarrow  \mathrm{e}^{\mathrm{i} (\varphi(i) - \varphi(j))} \mc{B}_{i, j}.
\end{align}
The ansatz \eqref{eq:subeqns} is characterized by $\mc{A}_{i,i+\hat{x}} = \mc{A}_1$ and $\mc{A}_{i,i+\hat{y}} = (-1)^{i_x+i_y} \mc{A}_1$, whereas that of Ref.~\onlinecite{yang2016schwinger} has $\mc{A}_{i,i+\hat{x}} = (-1)^{i_y} \mc{A}_1$ and $\mc{A}_{i,i+\hat{y}} = - \mc{A}_1$. If the two are to be related by a gauge transformation, then the phase $\varphi (j)$ must satisfy
\begin{align}
\varphi(j_x,j_y) + \varphi(j_x+1,j_y) = \pi j_y,\,\quad
\varphi(j_x,j_y) + \varphi(j_x,j_y+1) = \pi(j_x+j_y+1).
\end{align}
Both these equations hold modulo $2\pi$ and their solution is $\varphi(j_x,j_y) = \pi \left( -2 j_x^2 + 2 j_y + 1 \right)/4$. Applying this transformation shifts the one-orbital dispersion minima, which are inherently gauge dependent: with the earlier gauge choice, the minima were positioned at $(\pm \pi/2,0)$ but in the new gauge, they are at $\pm(\pi/2,\pi/2)$, as expected from Ref.~\onlinecite{yang2016schwinger} in the limit where all terms but $\mc{A}_1$ are zero. Proceeding beyond this special case, we can similarly transform the remaining ($\mc{A}_2$, $\mc{B}_1$, $\mc{B}_2$) terms in Eq.~\eqref{FullAnsatz1} according to \equref{eq:GT}, and the minimal ansatz which gives quantized Chern bands in this gauge reads: 
\begin{alignat}{2}
\nonumber
\mc{A}_{i,i+\hat{x}} &= (-1)^{i_y} \mc{A}_1,\,\, &&\mc{A}_{i,i+\hat{y}} = - \mc{A}_1, \\\mc{B}_{i,i+\hat{x}} &= -(-1)^{i_x} \mc{B}_1,\,\, &&\mc{B}_{i,i+\hat{y}} =  \mc{B}_1(-1)^{i_x+i_y},\,\,
\mc{B}_{i,i+\hat{x} + \hat{y}} =  \mc{B}_{i,i+\hat{x}-\hat{y}} =  -\mathrm{i} \mc{B}_2 (-1)^{i_y}.
\label{eq:new_ansatz}
\end{alignat}
As Fig.~\ref{fig:YW} corroborates, the minima for the lowest-energy spinon band remain at $\pm(\pi/2,\pi/2)$ even on turning on $\mc{B}_1$ and $\mc{B}_2$ [cf.~Fig.~\ref{fig:1OB3}]. 

\begin{figure}[htb]
\begin{minipage}{0.3\linewidth}
\subfigure[]{\label{fig:YW}\includegraphics[width=\linewidth]{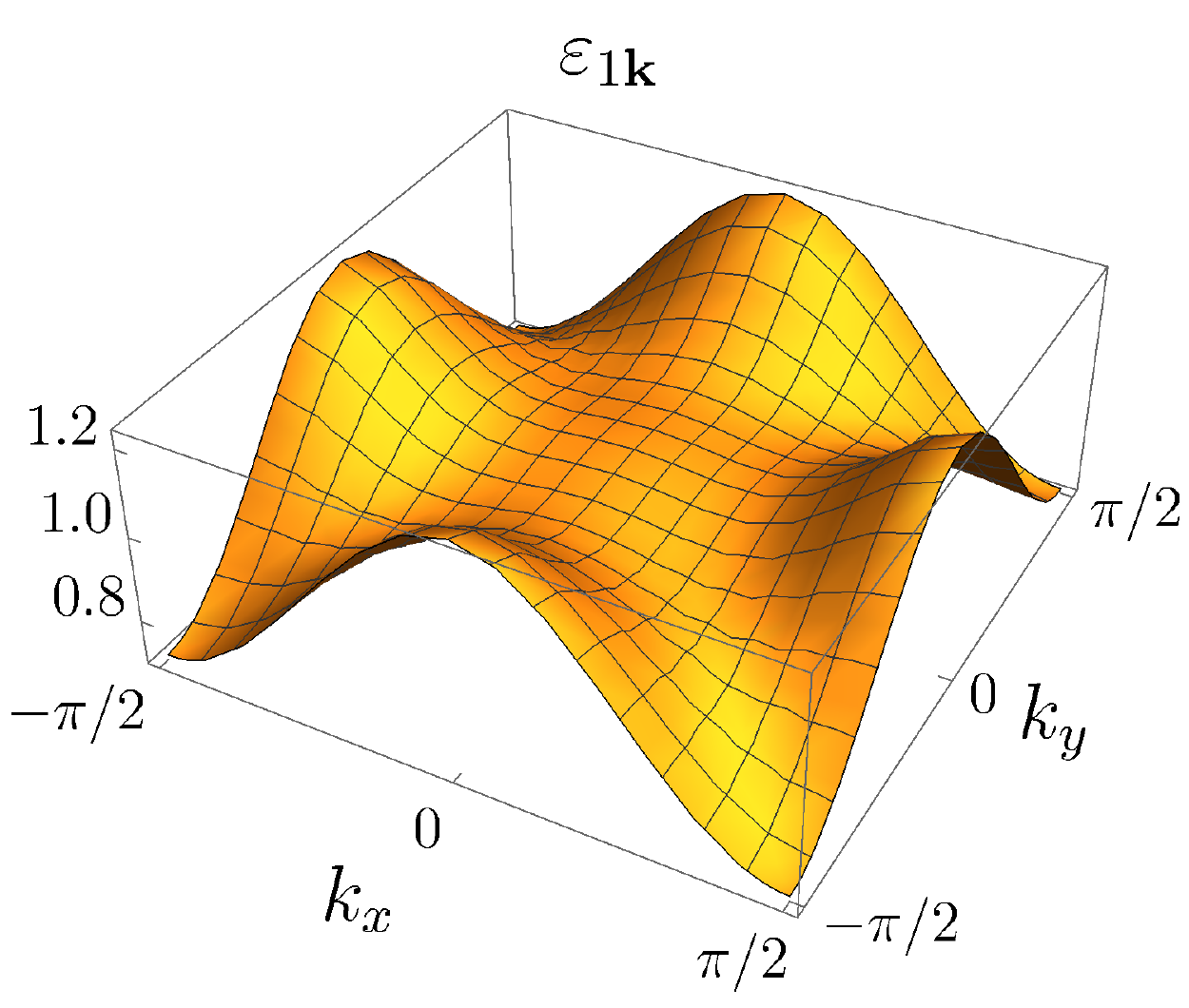}}
\end{minipage}
\begin{minipage}{0.65\linewidth}
\subfigure[]{\label{fig:Mag3}\includegraphics[width=0.49\linewidth]{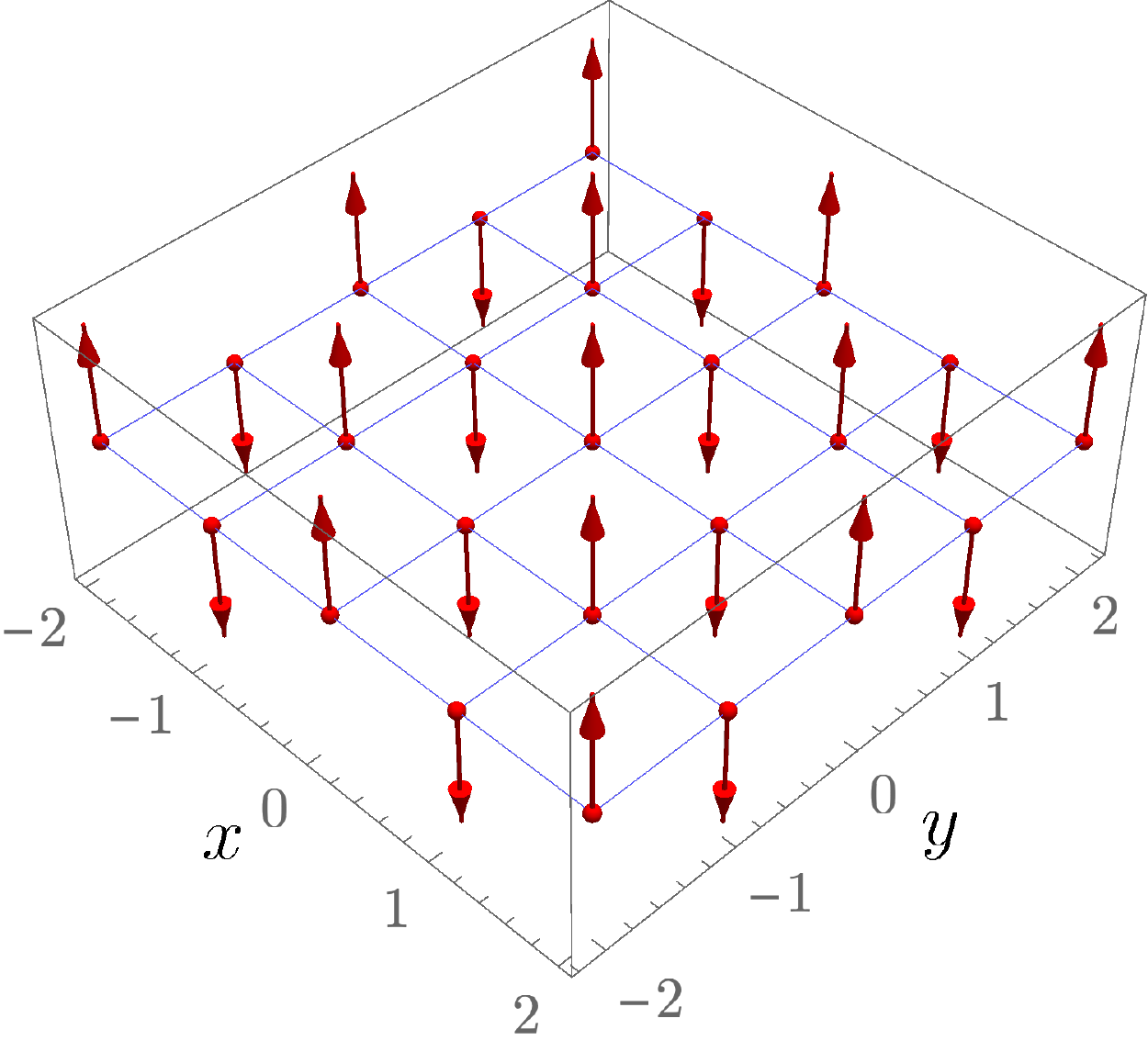}}
\subfigure[]{\label{fig:Mag4}\includegraphics[width=0.49\linewidth]{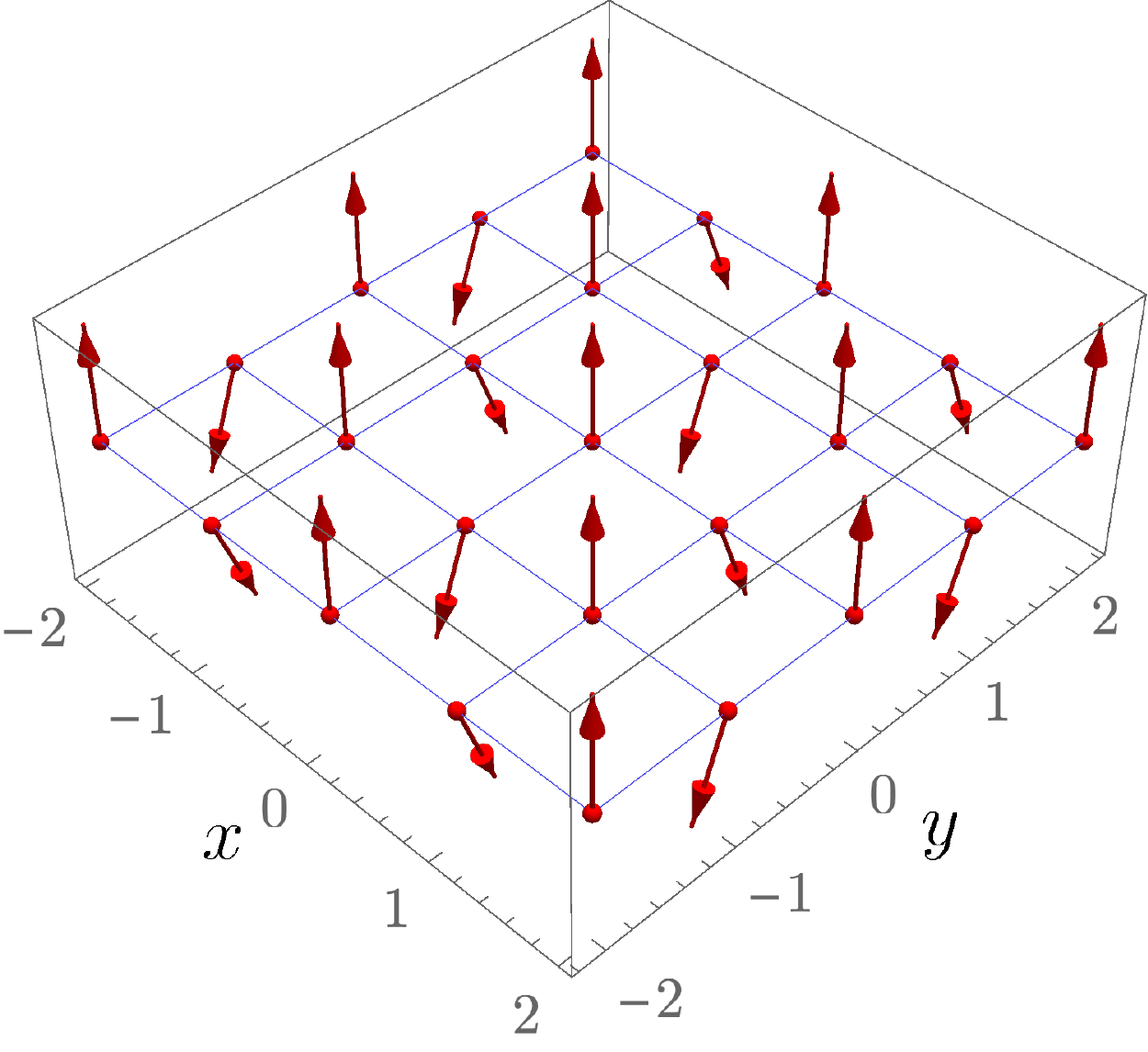}}
\subfigure[]{\label{fig:Mag1}\includegraphics[width=0.49\linewidth]{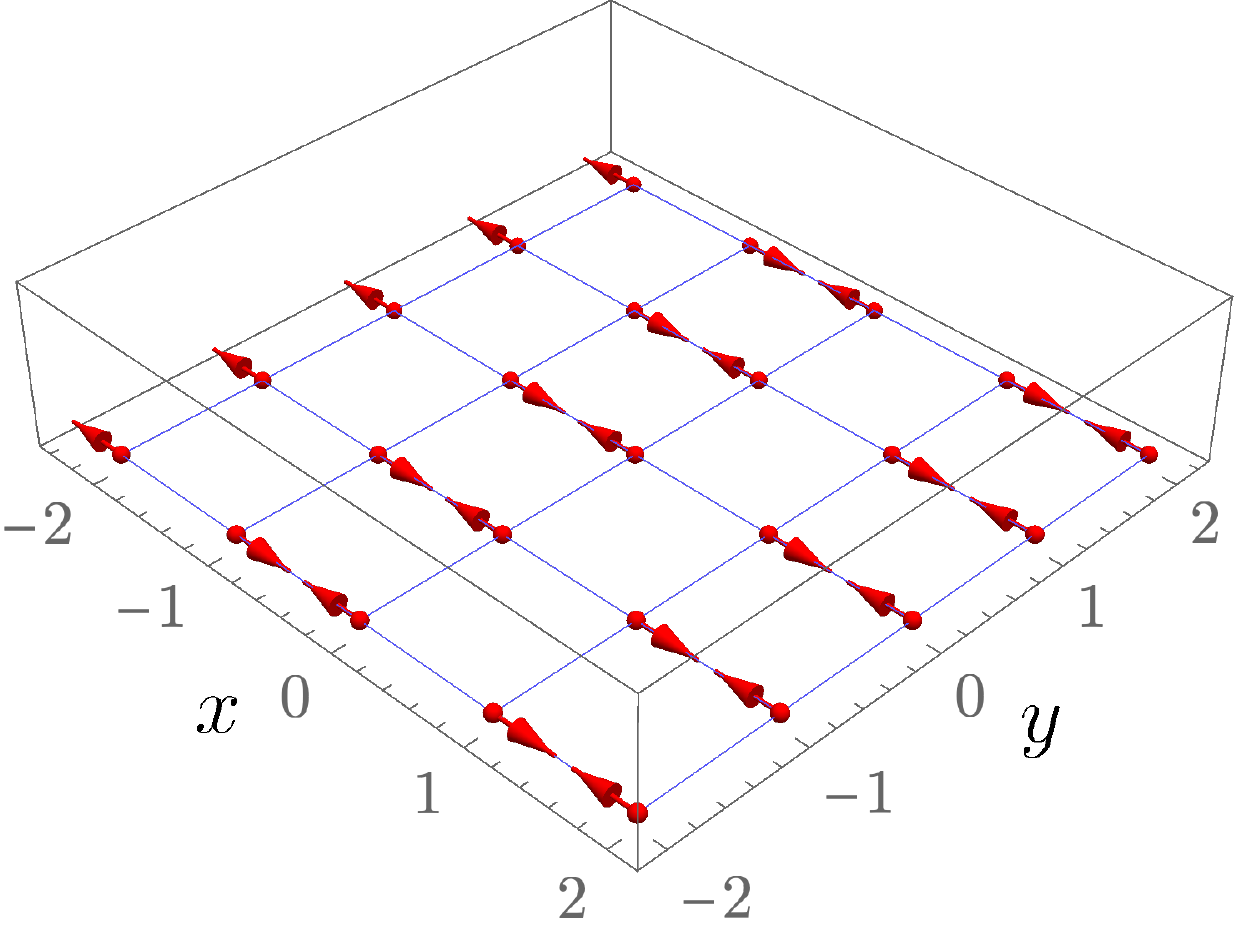}}
\subfigure[]{\label{fig:Mag2}\includegraphics[width=0.49\linewidth]{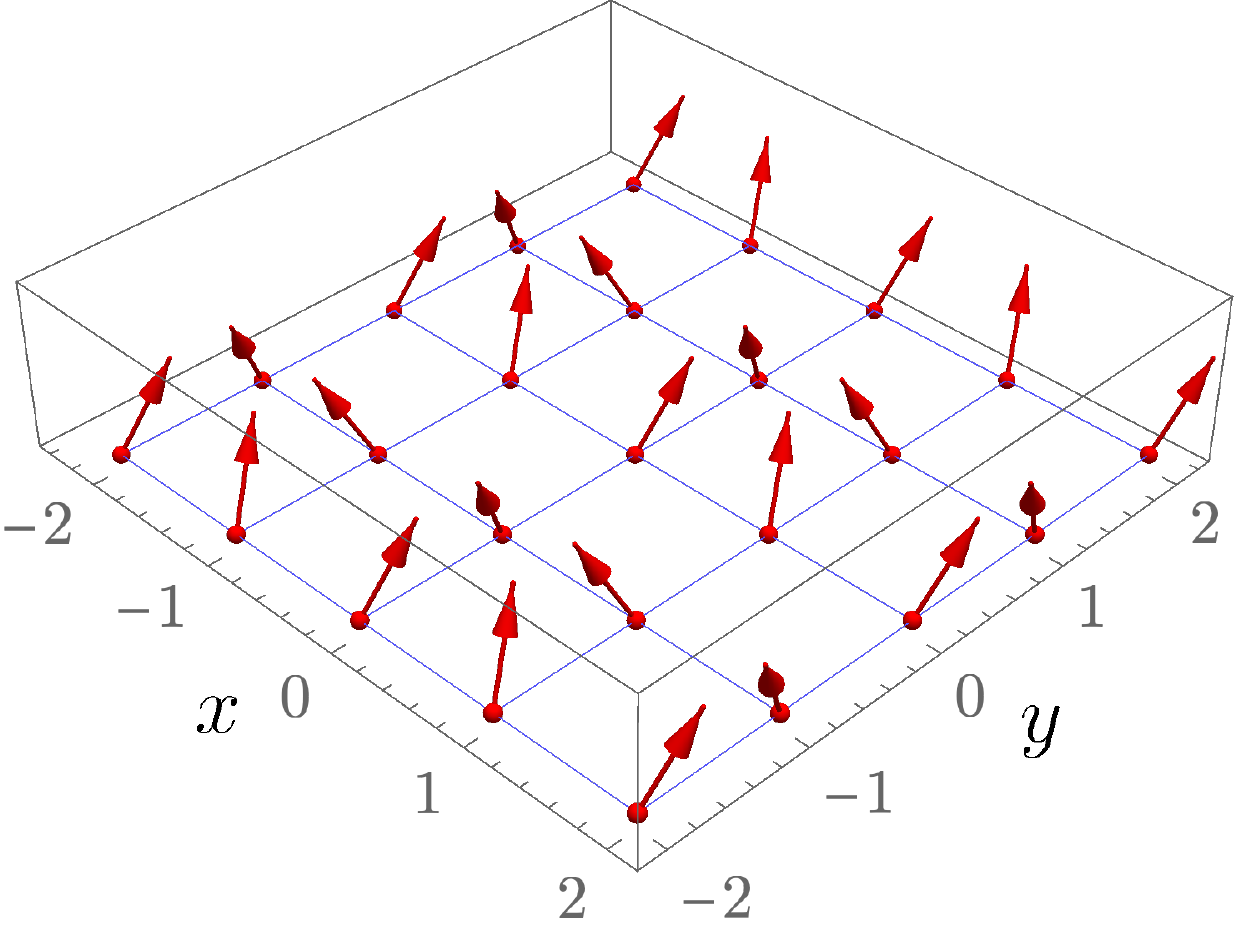}}
   \end{minipage}
    \caption{(a) Dispersion of the $n=1$ spinon band corresponding to the Schwinger boson ansatz for the one-orbital model in Eq.~\eqref{eq:new_ansatz} with $J =1$, $\mc{A}_1 = 1$, $\mc{B}_1 = 0.5$, $\mc{B}_2 =0.5$, $\lambda = 2.0$, and $B_z=0$. In this gauge, the minima are always at $\pm (\pi/2,\pi/2)$. (b) The N\'{e}el state is obtained upon boson condensation with $\xi = 1$ (implying $\mc{B}_1 = 0$) and only one of the $z^{}_i$ (taken to be $z_1$ here) nonzero. (c) This state can be perturbed by increasing $\xi$ which equals $1.05$ here, thus setting $\mc{B}_1 \approx 0.32 \mc{A}_1$. (d) The magnetically ordered state with the complex coefficients chosen to be $\{ z_1, z_2, z_3, z_4 \} = \{z,  \mathrm{i}z, 0, 0 \}$, and (e)  $\{z,  -\mathrm{i}z, 0, 0 \}$. The magnetic moment is uniform at all sites for the states exhibited, and the vector plotted in each figure is, for clarity, $(S_x/2, S_y/2, S_z)$.}
 \end{figure}

We return to our original gauge choice where the computation of magnetic order is more tractable. Noting that the spin-liquid state for the one-orbital model reduces to that in Ref.~\onlinecite{yang2016schwinger} in the limit of only $\mc{A}_1 \neq 0$, we first set $ \xi = 1$ (as dictated by the gap-closing condition with $\mc{B}_1 = 0$). Upon calculating the spin expectation values using the boson-condensation procedure, we find that the ordered moments on the four sites of a plaquette add to zero, i.e.
\begin{align}
\sum_{\mu = \alpha, \beta} \sum_{a = e,o} \langle \bm{S}_{\mu a} \rangle = 0,
\end{align}
which is precisely the four-sublattice ordered state in Ref.~\onlinecite{yang2016schwinger}. A particular instance thereof is the N\'{e}el state which is obtained when the coefficients are chosen such that only one of the four $z_i$ is nonzero. For general $ \xi > 1$, the spinors $X_{\mu a}$ are
\begin{alignat}{2}
X_{\alpha e} =& \begin{pmatrix}  \xi (\mathrm{e}^{\mathrm{i}\pi/4}z_1 +  \mathrm{e}^{-\mathrm{i}\pi/4}z_3)  + \mathrm{i} \sqrt{ \xi^2-1}( z_2 + z_4)  \\  z_4^* + z_2^*
\end{pmatrix}, 
&&X_{\alpha o} = \begin{pmatrix}   \xi (\mathrm{e}^{\mathrm{i}\pi/4}z_1 -  \mathrm{e}^{-\mathrm{i}\pi/4}z_3)  + \mathrm{i} \sqrt{ \xi^2-1}  (z_2 -  z_4) \\  z_4^* - z_2^*
\end{pmatrix}, \\ \nonumber
X_{\beta e} =& \begin{pmatrix} 
-  \xi( \mathrm{e}^{-\mathrm{i}\pi/4}z_2 + \mathrm{e}^{\mathrm{i}\pi/4}z_4)  +  \mathrm{i} \sqrt{ \xi^2-1} ( z_1 + z_3)   \\
 z_3^* + z_1^*
\end{pmatrix}, \,
&&X_{\beta o} = \begin{pmatrix} 
  \xi (- \mathrm{e}^{-\mathrm{i}\pi/4}z_2 + \mathrm{e}^{\mathrm{i}\pi/4}z_4) + \mathrm{i} \sqrt{ \xi^2-1}(z_1- z_3)   \\
 z_3^* - z_1^*
\end{pmatrix}.
\end{alignat}
Akin to the analysis above, we again compute the values of the ordered moment at each site but the analytical expressions in this case prove to be unwieldy. Specifically, $S^z_{\mu a}$ takes the form
\begin{align}
S^z_{\alpha e} =  \xi^2\left(\lvert z_1- \mathrm{i} z_3\rvert^2 + \lvert z_2 + z_4|^2\right) - 2 \lvert z_2 + z_4|^2 + 2 \xi \sqrt{ \xi^2 -1} \text{ Im}\left[ (z_2^* + z_4^*)(z_1 \mathrm{e}^{\mathrm{i}\pi/4} + z_3 \mathrm{e}^{-\mathrm{i}\pi/4}) \right], \\ \nonumber
S^z_{\alpha o} =  \xi^2\left(\lvert z_1 + \mathrm{i} z_3\rvert^2 + \lvert z_2 - z_4|^2\right) - 2 \lvert z_2 - z_4|^2 + 2 \xi \sqrt{ \xi^2 -1} \text{ Im}\left[ (z_2^* - z_4^*)(z_1 \mathrm{e}^{\mathrm{i}\pi/4} - z_3 \mathrm{e}^{-\mathrm{i}\pi/4}) \right], \\ \nonumber
S^z_{\beta e} =  \xi^2\left(\lvert z_2 +  \mathrm{i} z_4\rvert^2 + \lvert z_1 + z_3|^2\right) - 2 \lvert z_1 + z_3|^2 + 2 \xi \sqrt{ \xi^2 -1} \text{ Im}\left[ (z_1 + z_3)(z_2^* \mathrm{e}^{\mathrm{i}\pi/4} + z_4^* \mathrm{e}^{-\mathrm{i}\pi/4}) \right], \\ \nonumber
S^z_{\beta o} =  \xi^2\left(\lvert z_2 -  \mathrm{i} z_4\rvert^2 + \lvert z_1 - z_3|^2\right) - 2 \lvert z_1 - z_3|^2 + 2 \xi \sqrt{ \xi^2 -1} \text{ Im}\left[ (z_1 - z_3)(z_2^* \mathrm{e}^{\mathrm{i}\pi/4} - z_4^* \mathrm{e}^{-\mathrm{i}\pi/4}) \right] .
\end{align}
As can be seen, for general complex values $z_i$, there is no simple relation between the $z$ components. Further, 
\begin{equation}
\sum_{\mu = \alpha, \beta}\sum_{a = e,o} \langle S^z_{\mu,a} \rangle = 4( \xi^2 -1) \sum_{i=1}^{4}\lvert z_i|^2 + 4  \xi\sqrt{ \xi^2-1} \text{ Im}\left[z_1 z_2^* \mathrm{e}^{\mathrm{i} \pi/4} + z_3 z_4^* \mathrm{e}^{-\mathrm{i} \pi/4} \right] 
\end{equation} 
vanishes only for $ \xi = 1$. Therefore, the sum of ordered moments on the four sites of a plaquette is nonzero, and the spin order parameter can be parametrized as
\begin{equation}
\langle \bm{S}(j) \rangle = \n_{(0,0)} + (-1)^{j_x} \n_{(\pi,0)} + (-1)^{j_y} \n_{(0,\pi)} + (-1)^{j_x+j_y}  \n_{(\pi,\pi)},
\end{equation}
where we have defined
\begin{align}
 \nonumber \n_{(0,0)} &=  \frac{1}{4}\bigg( \langle \bm{S}_{\alpha e} \rangle + \langle \bm{S}_{\alpha o} \rangle + \langle \bm{S}_{\beta e} \rangle + \langle \bm{S}_{\beta o} \rangle \bigg), \,
&& \n_{(\pi,0)} =  \frac{1}{4}\bigg( \langle \bm{S}_{\alpha e} \rangle - \langle \bm{S}_{\alpha o} \rangle + \langle \bm{S}_{\beta e} \rangle - \langle \bm{S}_{\beta o} \rangle \bigg), \\
\n_{(0,\pi)} &=  \frac{1}{4}\bigg( \langle \bm{S}_{\alpha e} \rangle - \langle \bm{S}_{\alpha o} \rangle - \langle \bm{S}_{\beta e} \rangle + \langle \bm{S}_{\beta o} \rangle \bigg), \,
&&\n_{(\pi,\pi)} =  \frac{1}{4}\bigg( \langle \bm{S}_{\alpha e} \rangle + \langle \bm{S}_{\alpha o} \rangle - \langle \bm{S}_{\beta e} \rangle - \langle \bm{S}_{\beta o} \rangle \bigg).
\end{align}
It is noteworthy that $\n_{(0,0)} = 0$ exactly corresponds to the solution of Ref.~\onlinecite{yang2016schwinger} with zero average moment on a plaquette. The most general ordered state breaks $C_4$ and lattice translation ($T_x$ and $T_y$) symmetries but preserves the reflections $\mc{R}_x$ and $\mc{R}_y$; of course, it also breaks time reversal and SRI. While the moments on the four sites of each plaquette are generically distinct, previously studied states on the square lattice, like the N\'{e}el, the canted N\'{e}el, or the tetrahedral umbrella state \cite{hickey2017emergence} are not necessarily ruled out. If the structure of the condensate is such that $\n_{(\pi,\pi)}$ is large in magnitude compared to $\n_{(0,0)},\n_{(\pi,0)}$, and $\n_{(0,\pi)}$, the magnetically ordered state can be thought of as a perturbation to the N\'{e}el state, an example of which is sketched in Fig.~\ref{fig:Mag4} for $z_i = 0 \, \forall\, i \ne 1 $. The magnitude of the ordered moments is uniform at all lattice sites, i.e. $X^\dagger_{\mu a} X^{}_{\mu a} = \mathrm{constant}$, if we choose such a solution for the $z_i$. One can also impose this requirement of uniformity when more than one coefficient is nonzero. Endowed with this constraint, there are four solutions, which are $\{ z_1, z_2, z_3, z_4 \} = \{z, \pm \mathrm{i}z, 0, 0 \}$ or $ \{0, 0, z, \pm \mathrm{i}z \}$. The two associated symmetry-inequivalent ordered states are shown in Figs.~\ref{fig:Mag1} and \ref{fig:Mag2}.

}
\end{widetext}

\section{SBMFT with Dzyaloshinskii-Moriya interactions}
\label{sec:DM_Calc}

In this appendix, we continue along the lines of Sec.~\ref{sec:DM} to develop the mean-field Hamiltonian for the nearest-neighbor Heisenberg antiferromagnet with additional Dzyaloshinskii-Moriya couplings. Pursuant to Eq.~\eqref{eq:DM_identity}, the mean-field approximation for the in-plane DM term is
\begin{alignat}{1}
&\nonumber H_{\textsc{mf}}^{(3)} = \frac{D_{\parallel}}{2} \sum_{\langle i, j \rangle} \left(\mc{B}^*_{i,j} \hat{\mc{C}}_{i,j} + \hat{\mc{C}}^\dagger_{i,j} \mc{B}_{i,j} + \mc{A}^*_{i,j} \hat{\mc{D}}_{i,j} + \hat{\mc{D}}^\dagger_{i,j} \mc{A}_{i,j} \right)\\
\label{eq:H3mf}
&=  \frac{D_{\parallel}}{2} \sum_{\langle i, j \rangle}-\frac{\mathrm{i}}{2} d_{ij} \mathrm{e}^{\mathrm{i} \sigma \theta_{ij}} \left(\mc{B}^*_{i,j} b^\dagger_{j -\sigma} b_{i \sigma}  + \sigma \mc{A}^*_{i,j}b_{i \sigma} b_{j \sigma}\right)+ \mathrm{H.c.}.
\end{alignat}
The total mean-field Hamiltonian $H_{\textsc{mf}}$ is now a sum of Eqs.~\eqref{eq:H1mf}, \eqref{eq:H2mf}, and \eqref{eq:H3mf}. All things considered, $H_{\mathrm{spin}}$ bears the mean-field momentum-space representation:
\begin{alignat}{2}
\label{eq:Hk}
\nonumber H_{\textsc{mf}}^{(1)} &= \sum_{{\bf k} \,\sigma\,\mu}&& \bigg[ J_\mu \mathrm{e}^{-\mathrm{i}k_\mu} \bigg(\frac{\mc{B}^*}{2} \, b^\dagger_{{\bf k} \sigma} \,b^{}_{{\bf k} \sigma} - \sigma \frac{\mc{A}^*}{2} \, b^{}_{{\bf k} \sigma}\, b^{}_{-{\bf k} -\sigma}  \bigg) + \mathrm{H.c.}\\
\nonumber & &&+ \lambda\, b^\dagger_{{\bf k} \sigma}\, b^{}_{{\bf k} \sigma} \bigg] - 2 N_s \lambda S + 2 N_s J \left (\lvert \mc{A} \rvert^2 - \lvert \mc{B} \rvert^2 \right),\\
H_{\textsc{mf}}^{(2)} &= -\frac{B}{2} &&\sum_{{\bf k} \,\sigma} \sigma\,b^\dagger_{{\bf k} \sigma}\, b^{}_{{\bf k} \sigma},\\
\nonumber H_{\textsc{mf}}^{(3)} &= \frac{D^{\textsc{m}}}{4}  &&\sum_{{\bf k} \,\sigma} \left [ -\mathrm{i} \,\bar{\mc{E}}_\sigma  \left( \mc{B}^*  b^\dagger_{{\bf k} -\sigma}\, b^{}_{{\bf k} \sigma} + \sigma\,\mc{A}^* b^{}_{{\bf k} \sigma}\, b^{}_{-{\bf k} \sigma}\right) +\mathrm{H.c.}\right ].
\end{alignat}
For the sake of notational brevity, we work with the shorthand $\mc{E}_\sigma \equiv (\mathrm{e}^{\mathrm{i}\,k_x} + \mathrm{i}\,\sigma\, \mathrm{e}^{\mathrm{i}\,k_y})$ and overhead bars connote the same expressions but with the replacement $\mathbf{k} \rightarrow -\mathbf{k}$; thus, $(\mc{E}_+)^* = \bar{\mc{E}}_-$. Upon expanding and explicitly summing over $\sigma = \uparrow, \downarrow$ in Eq.~\eqref{eq:Hk}, the full Hamiltonian is expressible, as before, as  $H_{\textsc{mf}} = \sum_{\k} (\Psi_{\k}^\dagger\, \mathcal{H}({\k})\, \Psi_{\k})/2 $ with the spinor $\Psi^\dagger_{\bf k} \equiv \left(b^\dagger_{{\bf k} \uparrow}\, b^\dagger_{{\bf k} \downarrow} \, b_{-{\bf k} \uparrow}\,  b_{-{\bf k} \downarrow}  \right)$, and the kernel
\begin{widetext}
\begin{equation}
\mathcal{H}({\bf k}) = \left(
\begin{array}{cccc}
\left(\mc{B}\, J_\mu \mathrm{e}^{\mathrm{i} k_\mu}\right)^r +\left(\lambda-\frac{B}{2}\right) &  \frac{\mathrm{i}}{4} \,D_{\parallel}\left(\mc{B}\, \mc{E}_- - \mc{B}^*\,\bar{\mc{E}}_-\right)  & \frac{\mathrm{i}}{2} D_{\parallel} \mc{A}\,\mc{E}_- & -J\,\mc{A} \,E_+ \\
\frac{\mathrm{i}}{4} \,D_{\parallel}\left(\mc{B}\, \mc{E}_+ - \mc{B}^*\,\bar{\mc{E}}_+\right) & \left(\mc{B}\, J_\mu \mathrm{e}^{\mathrm{i} k_\mu}\right)^r+ \left(\lambda + \frac{B}{2}\right)  & J \mc{A}\, E_+ & -\frac{\mathrm{i}}{2} D_{\parallel}\, \mc{A}\,\mc{E}_+  \\
-\frac{\mathrm{i}}{2} D_{\parallel}\, \mc{A}^*\,\bar{\mc{E}}_+ & J\,  \mc{A}^*\,\overline{E}_{+} & \left(\mc{B}\, J_\mu \mathrm{e}^{-\mathrm{i} k_\mu}\right)^r +\left(\lambda -\frac{B}{2}\right)  &  \frac{\mathrm{i}}{4} \,D_{\parallel}\left(\mc{B}\, \bar{\mc{E}}_+ - \mc{B}^*\,\mc{E}_+\right) \\
-J\, \mc{A}^*\,\overline{E}_{+} & \frac{\mathrm{i}}{2} D_{\parallel}\mc{A}^* \bar{\mc{E}}_- & \frac{\mathrm{i}}{4} \,D_{\parallel}\left(\mc{B}\, \bar{\mc{E}}_- - \mc{B}^*\,\mc{E}_-\right)  & \left(\mc{B}\, J_\mu \mathrm{e}^{-\mathrm{i} k_\mu}\right)^r+\left(\lambda +\frac{B}{2}\right) \\
\end{array}
\right),
\end{equation}
\end{widetext}
where the superscript $r$ stands for the real part; $H_{\textsc{mf}}$ further includes another constant piece, which we ignore. Diagonalizing with the paraunitary matrix $\TM_\mathbf{k}$ gives the full information of the dispersions for the volume-mode bands and some representative energy dispersions are shown in Fig.~\ref{fig:Evsk3D}.

\begin{widetext}
{\setstretch{1.075}
\section{Three-orbital model}
\label{3OrbitalModel}

The three-orbital CuO$_2$ model---with the broken time-reversal and reflection symmetries of pattern D---allows for nonzero loop currents unlike its one-orbital counterpart \cite{scheurer2018orbital} studied in Sec.~\ref{sec:1O}, and offers the added advantage of an explicitly translation-invariant ansatz. In this appendix, we illustrate that the three-orbital model also shows a large thermal Hall conductivity in the presence of a magnetic field, analogous to the one-orbital model, with identical broken symmetries as in Sec.~\ref{AnsatzWithChernNumbers}. 

\begin{figure}[htb]
    \centering
    \subfigure[]{\label{fig:3Oa}\includegraphics[width = 0.3\linewidth]{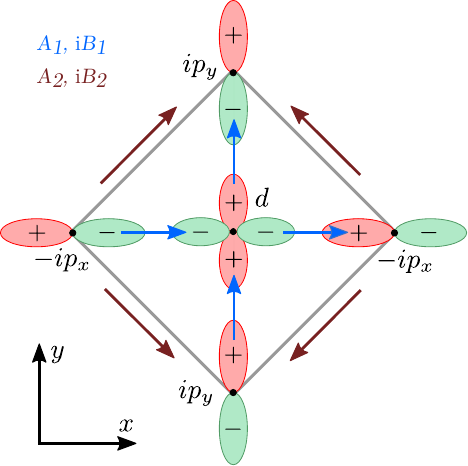}}\qquad\qquad
    \subfigure[]{\label{fig:3Ob}\includegraphics[width = 0.3\linewidth]{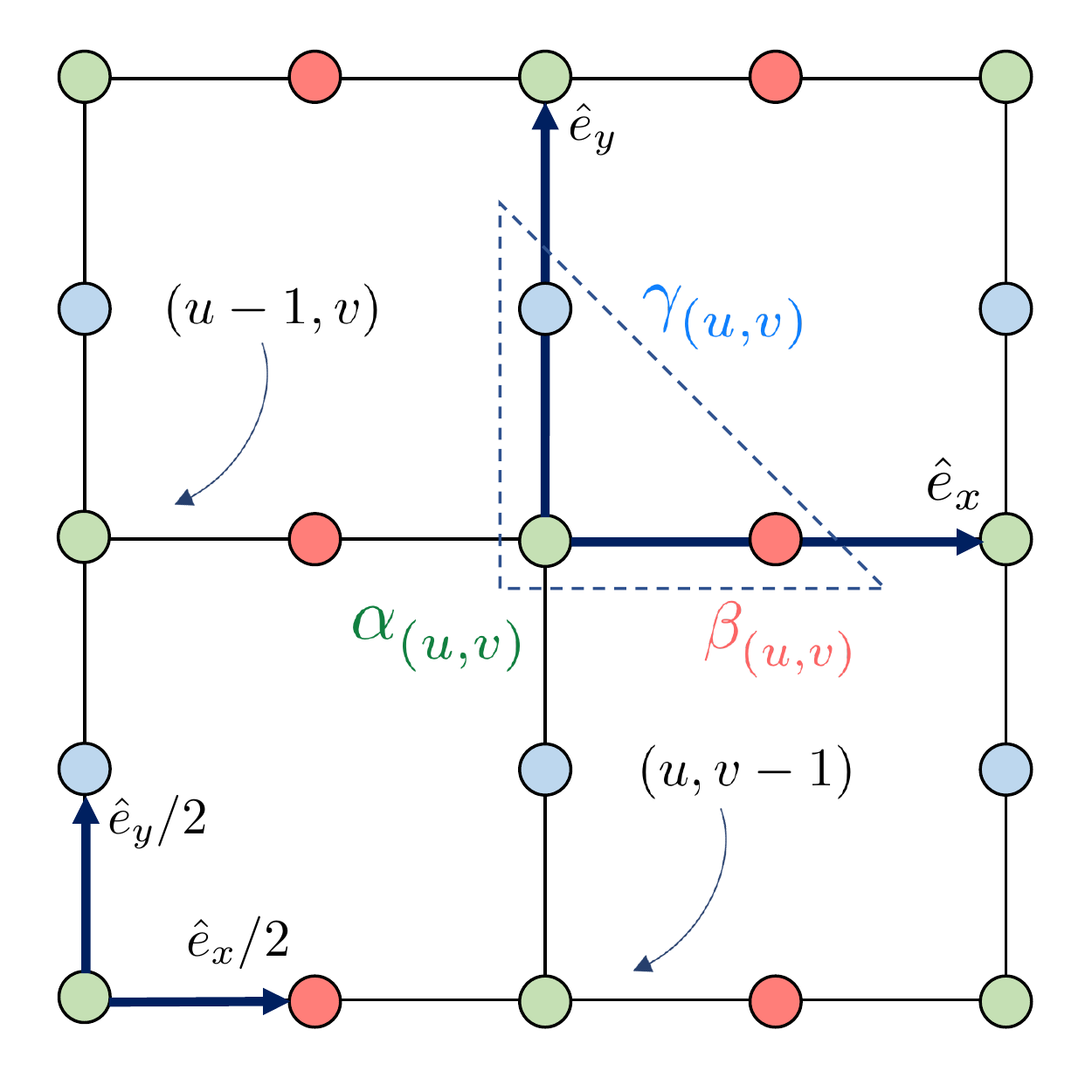}}
    \caption{(a) Schwinger boson mean-field ansatz for the three-orbital model. Since the ansatz is explicitly translation invariant, we show only one Cu atom and its four neighboring oxygen atoms. The arrows indicate the directionality of the bonds, which is required for specifying $\mathcal{A}_{i,j}$ (odd under $i \leftrightarrow j$) and complex $\mc{B}_{i,j}$ (as $\mc{B}^*_{i,j}=\mc{B}_{j,i}$). (b) The lattice conventions employed for this model. A single unit cell, shown here with dotted lines, consists of three sites, labeled $\alpha$, $\beta$, and $\gamma$.}
    \label{SBAnsatzPatternD}
\end{figure}

Let us consider the Schwinger-boson ansatz for this model, illustrated in Fig.~\ref{fig:3Oa}. More explicitly, in the mean-field Hamiltonian \eqref{eq:H1O}, the only bond operator expectation values are
\begin{alignat}{2}
    \mc{A}_{j,j\pm\frac{\hat{e}_\mu}{2}}=\pm \mc{A}_1, \quad \mc{B}_{j,j\pm\frac{\hat{e}_\mu}{2}}=\pm \mathrm{i}\, \mc{B}_1,\quad \mc{A}_{j\pm\frac{\hat{x}}{2},j+\frac{\hat{y}}{2}}=\mc{A}_{j\pm\frac{\hat{x}}{2},j-\frac{\hat{y}}{2}}=\mc{A}_2,\, \quad \mc{B}_{j\pm\frac{\hat{x}}{2},j+\frac{\hat{y}}{2}}=\mc{B}_{j\pm\frac{\hat{x}}{2},j-\frac{\hat{y}}{2}}= \mathrm{i}\,\mc{B}_2,
\end{alignat}
where $\mc{A}_{j, k},i\mc{B}_{j,k}\in\mathbb{R}$. The basis vectors of the direct lattice are $\hat{e}_\mu; \, \mu=x,y$, and we adopt the convention that integer-valued (half-integer-valued) lattice indices refer to copper (oxygen) sites (see Fig.~\ref{fig:3Ob}). The state with only $\mc{A}_1$ (or also $\mc{B}_1$) nonzero has the full symmetries of the square lattice but turning on $\mc{B}_2$ (and/or $\mc{A}_2$) breaks the symmetries down to $\frac{4}{m}m'm'$. 
Denoting the three sites of a unit cell, at position $(u,v)$, as $\alpha^{}_{(u,v)}$, $\beta^{}_{(u,v)}$, $\gamma^{}_{(u,v)}$ and expanding, the mean-field Hamiltonian is
\begin{alignat}{2}
\nonumber H_{\textsc{mf}} = &\sum_{(u,v),\, \sigma}&&\frac{J}{2} \bigg[\bigg ( \mathrm{i} \,\mc{B}_1\, \alpha^\dagger_{(u,v) \sigma} \beta^{}_{(u,v) \sigma} + \mathrm{i}\, \mc{B}_1 \,\alpha^\dagger_{(u,v) \sigma} \gamma^{}_{(u,v) \sigma} - \mathrm{i}\, \mc{B}_1\, \alpha^\dagger_{(u,v) \sigma} \beta^{}_{(u-1, v) \sigma} - \mathrm{i}\, \mc{B}_1 \,\alpha^\dagger_{(u,v) \sigma} \gamma^{}_{(u, v-1) \sigma} \bigg)\\
\nonumber& &&-\bigg(\mc{A}^*_1\, \sigma \,\alpha^{}_{(u,v) \sigma} \beta^{}_{(u,v) -\sigma} + \mc{A}^*_1 \,\sigma\, \alpha^{}_{(u,v) \sigma} \gamma^{}_{(u,v) -\sigma} -\mc{A}^*_1 \,\sigma\, \alpha^{}_{(u,v) \sigma} \beta^{}_{(u-1, v) -\sigma}-\mc{A}^*_1\, \sigma\, \alpha^{}_{(u,v) \sigma} \gamma^{}_{(u, v-1) -\sigma}\bigg)\\
\nonumber & &&-\bigg (\mc{A}^*_2\, \sigma\, \beta^{}_{(u,v) \sigma} \gamma^{}_{(u,v) -\sigma} + \mc{A}^*_2\, \sigma\, \beta^{}_{(u-1, v) \sigma} \gamma^{}_{(u,v) -\sigma} +\mc{A}^*_2\, \sigma\, \beta^{}_{(u-1, v) \sigma} \gamma^{}_{(u, v-1) -\sigma}+\mc{A}^*_2\, \sigma\, \beta^{}_{(u,v) \sigma} \gamma^{}_{(u, v-1) -\sigma}\bigg)\\
\nonumber & &&+ \bigg ( \mathrm{i}\, \mc{B}_2 \,\beta^\dagger_{(u,v) \sigma} \gamma^{}_{(u,v) \sigma} + \mathrm{i} \,\mc{B}_2\, \beta^\dagger_{(u-1, v) \sigma} \gamma^{}_{(u,v) \sigma} + \mathrm{i}\, \mc{B}_2 \,\beta^\dagger_{(u-1, v) \sigma} \gamma^{}_{(u, v-1) \sigma} + \mathrm{i}\, \mc{B}_2 \,\beta^\dagger_{(u,v) \sigma} \gamma^{}_{(u, v-1) \sigma} \bigg) +\mathrm{H.c.} \bigg]\\
+ & \sum_{(u,v),\, \sigma} &&\lambda\, \bigg(\alpha^\dagger_{(u,v) \sigma} \alpha^{}_{(u,v) \sigma} + \beta^\dagger_{(u,v) \sigma} \beta^{}_{(u,v) \sigma} + \gamma^\dagger_{(u,v) \sigma} \gamma^{}_{(u,v) \sigma}  - 3 S\bigg).
\end{alignat}

After a Fourier transform to momentum space, this reads as (up to constants)
\begin{alignat}{2}
\nonumber H_{\textsc{mf}} = \sum_{\mathbf{k} \sigma} \bigg[&-J \mc{B}_1 \bigg(\alpha^\dagger_{\mathbf{k} \sigma} \beta_{\mathbf{k} \sigma} \,\mc{S}_x + \alpha^\dagger_{\mathbf{k} \sigma} \gamma_{\mathbf{k} \sigma} \,\mc{S}_y \bigg)
+ \mathrm{i}\, J \mc{A}^*_1\,\sigma \bigg(\alpha_{\mathbf{k} \sigma} \beta_{-\mathbf{k} -\sigma} \, \mc{S}_x + \alpha_{\mathbf{k} \sigma} \gamma_{-\mathbf{k} -\sigma}\, \mc{S}_y \bigg) + \mathrm{H.c.}\\
&+2J\,\mc{C}_x \mc{C}_y \left(\mathrm{i} \,\mc{B}_2 \beta^\dagger_{\mathbf{k} \sigma} \gamma_{\mathbf{k}  \sigma} - \mc{A}^*_2 \,\sigma\,\beta_{\mathbf{k} \sigma} \gamma_{-\mathbf{k}  -\sigma}  + \mathrm{H.c.}\right) + \lambda \bigg(\alpha^\dagger_{\mathbf{k} \sigma} \alpha_{\mathbf{k} \sigma} + \beta^\dagger_{\mathbf{k} \sigma} \beta_{\mathbf{k} \sigma} + \gamma^\dagger_{\mathbf{k} \sigma} \gamma_{\mathbf{k} \sigma} - 3S \bigg) \bigg],
\end{alignat}
where we use the shorthand $\mc{C}_\mu (\mc{S}_\mu) \equiv \cos\,(\sin)\, \frac{k_\mu}{2}$. Adding on an external magnetic field introduces the Zeeman term of Eq.~\eqref{eq:H2mf} and subsequently, the Hamiltonian can be expressed as
\begin{equation}
\label{eq:Kernel}
H_{\textsc{mf}} = \sum_{\bf k} \Psi_{\bf k}^\dagger\, \mathcal{H}({\bf k})\, \Psi_{\bf k}; \,\,\Psi^\dagger_{\bf k} \equiv \left(\alpha^\dagger_{{\bf k} \uparrow}\, \beta^\dagger_{{\bf k} \uparrow}\,\gamma^\dagger_{{\bf k} \uparrow} \, \alpha_{-{\bf k} \uparrow}\,  \beta_{-{\bf k} \downarrow} \,\gamma_{-{\bf k} \downarrow}  \right) ,
\end{equation}
with the kernel
\begin{equation}
\mc{H} (\mathbf{k})= \frac{1}{2}\left(
\begin{array}{cccccc}
 2 \lambda -B_z& -2 J\,\mc{B}_1\,  \mc{S}_x & -2 J\,\mc{B}_1\,  \mc{S}_y & 0 & -2 \mathrm{i} \,J\,\mc{A}_1\,  \mc{S}_x & -2 \mathrm{i}\, J\,  \mc{A}_1 \, \mc{S}_y \\
 -2 J\,\mc{B}_1\,  \mc{S}_x & 2 \lambda -B_z& 4 \mathrm{i} \,J\,\mc{B}_2 \,\mc{C}_x \,\mc{C}_y  & -2 \mathrm{i} \,J\,\mc{A}_1\,  \mc{S}_x & 0 & -4 J\, \mc{A}_2\, \mc{C}_x\, \mc{C}_y  \\
 -2 J\, \mc{B}_1\, \mc{S}_y & -4 \mathrm{i}\,J\, \mc{B}_2\, \mc{C}_x\, \mc{C}_y & 2 \lambda -B_z& -2 \mathrm{i} \,J\,\mc{A}_1 \,\mc{S}_y & 4 J\,\mc{A}_2\, \mc{C}_x\, \mc{C}_y & 0 \\
 0 & 2 \mathrm{i} J\,\mc{A}_1\, \mc{S}_x & 2 \mathrm{i}\,J\, \mc{A}_1\, \mc{S}_y & B_z+2 \lambda  & 2 J\,\mc{B}_1 \,\mc{S}_x & 2 J\,\mc{B}_1 \,\mc{S}_y \\
 2 \mathrm{i}\,J \, \mc{A}_1\, \mc{S}_x & 0 & 4 J\, \mc{A}_2\, \mc{C}_x\, \mc{C}_y & 2J\, \mc{B}_1\,  \mc{S}_x & B_z+2 \lambda  & -4 \mathrm{i}\,J\, \mc{B}_2\, \mc{C}_x\, \mc{C}_y \\
 2 \mathrm{i} \,J\,\mc{A}_1 \, \mc{S}_y & -4 J\, \mc{A}_2\, \mc{C}_x\, \mc{C}_y & 0 & 2 J\,\mc{B}_1  \,\mc{S}_y & 4 \mathrm{i}\, J\,\mc{B}_2\, \mc{C}_x\, \mc{C}_y & B_z+2 \lambda  \\
\end{array}
\right).
\end{equation}

\begin{figure}[bht]
\subfigure[]{\label{fig:3OB1}\includegraphics[width=0.245\linewidth]{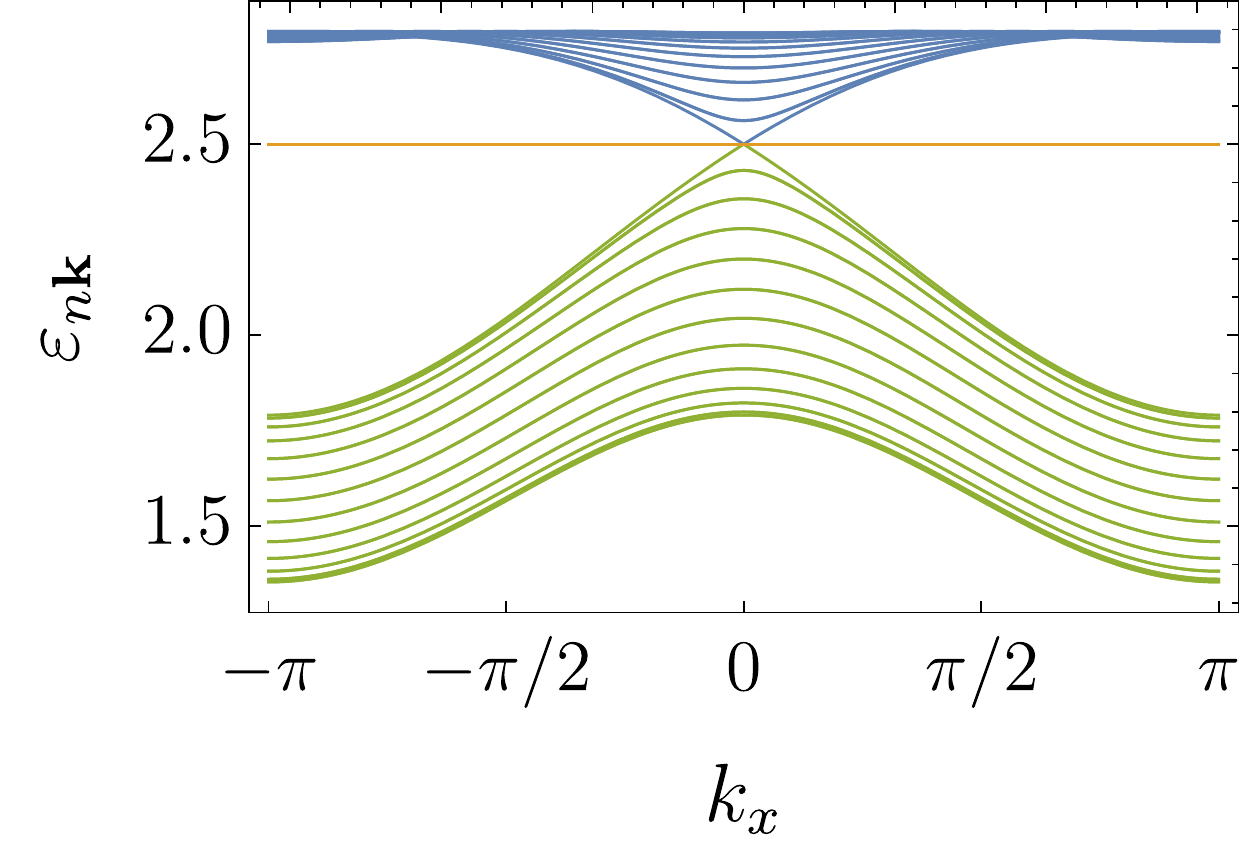}}
\subfigure[]{\label{fig:3OB2}\includegraphics[width=0.245\linewidth]{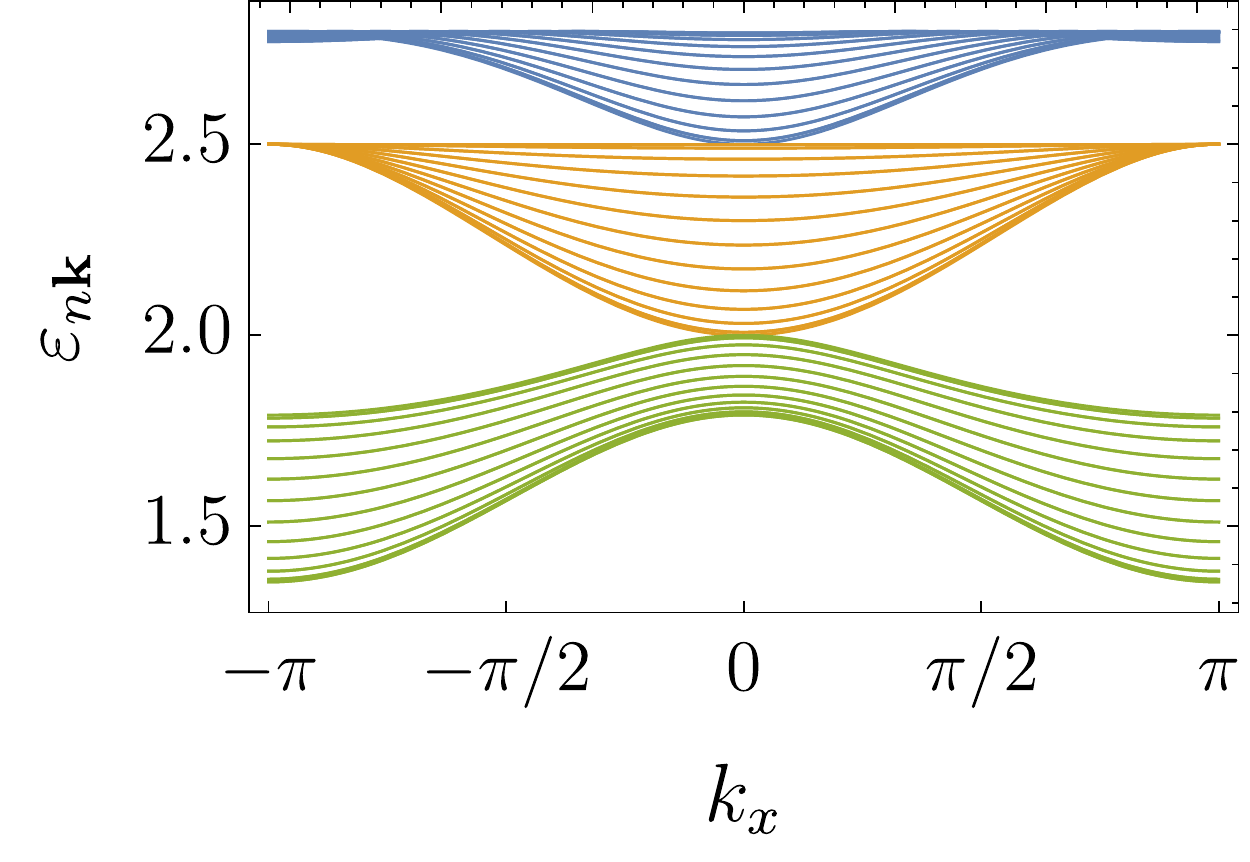}}
\subfigure[]{\label{fig:3OB3}\includegraphics[width=0.245\linewidth]{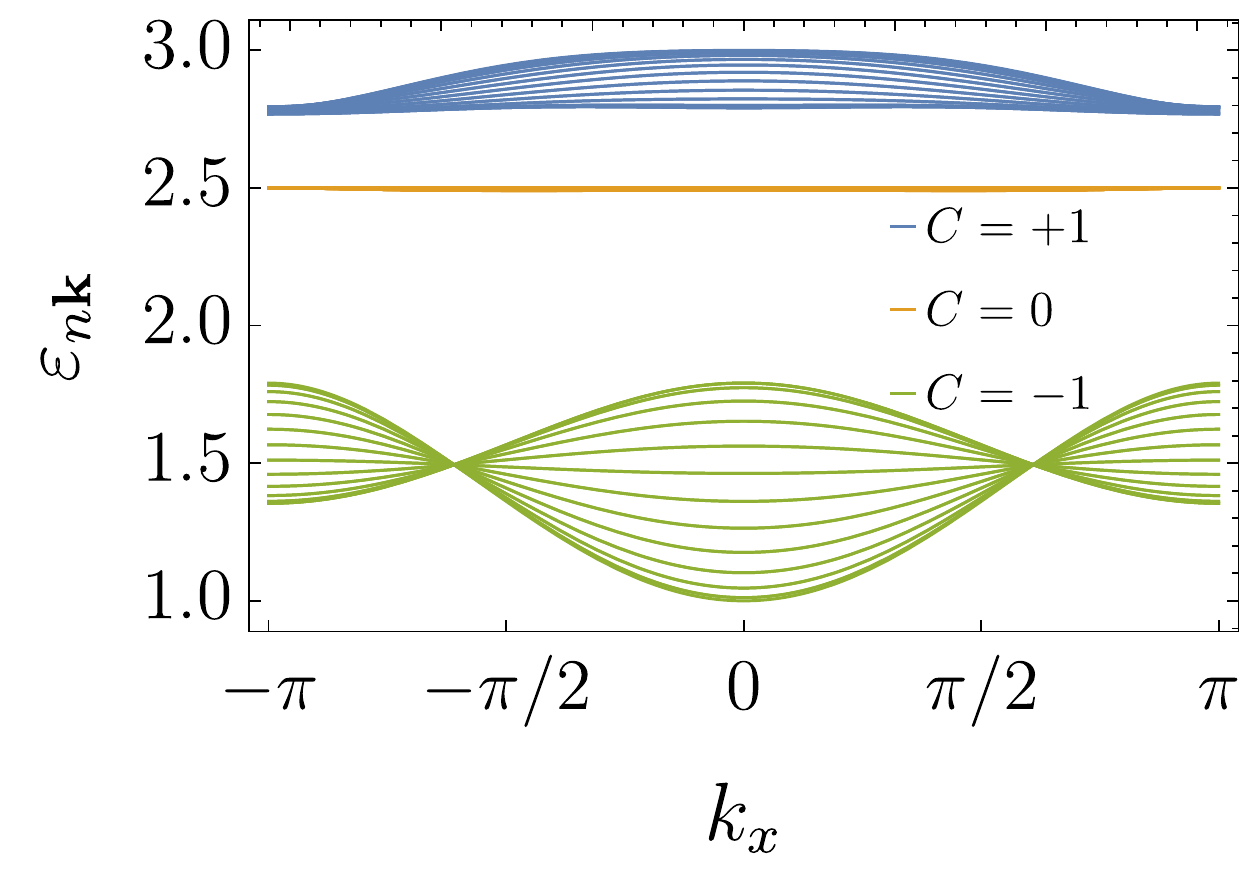}}
\subfigure[]{\label{fig:3OB4}\includegraphics[width=0.245\linewidth]{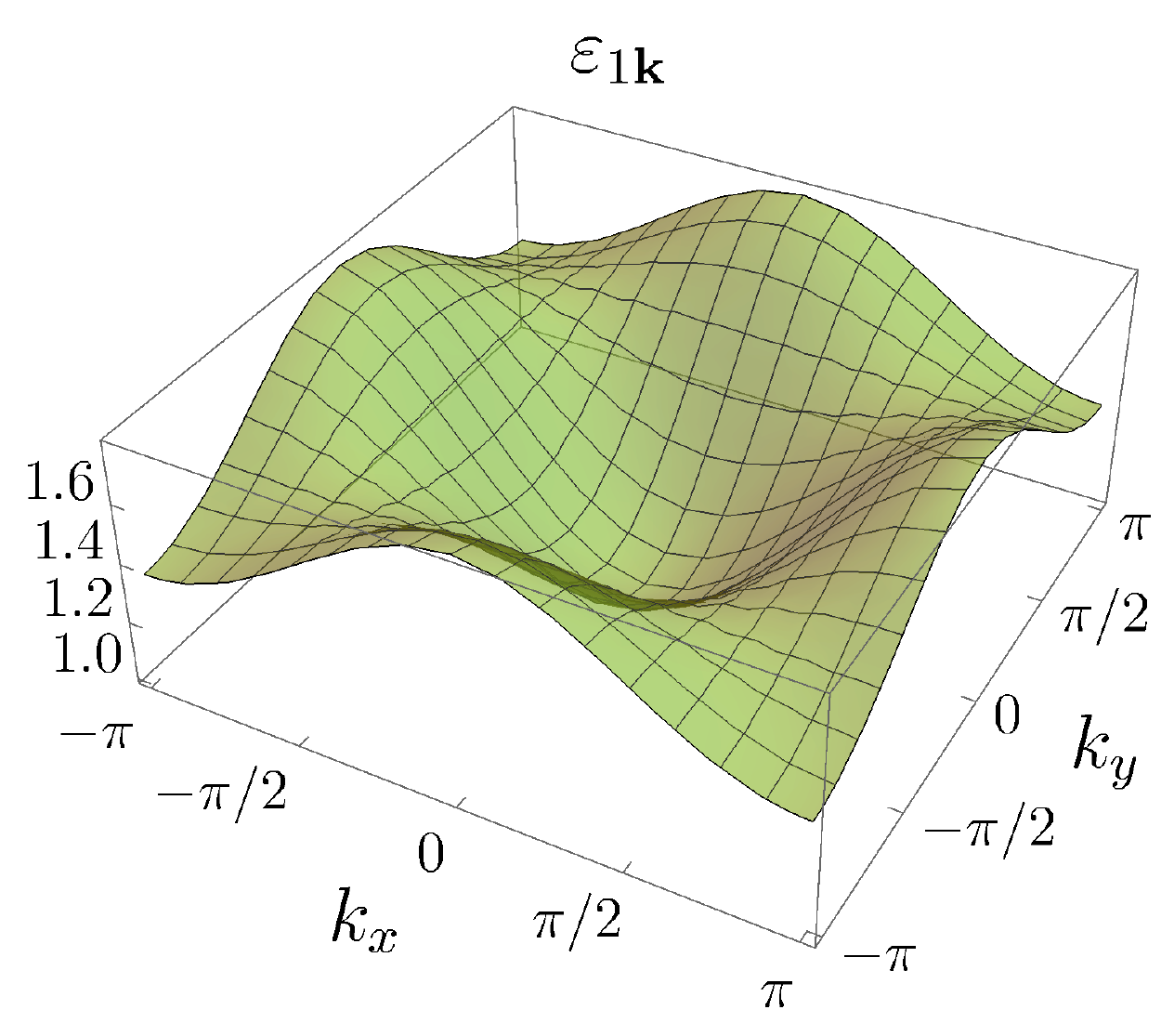}}
    \caption{Schwinger boson band structure for three of the six different particle bands with $J =1$, $\mc{A}_1 = 1$, $\mc{B}_1 = 0.5$, $\lambda = 2.5$, and $B_z=0$; the other bands are degenerate at zero field and are not shown. The remaining parameters are chosen as follows: (a) $\mc{A}_2 =0$, $\mc{B}_2 =0$; (b) $\mc{A}_2 =0.75 $, $\mc{B}_2 =0$; (c) $\mc{A}_2 =0.75$, $\mc{B}_2 =0.5$. Only with $\mc{B}_2 \ne 0$ are the upper bands prevented from touching; all the bands then acquire well-defined Chern numbers. The bands that are the degenerate counterparts of the ones shown have the same Chern numbers. (d) The dispersion for the lowest-energy band exhibits minima at ${\bf k} = (0, 0)$, signaling ferromagnetic order in the spin correlations.}
    \label{fig:3BandsPD}
\end{figure} 

This mean-field Hamiltonian can now be easily diagonalized, employing the standard methods formulated above---the resultant band structure is sketched in Fig.~\ref{fig:3BandsPD}.

\begin{figure}[htb]
\subfigure[]{\label{fig:3OC}\raisebox{5mm}{\includegraphics[height=0.23\linewidth]{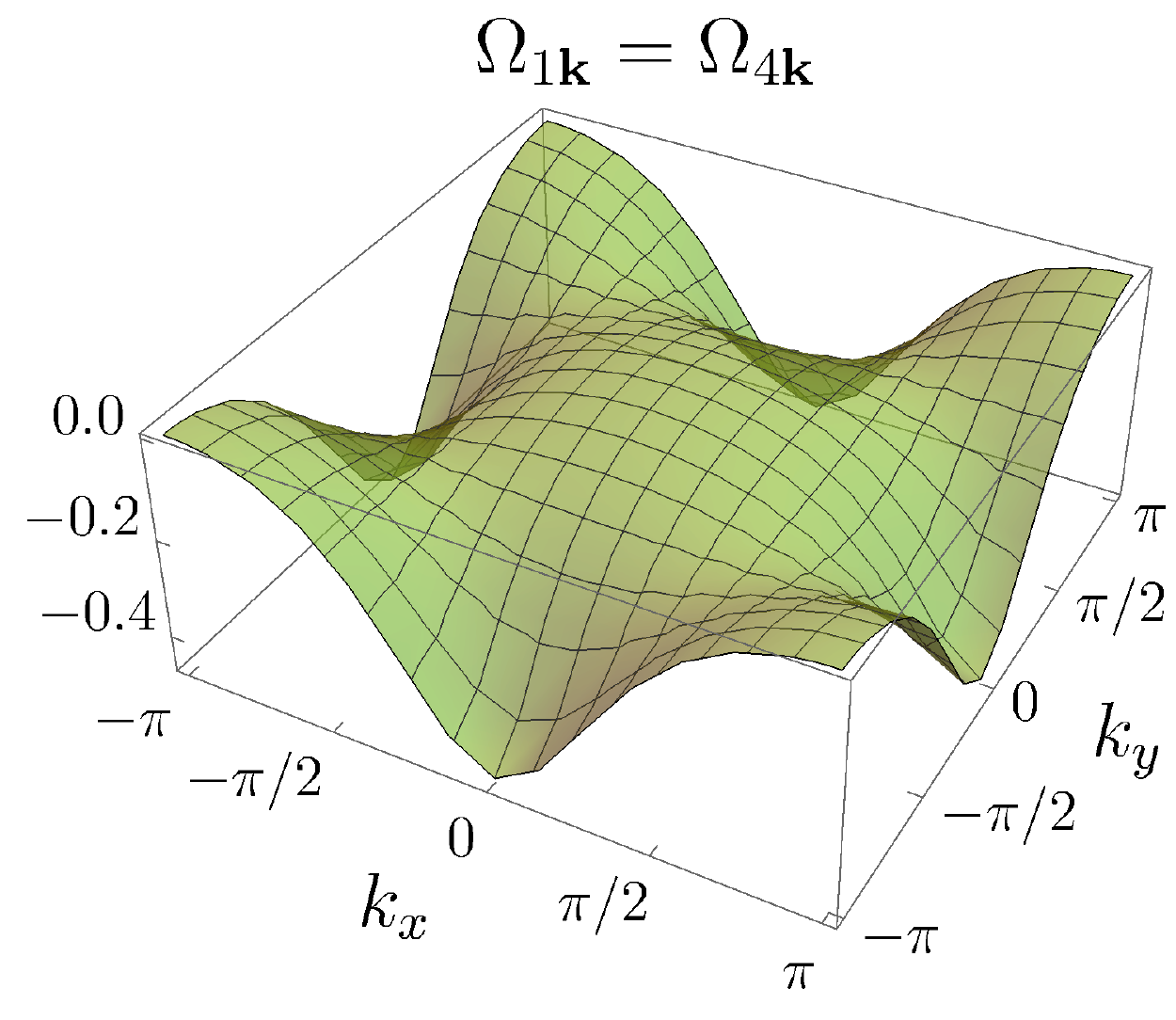}}
\raisebox{5mm}{\includegraphics[height=0.23\linewidth]{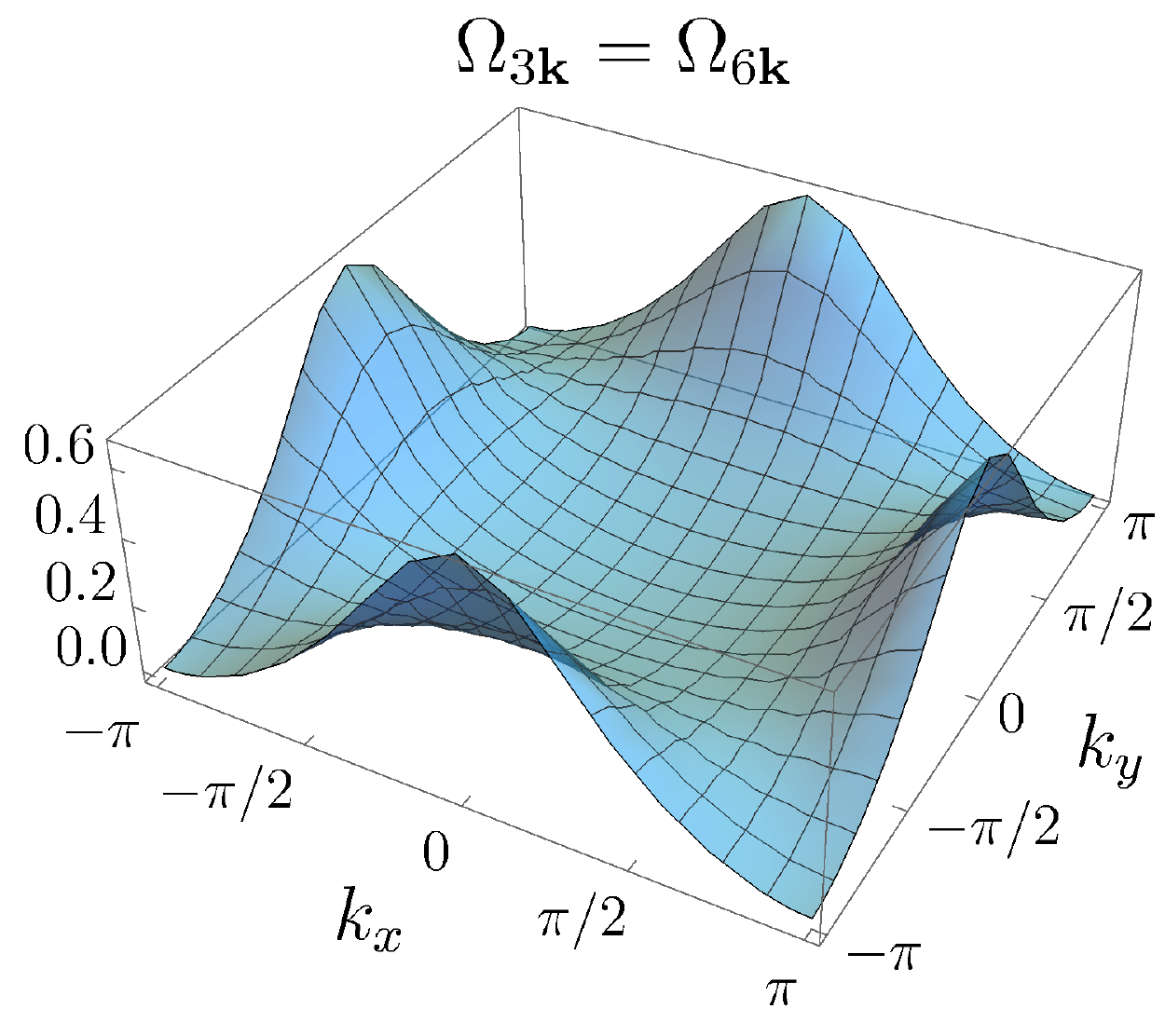}}}\quad
\subfigure[]{\includegraphics[height=0.26\linewidth]{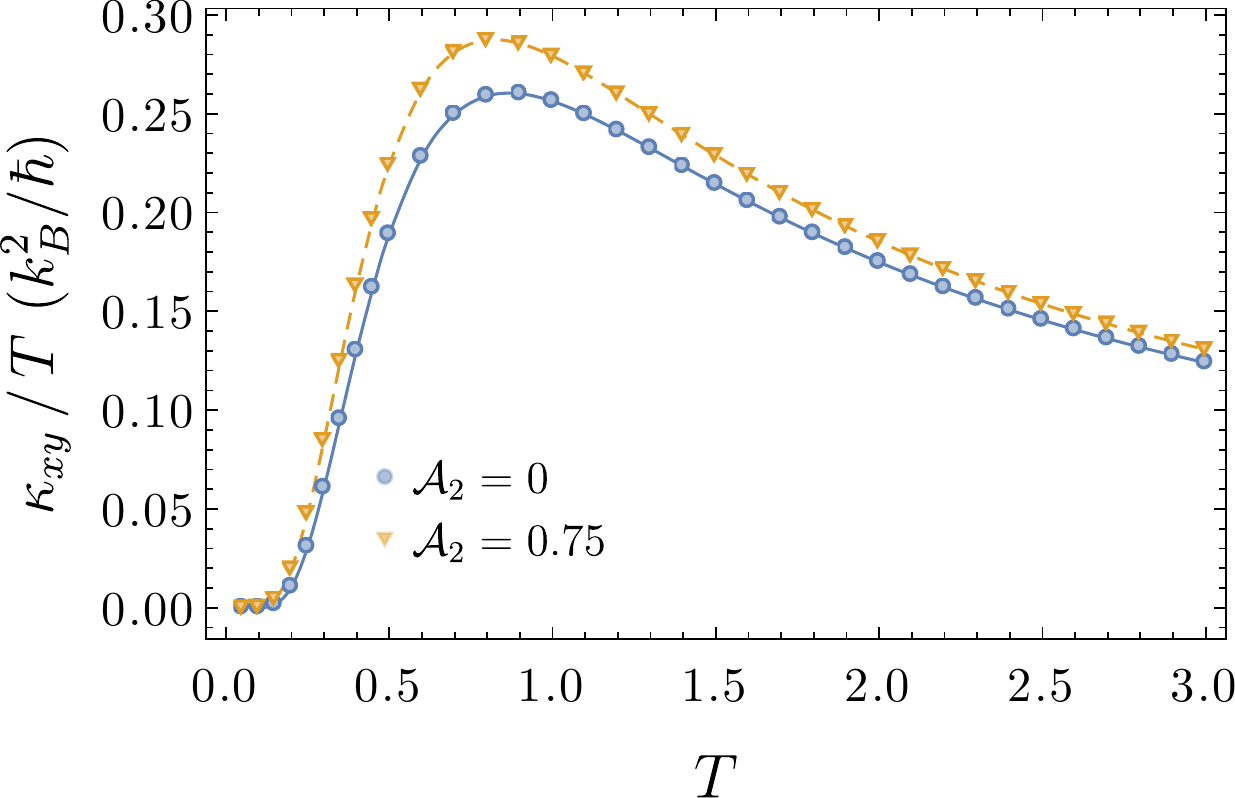}}
\subfigure[]{\includegraphics[height=0.26\linewidth,trim={3cm 19cm 6.5cm 2.5cm},clip]{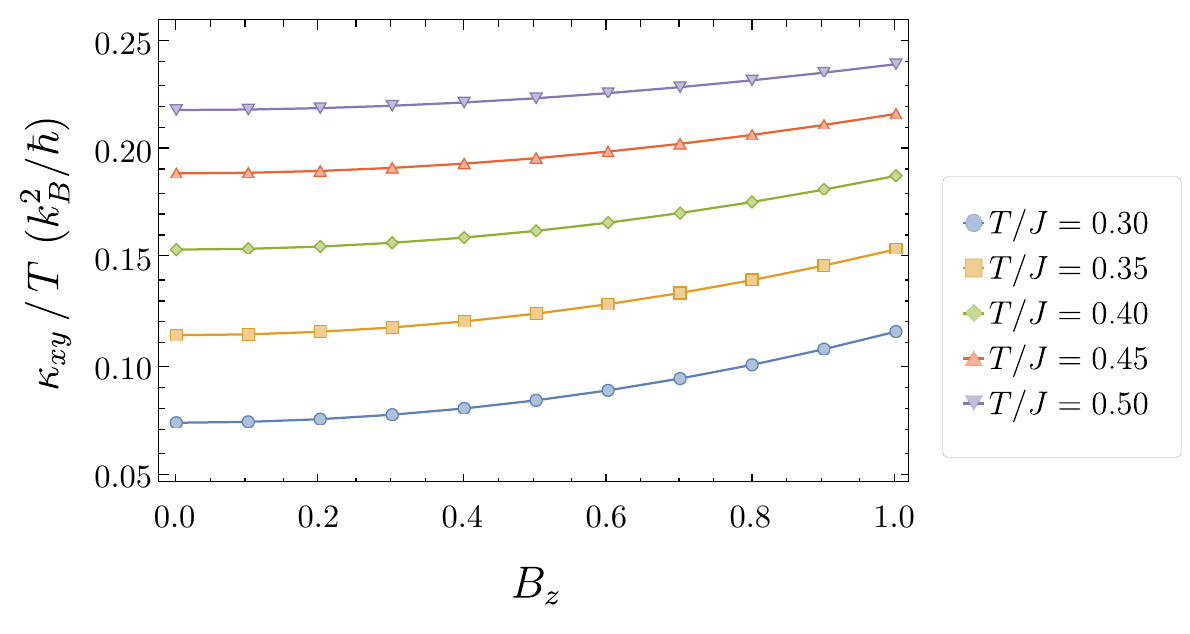}}
\subfigure[]{\includegraphics[height=0.26\linewidth,trim={3cm 19cm 6.5cm 2.5cm},clip]{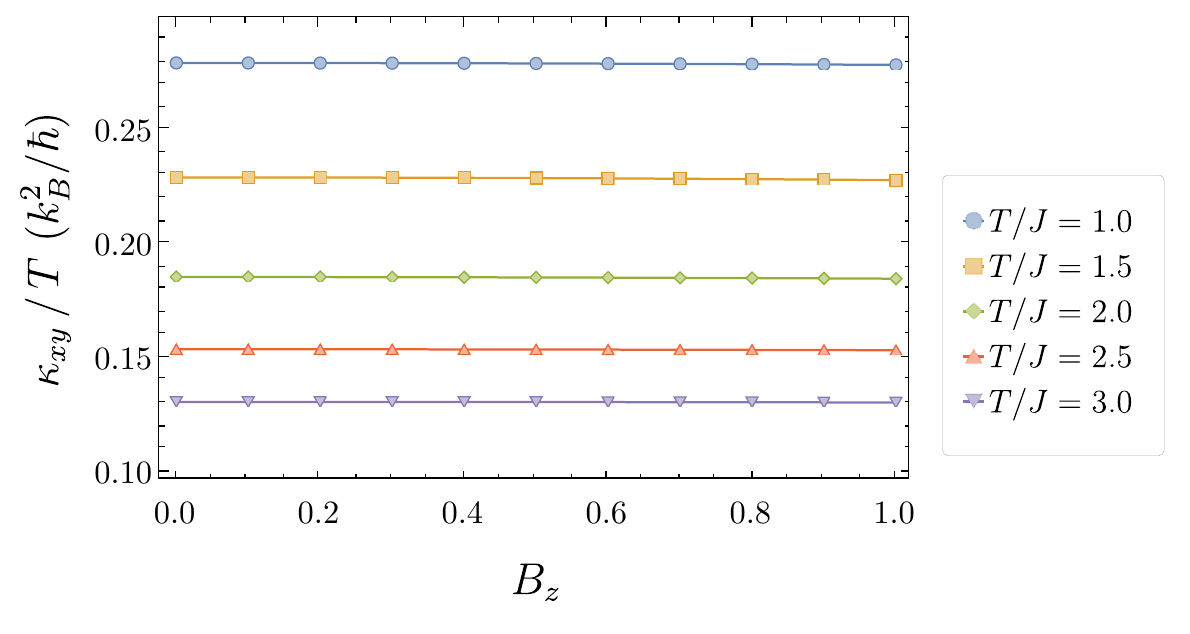}}
\caption{(a) Berry curvatures of the particle bands with nonzero Chern numbers  in the three-orbital model, with the parameters $J =1$, $\mc{A}_1 = 1$, $\mc{A}_2=0.75$, $\mc{B}_1 = \mc{B}_2 = 0.5$, $B_z = 0$, and $\lambda = 2.5$. (b) The thermal Hall conductivity as a function of temperature at fixed $B_z = 0.5$ for two (arbitrarily chosen) values of $\mc{A}_2$, indicating that the strength of the thermal Hall signal does vary with $\mc{A}_2$ even though the Chern numbers do not. (c--d) The magnetic field dependence of the conductivity for different temperatures with $\mc{A}_2=0.75$.}
    \label{fig:3OK_xyPD}
\end{figure}

In a like manner, from the paraunitary matrix $\TM_{\bf k}$, one can once again calculate the Berry curvature for these bands [Fig.~\ref{fig:3OC}] using the partition $H_1 = \{ \mathbf{k}: k_y < 0 \}$ and $H_2 = \{ \mathbf{k}: k_y \ge 0 \}$. The caveat is that the expression for the thermal Hall conductivity in Eq.~\eqref{eq:k_xy} is formulated exclusively in terms of particle bands whereas our choice of the six-component spinor in Eq.~\eqref{eq:Kernel} eliminates the trivial particle-hole duplication, leaving us with three particle and three hole bands. Exploiting the relation \eqref{eq:relate} between the curvatures of the particle and hole bands, Eq.~\eqref{eq:k_xy} can be brought to the more implementable form
\begin{equation}
\kappa^{}_{xy} = - \frac{k_B^2\,T}{\hbar\, V} \sum_{\mathbf{k}} \left[ \sum_{n\, \in \,\mathrm{particle}}\left \{ c_2 \left[ n_B \left(\varepsilon_{n \mathbf{k}} \right)\right] - \frac{\pi^2}{3}\right\} \Omega_{n \mathbf{k}} - \sum_{n \,\in\, \mathrm{hole}} \left \{ c_2 \left[ n_B \left(\varepsilon_{n -\mathbf{k}} \right)\right] - \frac{\pi^2}{3}\right\} \Omega_{n -\mathbf{k}} \right].
\end{equation}
Summing over all six bands, the net conductivity in Fig.~\ref{fig:3OK_xyPD} is observed to be three orders of magnitude greater than in the model with Dzyaloshinskii-Moriya interactions alone. The behaviors at both high and low temperatures resemble that for the one-orbital model in Fig.~\ref{fig:1OTH} and is owed to origins similar to the discussion in  Sec.~\ref{sec:asymptotic}. Furthermore, we again find an anomalous contribution.

}
\end{widetext}

\bibliographystyle{apsrev4-1_custom}
\bibliography{thermalHallSB}

\end{document}